\documentclass[final, review, 12pt]{elsarticle}

\usepackage{graphicx}
\usepackage{amssymb}
\usepackage{color}

\usepackage{lineno}
\usepackage{amsmath}
\usepackage{textgreek}
\usepackage{array}
\usepackage{tabularx}
\usepackage{lscape}
\usepackage{multirow}
\usepackage{footnote}
\usepackage[table]{xcolor}
\usepackage{paralist}
\usepackage[font={scriptsize,it}]{caption}
\usepackage{sidecap} 
\usepackage{hyperref}

\biboptions{comma,square}

\journal{Icarus}

\begin{document}

\begin{frontmatter}


\title{Reflectance study of ice and Mars soil simulant associations -- II. CO$_2$ and H$_2$O ice.}

\tnotetext[label1]{\url{https://doi.org/10.1016/j.icarus.2022.115116}}


\author{Zuri\~ne Yoldi\textsuperscript{1}*}
\author{Antoine Pommerol\textsuperscript{1}}
\author{Olivier Poch\textsuperscript{1,2}}
\author{Nicolas Thomas\textsuperscript{1}}

\address{\textsuperscript{1}Physikalisches Institut, Universit\"at Bern and NCCR PlanetS, Sidlerstrasse 5, 3012 Bern, Switzerland.}
\address{\textsuperscript{2}Univ. Grenoble Alpes, CNRS, IPAG, 38000 Grenoble, France}
\address{\textsuperscript{*} Now at the section for the Physics of Ice, Climate and Earth at the Niels Bohr Institute, University of Copenhagen.}

\begin{abstract}
We measure the visible and near-infrared reflectance of icy analogues of the Martian surface made of CO$_2$ ice associated in different ways with H$_2$O ice and the regolith simulant JSC Mars-1. Such experimental results obtained with well-controlled samples in the laboratory are precious to interpret quantitatively the imaging and spectral data collected by various Mars orbiters, landers and rovers. Producing and maintaining well-characterized icy samples while acquiring spectro-photometric measurements is however challenging and we discuss some of the difficulties encountered in preparing and measuring our samples. We present the results in the form of photometric and spectral criteria computed from the spectra and plotted as a function of the composition and physical properties of the samples. Consistent with previous studies, we find that when intimately mixed with other materials, including water ice, CO$_2$ ice becomes rapidly undetectable due to its low absorptivity. As low as 5 wt.\% of fine-grained H$_2$O ice is enough to mask entirely the signatures of CO$_2$. Similarly, sublimation experiments performed with ternary mixtures of CO$_2$ ice, H$_2$O ice and JSC Mars-1 show that water, even when present as a minor component (3 wt.\%), determines the texture and evolution of the mixtures. We assess the ability of various combinations of spectral parameters to identify samples with H$_2$O, CO$_2$, JSC Mars-1, or various mixtures from their reflectance and orient our study to helping interpret ice and soil reflectance spectra from the Martian surface. From the laboratory spectra, we simulate the colour signal generated by the CaSSIS instrument to allow for direct comparisons with results from this instrument and provide to databases the necessary spectral data to perform the same operations with other instruments. 
\end{abstract}

\begin{keyword}
Reflectance \sep CO$_2$ ice \sep Mixture \sep Mars \sep H$_2$O ice \sep Polar Caps


\end{keyword}

\end{frontmatter}


\section{Introduction} \label{S:Introduction}

\subsection{CO$_2$ ice and the martian CO$_2$ cycle} \label{S:Intro_a} 

Solid carbon dioxide (CO$_2$ ice) is present at the surface of various bodies of the Solar System, such as comet 67P-CG \citep{Filacchione:2016}, Triton \citep{Cruikshank:2016} or Mars. In the case of Mars, where it represents 96$\%$ of the atmosphere \citep{Mahaffy:2013}, CO$_2$ drives the condensation flow \citep{Pollack:1990}: as a result of the current obliquity of Mars ($\sim$25$^\circ$), its hemispheres do not receive the same amount of sunlight, which results in seasonal differences of temperature and pressure conditions in each hemisphere. As a consequence, a large fraction of the atmosphere moves seasonally from the summer hemisphere to the winter hemisphere and back. Almost a quarter of the atmosphere condenses in winter \citep{Hourdin:1995}, sublimates during spring and migrates to the other hemisphere, where it condenses again to continue the cycle. Beside the CO$_2$ cycle \citep{James:1992}, H$_2$O \citep{Jakosky:1985} and dust \citep{Kahn:1990} cycles happen yearly within the Martian atmosphere. CO$_2$ frost has also been detected at all latitudes on Mars \citep{Piqueux:2016}.

In 1971, \citet{Neugebauer:1971} confirmed with Mariner 6 and 7 data the predictions previously made by \citet{Leighton:1966}: the southern seasonal cap was made of CO$_2$.  Since then, the community has grown a broad knowledge of the CO$_2$ and H$_2$O condensation/sublimation processes by monitoring the high-latitude regions of Mars with multiple techniques (e.g., gravimetry, gamma-ray and neutron spectrometer, laser altimetry, reflectance spectroscopy or temperature measurements). We know that at the martian mid-to-high latitudes, water ice starts condensing as autumn arrives, first as night-time frost and later as seasonal frost. Once in autumn, the condensation temperature of CO$_2$ is attained at the surface (150K at 6 mbar), causing CO$_2$ to directly condense on the regolith, or in the atmosphere, causing solidified CO$_2$ to deposit as frost. In the latter case, CO$_2$ can nucleate around atmospheric particles of water ice or dust, resulting in the cleaning of the atmosphere and contaminated CO$_2$ ice at the surface. As spring starts, the first light from the Sun causes the seasonal deposits to sublimate, first the CO$_2$ and then H$_2$O. By mid-summer, all the seasonal deposits have sublimated.

\subsection{Parameters controlling the reflectance - Models} \label{S:Intro_b}

Comparisons of the optical constants of H$_2$O and CO$_2$ ices \citep{Warren:1984, Warren:1986, Hansen:1997, Hansen:2005} already suggest that their spectro-photometric properties will be relatively similar with very low absorption in the visible spectral range and intense and wide absorption bands in the near- to mid-infrared. Different types of radiative transfer models and numerical simulations \citep{Bohren:1974, Wiscombe:1980, Hapke1993, Singh:2016, khuller:2021} can then be used to compute the reflectance of a parameterized surface made of the ices, pure or mixed with contaminants, from their optical constants. The relevant parameters are related to both the individual grains (size, shape, surface texture, internal defects and inclusions...) and the macroscopic properties of the surface (density, roughness...). Because of the similarity of their optical properties, a lot of what is known about water ice can be extrapolated to carbon dioxide ice while one keeping an eye on the known differences.

The physical size of the ice articles plays a very large role in the reflectance of icy surfaces. In the visible spectral range, fine grained ($\lesssim$ 100 \textmu m) frost and snow made of both CO$_2$ and H$_2$O in their pure form present a very high reflectance and thin ($\lesssim$ 1 cm) layers are sufficient to hide the substrate and reflect nearly 100$\%$ of the incident light. The bidirectional reflectance behaviour of water snow is also observed to be nearly lambertian, i.e. the measured radiance does not depend on the position of the observer and shows a cosine dependence on the light incidence angle. Coarse grained and compact ice however can transmit the light to large depth and the effects of absorption can be observed because of the long optical path. This gives rise for instance to the blue colour of the compact ice in glaciers as the absorption index of H$_2$O increases with wavelength. In the near- and mid- infrared spectra of both H$_2$O and CO$_2$, the depths of the absorption bands are strongly influenced by particle size and the most intense bands saturate for coarse grained and compact ice. Note that  CO$_2$ displays sharper and less intense infrared absorption bands than H$_2$O overall resulting in a significantly higher albedo for a given particle size \citep{Singh:2016}.

Beside particle size, the other parameter which strongly influences the reflectance of icy surfaces is the presence of contaminants, even in trace amount and the way they are aggregated with the ice (intimate or geographic mixtures, coating at the surface of the grains...). It has already been shown in a variety of contexts that tiny amounts of dark contaminants are sufficient to hide the photometric signatures of water ice. For instance, \citet{Yoldi:2015} show experimentally that intimate mixtures of fine grained basalt powder and coarser grained water ice do not appear brighter than the pure basalt powder even with three times more ice than basalt in the mixture. \citet{khuller:2021} model binary mixtures of water ice and Martian dust and find that less than 1 wt$\%$ of dust mixed with fine grained ice can reduce its albedo by a factor of 10. 

\subsection{Observations of CO$_2$ ice in the southern hemisphere} \label{S:Intro_c}

In the southern Martian hemisphere, a small and thin ($<$ 20 m) permanent cap of CO$_2$ ice \citep{Thomas:2005} covers part of a much larger km-thick unit known as the South Polar Layered Deposits (SPLD). This unit is made essentially of water ice mixed with dust \citep{Titus:2003, Plaut:2007} but some ancient and massive deposits of carbon dioxide ice have also been found \citep{Bierson:2016}. The water ice is exposed at a few places, notably along scarps \citep{Titus:2003, Bibring:2004} but is otherwise covered by a desiccated dust layer.

In winter, seasonal ice deposits extend well beyond the limits of the SPLD to reach a latitude of about 50 degrees and recede during spring until only the permanent CO$_2$ cap is left in Summer. \citet{Calvin:1994} and later on \citet{Kieffer:2000} studied the evolution of the southern seasonal cap to discover that it was controlled by CO$_2$ grain size. They found both long and short optical path lengths, compatible with ice slabs (metres) and smaller (centimetres) grains of CO$_2$ ice, respectively. The size of the CO$_2$ ice grains is the primary cause for albedo variations. The presence of translucent slab ice in the so-called cryptic region at the beginning of spring is also hypothesised to be the reason for the intense jet activity observed and attributed to solid state greenhouse effect within the layer of translucent CO$_2$ ice \citep{Kieffer:2007}.

 \citet{Langevin:2007} used the Observatoire pour la Min{\'e}ralogie, l'Eau, les Glaces, et l'Activit{\'e} (OMEGA) on Mars Express to monitor the evolution of the southern seasonal cap from the southern winter solstice to mid-summer. The observations were consistent with the presence of transparent CO$_2$ slabs but also locally with events related to the sublimation and re-condensation of water ice. With the Compact Reconnaissance Imaging Spectrometer for Mars (CRISM, \citet{murchie:2007}) and the Mars Color Imager (MARCI, \citet{Bell:2009}), on Mars Reconnaissance Orbiter, \citet{Brown:2010} confirmed the presence of pure water ice deposits during the final stages of the recession of the southern seasonal cap. They also inferred CO$_2$ grain sizes of up to 70 $\pm$  10 mm before the springtime sublimation of the cap, and of 2.5-5 $\pm$ 1 mm in the residual cap. Using CRISM and the High Resolution Imaging Science Experiment (HiRISE, \citet{McEwen:2007}), \citet{Pommerol:2011} monitored the evolution of the south polar layered deposits from spring to summer. Around L$_s$ $sim$ 250 degrees, shortly before the southern summer, they observe an increase of both the albedo and the strength of the CO$_2$ ice bnd strength, which they attribute to a cleaning of the icy deposits as they sublimate. 

\citet{Brown:2014} studied, with CRISM, the origin of the water ice signature that had been repeatedly noticed at the southern cap during summer. They linked this signature with the deposition of atmospheric H$_2$O ice on the cap. They assessed this deposition to an equivalent H$_2$O layer of 0.2 mm and 70$\%$ porosity (0.013 $g/cm^2$). 

\subsection{Observations of CO$_2$ ice in the northern hemisphere} \label{S:Intro_d}

Contrary to the southern hemisphere where the SPLD are almost entirely buried below an optically thick layer of dust and the small CO$_2$ cap, the Northern Polar layered Deposits (NPLD) expose a vast surface of water ice to the atmosphere. This has important consequences for seasonal evolution of northern polar regions. While overall the seasonal processes are relatively similar in both hemisphere, significant differences exist. In particular, the seasonal deposits in the northern hemisphere contain much more water ice than in the southern hemisphere. Using THEMIS \citep{Christensen:2004} and TES \citep{Christensen:2001} data, \citet{Wagstaff:2008} identified and characterized the seasonal water ice annulus first hypothesised by \citep{Houben:1997}. 
\citet{Langevin:2005} monitored with OMEGA the composition of the northern latitudes of Mars. They linked albedo changes in early summer with an increase in the size of the ice grains. They also assessed the dust content within the ice to be low (a maximum of 5$\%$ if intimate mixtures and $<<$1$\%$ if intramixtures). \citet{Appere:2011} used the same instrument to study the retreat of the north seasonal deposits during two consecutive Martian years, and mapped the temporal and spatial distributions of CO$_2$ and H$_2$O ice. They observed CO$_2$ winter condensation compatible with transparent slabs, surrounded by the water ice annulus. This annulus consisted first of thin frost and was later covered by water ice grains that were previously trapped in CO$_2$. This surficial water-rich layer dominated the reflectance spectra of the deposits around L$_S$=50$^{\circ}$, even though the CO$_2$ locally reappeared between L$_S$=50$^{\circ}$ and L$_S$=70$^{\circ}$.

\citet{Hansen:2013}, \citet{Portyankina:2013} and \citet{Pommerol:2013} studied dynamic processes in the northern seasonal polar cap by analysing HiRISE images. \citet{Hansen:2013} and \citet{Portyankina:2013} showed that sublimation process play an active role in the erosion of dunes. They conclue that the same model which explains dynamic southern sublimation processes \citep{Kieffer:2007} is also applicable to fans and dark spots over dunes fields observed in the northern hemisphere. \citet{Pommerol:2013} confirmed with CRISM data the local reappearances of the CO$_2$ spectral signatures in late spring first seen by \citet{Appere:2011} and hypothesize that the removal of surficial water frost by local winds explains this evolution. 

\subsection{Laboratory measurements of CO$_2$ ice and H$_2$O ice mixtures} \label{S:Intro_e}

Numerical models can be used to establish the properties of the polar surfaces (see section \ref{S:Intro_b}). Many experiments have also been conducted on thin films and single crystals to measure transmission spectra of pure CO$_2$ ice or mixtures with H$_2$O \citep{Quirico:1997, Oeberg:2007}. However, there is a lack of measured reflectance data with macroscopic and well-controlled and documented CO$_2$-ice samples, which are needed to calibrate the models and validate the interpretations. This is explained by the technical challenges that working with large amounts of CO$_2$ ice entails: for example, its fast sublimation and the over-pressure that it might induce if enclosed or its propensity to cold trap water frost from the atmosphere. We summarise here the most relevant experimental studies on the reflectance of CO$_2$ ice.

In the late 60s, \citet{kieffer:1968, Kieffer:1970} carried out the first laboratory studies on the spectral reflectance of CO$_2$ and H$_2$O frosts in the near- to mid- infrared (0.8 - 3.2 \textmu m). By introducing gas of controlled composition into a cold chamber, he condensed the water and carbon dioxide on the walls and measured their reflectance from 0.8 to 3.2 \textmu m. He observed a masking of the CO$_2$ signatures by H$_2$O, thus concluding that a CO$_2$-free reflectance spectrum does not discard the presence of CO$_2$. While the resolution and quality of the data in the (1.7 - 3.0 \textmu m) range are sufficient for comparison with orbital data, the spectral resolution in the (0.8 - 1.7) \textmu m range is very coarse and does not allow to resolve the fine CO$_2$ features. The reflectance was not measured in the visible range, shortward of 0.8 \textmu m. Despite these issues, this unique data set remains one of the best sources of information on the spectral reflectance of CO$_2$ ice, half a century after its collection.

Similar to \citet{Kieffer:1970}, \citet{Bernstein:2005} mixed carbon dioxide and water in their gaseous phases and condensed them to grow frost. They observed a possible enhancement of a CO$_2$ feature at 2.135 \textmu m in mixtures of CO$_2$ with other materials, which they pointed out as a potential indicator of the presence of CO$_2$ in mixtures. In the absence of pictures, it is understood that the samples assessed in their study are thin layers of ice. 

\citet{Oancea2012} measured infrared reflectance spectra of mixtures between CO$_2$ and H$_2$O to study specifically the spectroscopic signal of CO$_2$ clathrate hydrates. They identified two bands of the clathrate at 2.71 and 4.28 \textmu m which can be used for the identifications of such clathrates in remote-sensing data.

The team from IPAG (Institut de Plan\'etologie et d'Astrophysique de Grenoble) has also presented reflectance measurements of different mixtures of CO$_2$ and H$_2$O ices. \citet{Grisolle:2011} condensed various thicknesses of water frost on a sample of CO$_2$ snow and studied its reflectance between 1.3-1.7 \textmu m. \citet{Sylvain:2015} and \citet{Bernard:2020} have performed sublimation experiments with polycristalline translucent CO$_2$ ice and dust and have shown that dust sinking and slab cracking only produce a slight increase of the continuum reflectance (by 7 \%) and a slight decrease of the CO$_2$ signatures. They find however that compact polycrystalline CO$_2$ ice becomes brighter as it sublimates, probably as a result of surface scattering at the ice/gas interfaces of the joints boundaries as the ice becomes more porous. This mechanism is hypothesised to explain the brightening of the Martian seasonal polar caps observed in late Spring.  

Finally, \citet{Anya:2019} simulated atmospheric conditions of pressure and temperature close to the Martian ones and condensed atmospheric CO$_2$ on both metallic and dusty surfaces. They assessed the way in which the CO$_2$ condensed (e.g. flakes or slab), and confirmed the possibility of CO$_2$ to form centimeters-thick, highly translucent slabs of ice depending on the conditions of temperature and pressure. 

\subsection{Need for additional reflectance measurements} \label{S:Intro_f}
While the spectra published more than 50 years ago by \citet{kieffer:1968, Kieffer:1970} are of excellent quality and with sufficient spectral resolution at wavelengths larger than 1.6 \textmu m, the shorter wavelength near-infrared is only covered with a few spectral bands and the visible range was not measured. The spectra were also normalised to 1, providing no information of the absolute reflectance level. The first objective of this work is to extend the spectral range for similar samples over the entire visible (0.4-1.0 \textmu m) and near-infrared (1-2.5 \textmu m) spectral ranges with high signal to noise and spectral resolution and sampling. Another main objective is to broaden the variety of solid CO$_2$ samples available for reflectance studies. In addition to CO$_2$ slabs, we introduce ways of producing granulated CO$_2$ ice samples, which are key to address the range of grain sizes observed on the Martian surface. We present a study of the reflectance of CO$_2$ ice, both pure and intimately mixed with water ice or JSC Mars-1. We also produced ternary mixtures of CO$_2$ ice, H$_2$O ice and JSC Mars-1, and monitored their reflectance as the ices sublimated. 

Our range of analogues necessarily includes idealised and simplified associations of ices and dust such as  homogeneous intimate mixtures of the CO$_2$/H$_2$O ices and dust. It is unlikely that any process on Mars will produce --at a large scale-- very homogeneous mixtures but such well-controlled samples are ideal playgrounds for testing reflectance models. More complex samples are useful too and allow for more direct comparisons with the remote-sensing data but parameters are more difficult to control and the comparison with models is more challenging.

We discuss lessons learnt and remaining challenges to produce these samples. We then study strategic spectral criteria to characterise the results of our experiments, and propose methods to fully exploit our reflectance spectra. The spectra are also available in public databases for further/different analyses by interested readers. This paper is the continuation of the work published in \citet{Yoldi:2020}, which provided laboratory data and spectral analysis to help interpret water ice and soil reflectance spectra from the Martian surface. Toward the end of the Result and Discussion Sections, we consider all results from the two articles to provide insights to interpret remote-sensing colour and spectral information in terms of the composition and physical properties of Martian icy surfaces.

\section{Samples, instruments and methods}
\label{S:Methods}

\subsection{Samples} \label{M:samples}
\subsubsection{Ices} \label{M:ice}
\paragraph{H$_2$O ice}
\begin{figure}[!h]
    \centering\includegraphics[width=0.8\textwidth]{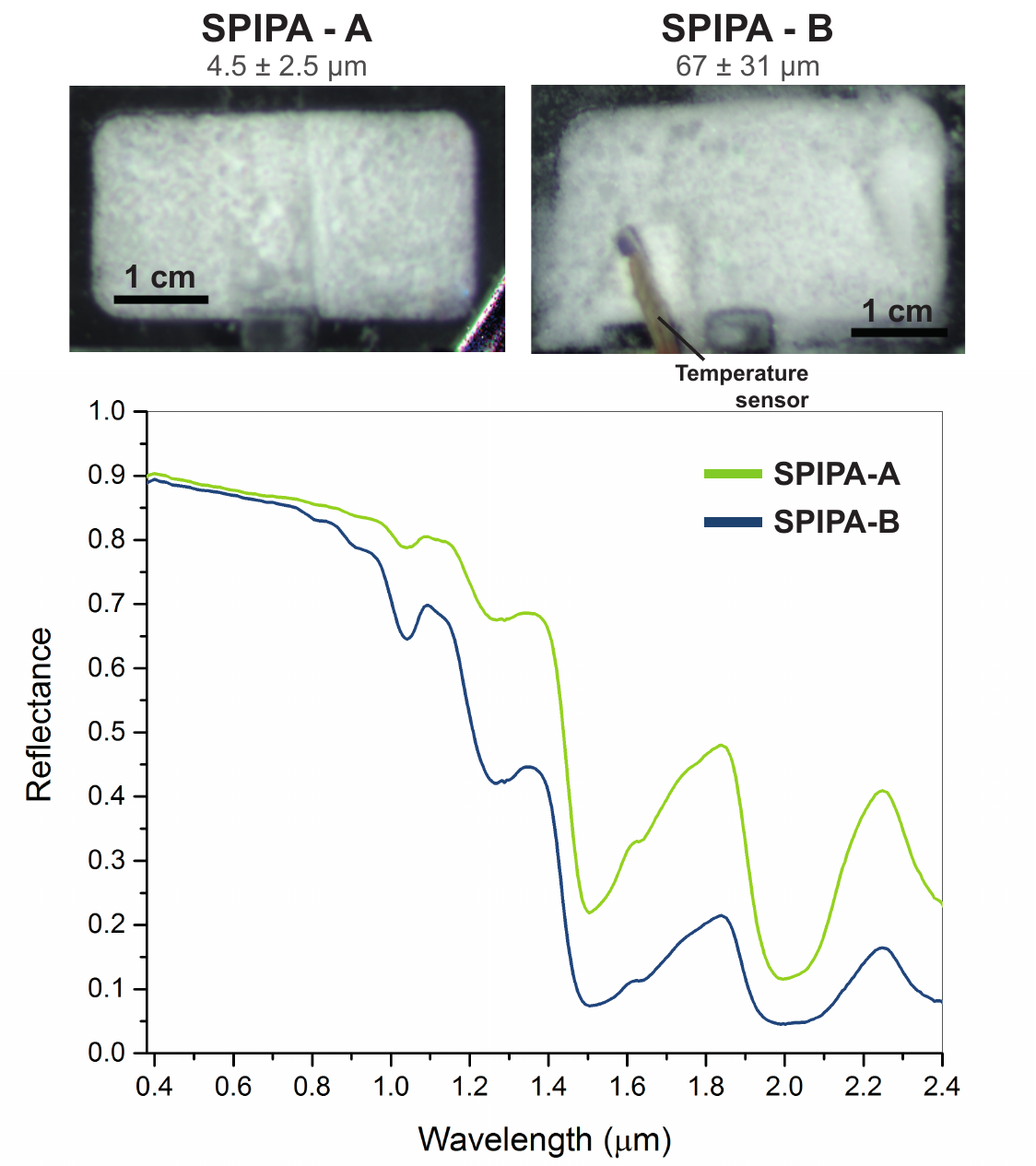}
    \caption[SPIPA-A and SPIPA-B spectra]{Top: Samples of fine-grained (4.5 \textmu m, SPIPA-A) H$_2$O (left) and coarser grained (67 \textmu m, SPIPA-B) H$_2$O (right) ices. Bottom: reflectance spectra of these two types of water ice.}
    \label{Fig:water_ice}
\end{figure}

We produced the granular water ice used in this study with the Setup for Production of Icy Planetary Analogues (SPIPA), at the Laboratory for Outflow Studies of Sublimating Materials (LOSSy) at the University of Bern. In this study, we work with fine-grained ice (mean diameter: 4.5$\pm$2.5 \textmu m) and coarser-grained ice (67$\pm$31 \textmu m), which we refer to as SPIPA-A and SPIPA-B respectively. Both SPIPA ices were produced by freezing nebulised, deionised water. We obtained different size distributions by using two different systems to nebulise the water. A detailed description of the production and characterisation of the SPIPA ices is available in \citet{Pommerol:2019}. Fig \ref{Fig:water_ice} shows pictures of the SPIPA samples, as well as their reflectance spectra. 

We also caused atmospheric water to condense onto a slab of CO$_2$ ice by exposing the slab to the atmosphere, outdoors, on a day with 93$\%$ of relative atmospheric humidity. Water frost did not condense homogeneously on the slab, which allowed us to study the spectra of different water frost thicknesses. 

\begin{figure}[!h]
    \centering\includegraphics[width=0.5\textwidth]{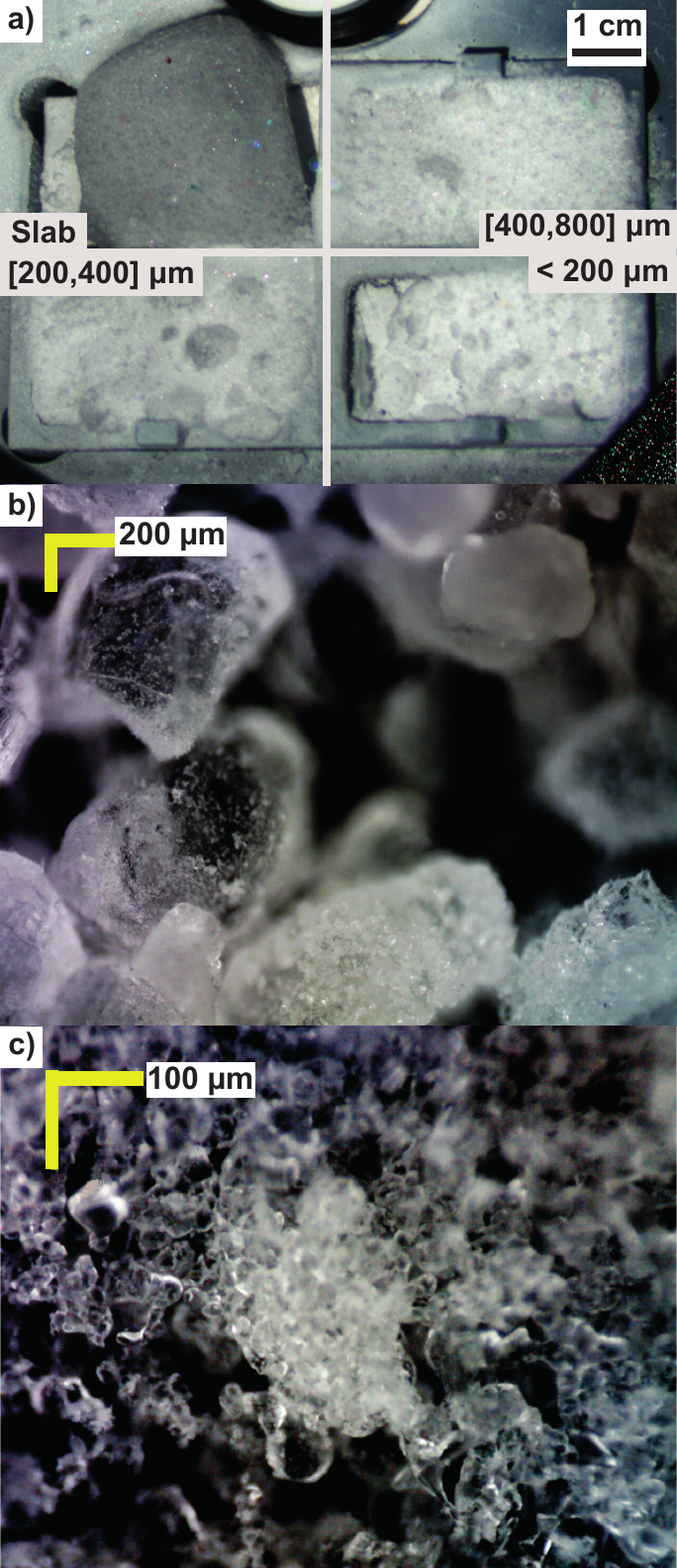}
    \caption[CO$_2$ ice characterisation]{a) Three size fractions of crushed CO$_2$ (400-800 \textmu m, 200-400 \textmu m and $<$ 200 \textmu m), along with a CO$_2$ slab. These RGB images are exported from the hyperspectral cubes and appear slightly blurry due to the wide spectral range of the measurement. The resolution is $\sim$ 50 \textmu m. b) Microscope image of the crushed CO$_2$ (400-800 \textmu m fraction). c) Microscope image of the CO$_2$ frost showing smooth and roundish particles with diameters ranging between 10 and 100 \textmu m. Many of the particles show clear signs of sintering, being connected by small necks.}
    \label{Fig:co2}
\end{figure}

\paragraph{CO$_2$ ice}
From frost \citep{Forget:1995} to thick and transparent slabs \citep{Kieffer:2006}, CO$_2$ ice is found in a variety of particle sizes on the Martian surface. In this study, we have used three types of pure CO$_2$ ice, which we call CO$_2$ frost, CO$_2$ slab and crushed CO$_2$ (Fig \ref{Fig:co2}). We detail below the process of production of each one of these samples. 

CO$_2$ frost was formed upon adiabatic cooling from pressurised CO$_2$ gas using a commercial CO$_2$ dry ice maker (Bel-Art$^{TM}$ SP Scienceware$^{TM}$ Frigimat$^{TM}$ Junior Dry Ice Make) and immediately stored in liquid nitrogen. Fig \ref{Fig:co2}c shows a representative microscope image of the CO$_2$ frost, from which we estimate the diameters of the individual CO$_2$ grains to range between about 10 \textmu m and 100 \textmu m upon formation. The individual grains show a tendency to sinter as they form. Smooth necks joining the grains are clearly visible under the microscope, which results in agglomerates of a few hundreds of micrometres. Unfortunately, the rapid sublimation of such fine grains of CO$_2$ at atmospheric pressure hampered further characterisation of these samples. 

We have also used commercial CO$_2$ ice pellets purchased from Carbagas AG (1.4 kg slices, Reference: I5891XXX) and delivered in wedge-shaped ingots of few centimetres of maximal thickness. The ice is compact and translucent but not transparent. Internal fractures and joints are visible and scatter the light internally (\ref{Fig:co2}a). Note that the bare black rims of the sample holder cannot be distinguished from the frost-covered white surface through the cm-thick slab ice. We refer to these pellets as slab CO$_2$ ice hereafter. We broke the slabs into small pieces to place them in the sample holder, as can be seen in Figure \ref{Fig:co2}a. We crushed some of the slabs and dry-sieved the resulting powder to obtain samples of CO$_2$ ice with various size distributions. We prepared three size distributions (Fig \ref{Fig:co2}a): particles between 400 and 800 \textmu m, particles between 200 and 400 \textmu m and particles smaller than 200 \textmu m. Fig \ref{Fig:co2}b shows an optical microscope picture of the fraction between 400 and 800 \textmu m. Unlike CO$_2$ frost, the crushed particles appear angular to sub-rounded because of the crushing process. Note also the fine-scale surface texture and the presence of internal fractures, pores or other types of defects within some of the particles (lower right corner) which effectively scatter the light

\subsubsection{Martian analogues} \label{M:analogues}
\begin{figure}[h]
    \centering\includegraphics[width=\textwidth]{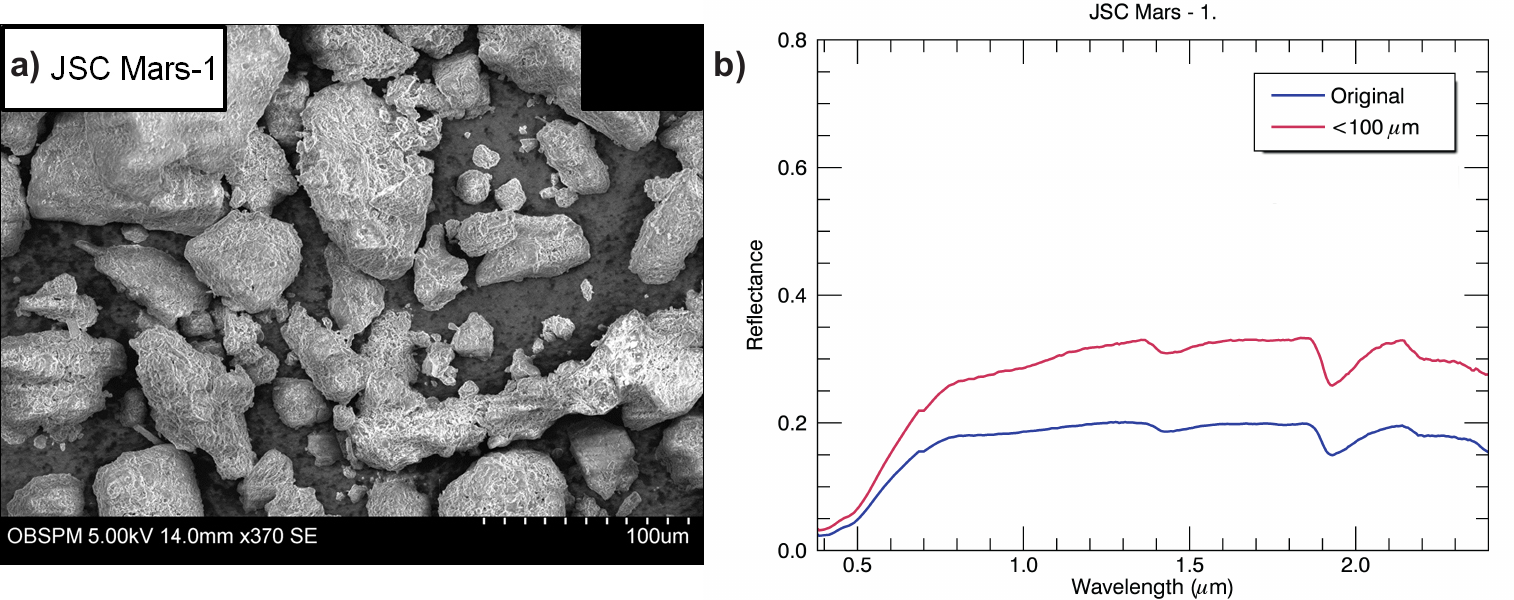}
    \caption[JSC Mars-1 characterization.]{a) SEM picture of a sieved sample of JSC Mars-1 (\textless80 \textmu m). Picture extracted from \citet{Yann:thesis}. b) Reflectance spectra of the original and sieved fraction (\textless100 \textmu m) of the JSC Mars-1}
    \label{Fig:JSCM1_spectra}
\end{figure}
As a Martian soil analogue, we have used the Martian regolith simulant JSC Mars-1 \citep{Allen:1997}, distributed by the Johnson Space Center. We already used this simulant in the first paper of this series \citep{Yoldi:2020}.

For part of this study, we have used the JSC Mars-1 as distributed; all the specifications regarding particle sizes, density, etc. are provided in \citet{Allen:1997}. We have also dry-sieved some of the original JSC Mars-1 to retrieve grains smaller than 100 \textmu m, which we used for the ternary mixtures. A Scanning Electron Microscope (SEM) picture of the fine grains and the reflectance spectra of both the original and the fine distributions are shown in Fig \ref{Fig:JSCM1_spectra}. Note that the size of the the fine Martian aeolian dust is estimated to be $\sim$ 2 \textmu m \citep{Wolff:2009, Goetz:2010} and while our fine fraction contains such very fine particles in large number, its mass distribution is dominated by the largest ones. In some contexts however, a broader size distribution is preferable. For instance, the jet activity observed over the cryptic region in the southern hemisphere and over dune fields in both hemispheres is known to mobilize and erode the Martian soil, depositing particles of all sizes over the seasonal ice.
 
\subsubsection{Mixtures} \label{M:associations}
\begin{figure}[h]
    \centering\includegraphics[width=\textwidth]{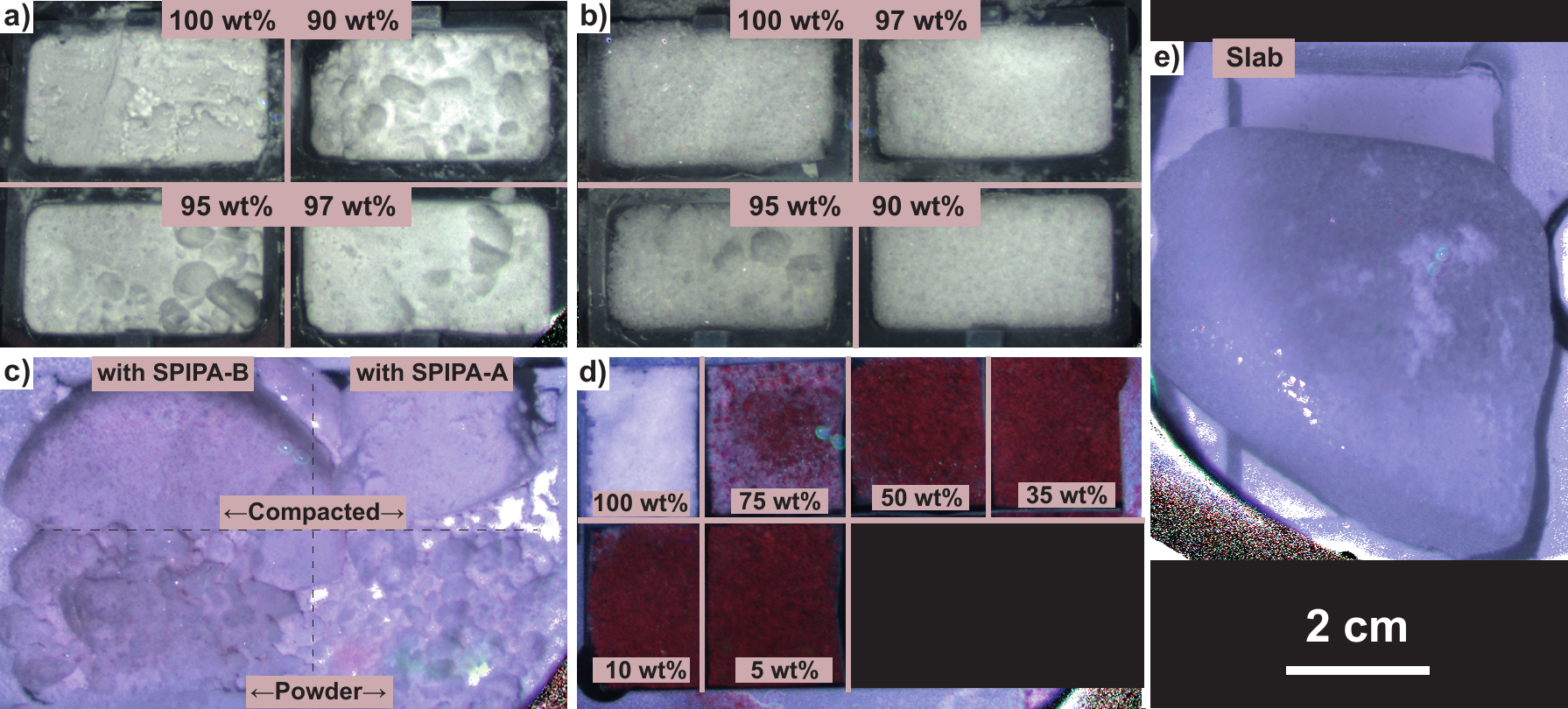}
    \caption[Mixtures with CO$_2$ ice.]{Mixtures with CO$_2$ ice. a) Various weight percentages of CO$_2$ frost (10-100 \textmu m) intimately mixed with fine grained ($\sim$ 4.5 \textmu m) water ice. b) Various weight percentages of crushed CO$_2$ (400-800 \textmu m) intimately mixed with coarser grained ($\sim$ 67 \textmu m ) water ice. c) Ternary mixtures of CO$_2$, fine grained H$_2$O (right) or coarser grained H$_2$O (left), and JSC Mars-1. On top, the mixtures have been compacted. d) Various weight percentages of crushed CO$_2$ (400-800 \textmu m) intimately mixed with JSC Mars-1. e) Slab of compact CO$_2$ ice with some water frost (white) on the right side.}
    \label{Fig:asso}
\end{figure}

Here, we explain the methodology followed to create the mixtures used in this study. Table \ref{Table:mixtures} provides an overview of the parameters and specifications of the samples.

\paragraph{Binary mixtures} 
We have developed and validated protocols to create intimate mixtures of water ice and refractory materials \citep{Pommerol:2019}. These protocols standardise the mixture production to guarantee the homogeneity and reproducibility of the samples. In short, we blend the ice and refractory materials with the help of a vortex shaker, alternating shaking and cooling intervals, during which we plunge the container with the samples into LN$_2$. The duration of these intervals is stipulated in the protocols. 

We have adapted those protocols to produce CO$_2$ icy samples. The main difficulty of the process is the different sublimation points of H$_2$O (273 K) and CO$_2$ (195 K) at 1 atm. To keep our working temperatures to a minimum, we kept the samples in aluminium bottles, which we plunged in LN$_2$. We took the bottles out of the nitrogen to weigh and mix the materials in intervals of maximum 15 seconds. We also kept the sample holders (2x4x2 cm) into LN$_2$ until filled with the samples, and the cylinder that contained the sample holder was filled with LN$_2$, as previously explained. The sample holders were filled inside a freezer to prevent water frost deposition.

We prepared the following intimate mixtures:

\begin{enumerate}
    \item CO$_2$-H$_2$O: we mixed both crushed CO$_2$ and CO$_2$ frost with both fine grained and coarser grained H$_2$O ice. We worked with small water weight percentages (10, 5 and 3 wt$\%$) since it has already been shown that the water signature dominates the reflectance spectra from small quantities \citep{Kieffer:1970, Singh:2016}. Fig \ref{Fig:asso}a shows intimate mixtures of CO$_2$ frost (10-100 \textmu m) and fine grained H$_2$O ice ($\sim$ 4 \textmu m). Fig \ref{Fig:asso}b shows intimate mixtures of crushed CO$_2$ and coarser grained H$_2$O ice. 
    \item CO$_2$--JSC Mars-1 intimate mixtures: by partitioning the sample holder, we obtained small volumes (2x2x2 cm) for multiple samples (Fig \ref{Fig:asso}d). This way, we managed to measure several samples under the same conditions. JSC Mars-1 was mixed with 75, 50, 35, 10 and 5 wt$\%$ of crushed-CO$_2$. 
\end{enumerate}

Even though the materials were homogeneously mixed, the reproducibility of the intimate mixtures of H$_2$O and CO$_2$ can be improved. For example, we observed the formation of pebble-like agglomerates for distinct grain sizes and water ice contents. Mixing materials in such ratios is challenging; the accuracy of water ice concentrations lower than 5 wt$\%$ is limited.

We also cannot exclude vertical variability within our samples. Either because of vertical segregation of particle sizes as we fill the holders with the particulate samples or as a result of thermo-physical evolutions of the samples (through sublimation, sintering...). Unfortunately, we have no means of assessing if vertical gradients of samples properties exist and their amplitude.

\paragraph{Ternary mixtures} \label{M:ternary_mixtures}
We have developed a new methodology to create homogeneous mixtures of dust, CO$_2$ frost and H$_2$O ice. These ternary mixtures will eventually be useful to simulate the volatile sublimation sequence observed during the retreat of the northern seasonal polar cap of Mars \citep{Langevin:2007, Appere:2011, Pommerol:2013}. For these mixtures, we used the fine fraction of JSC Mars-1 ($<$100 \textmu m) to simulate the dust grains suspended in the atmosphere that get trapped within the ice as CO$_2$ and H$_2$O condense. 

\begin{figure}[h]
    \centering\includegraphics[width=\textwidth]{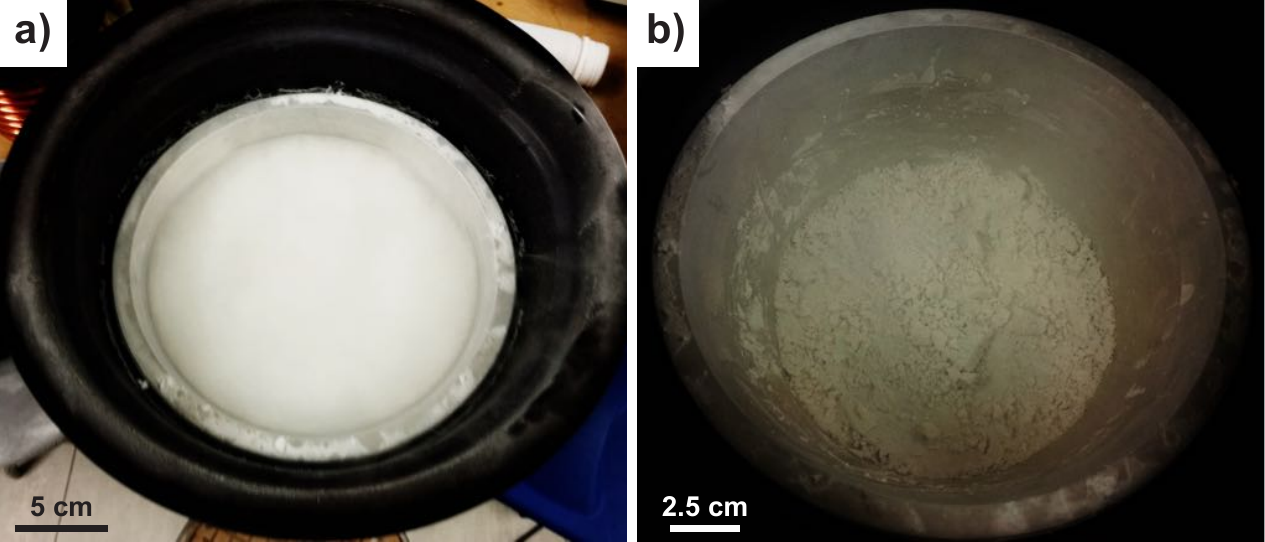}
    \caption[Production of ternary mixtures.]{Two steps of the production of ternary mixtures. a) CO$_2$ frost (10-100 \textmu m), coarser grained H$_2$O ice ($\sim$ 60 \textmu m) and JSC Mars-1 in liquid nitrogen. b) After sublimation of the liquid nitrogen, the components of the mixture settle down.}
    \label{Fig:cake_production}
\end{figure}

Figure \ref{Fig:cake_production} illustrates the process of production of the ternary mixtures. First, we produce the CO$_2$ frost, weigh it and immediately plunge it into LN$_2$. We produce the H$_2$O ice, and add it to the LN$_2$ and CO$_2$ mixture. We then add the dust to the mixture. At this point, the mixture looks as shown in Fig \ref{Fig:cake_production}a. Table \ref{Table:mixtures} shows the mass percentages used in each case. 

When the LN$_2$ has evaporated enough that the liquid surface is below the top of the powder, the mixture presents a texture similar to that of plaster. We stir softly but constantly to mix the components of the mixture while the remaining nitrogen evaporates. Once the LN$_2$ has evaporated entirely, the mix solidifies (Fig \ref{Fig:cake_production}b). Visual inspection confirms that the dust grains are homogeneously distributed within the ice. Further stirring resulted in the loosening of the sample, whereas the non-manipulated sample would tend to compact as the CO$_2$ particles -in our case, the main component- sintered.

\begin{figure}[h]
    \centering\includegraphics[width=\textwidth]{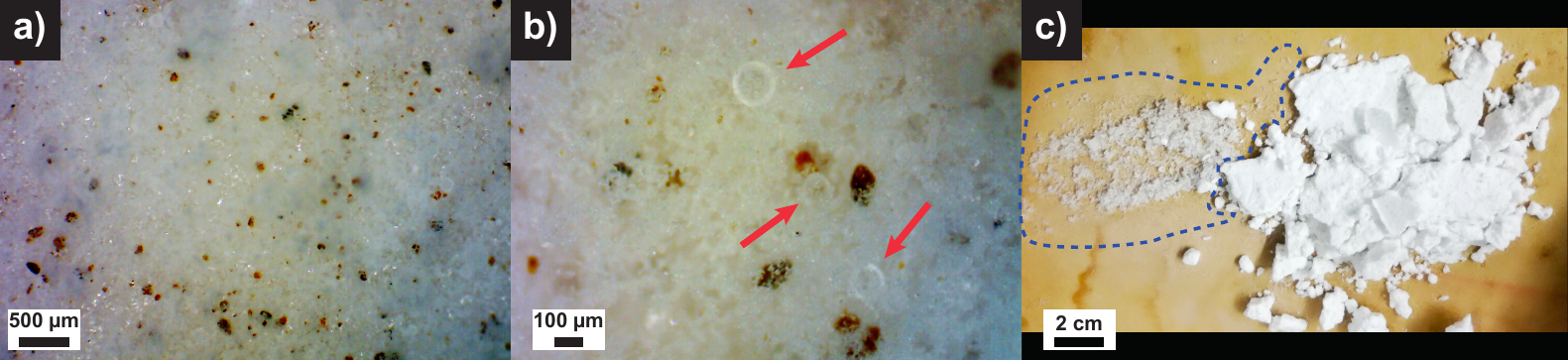}
    \caption[Sublimation of CO$_2$ ice in the ternary mixtures.]{Sublimation of CO$_2$ ice in the ternary mixtures. a) Microscope picture from the ternary mixture upon production. b) A few seconds afterwards, the first layer of CO$_2$ has sublimated, revealing the SPIPA-B particles in the mixture (marked with red arrows). c) The ternary mixture left on a table for a few minutes (left) and freshly spread (right). The sample on the left has lost all the CO$_2$ (whose initial contour is shown with the blue, dashed line), releasing the coarser CO$_2$ ice grains and JSC Mars-1. }
    \label{Fig:cake_sublimation}
\end{figure}

We present two different textures of the ternary mixtures: a powder-like one and a compressed one. The density of the samples was
\mbox{$\rho_{fine H_{2}O} = 0.47\ g/cm^3$} and 
\mbox{$\rho_{coarser H_{2}O} = 0.488\ g/cm^3$} for the powdered samples and \mbox{$\rho_{fine H_{2}O} = 0.89\ g/cm^3$} and
\mbox{$\rho_{coarser H_{2}O} = 0.846\ g/cm^3$} for the compacted ones. Fig \ref{Fig:cake_sublimation}a shows a microscope image of the powdered ternary mixture (with SPIPA-B). Seconds later, as CO$_2$ ice sublimated, the water particles became visible, as can be seen marked with red arrows in Fig \ref{Fig:cake_sublimation}b. Finally, Fig \ref{Fig:cake_sublimation}c shows the ternary mixture with SPIPA-B spread on a table; the sample on the left was spread on the table few minutes before the acquisition of the picture, and the sample on the right was spread right before acquiring the image. On the left, the carbon dioxide had completely sublimated, and only water ice and JSC Mars-1 is left.

Even though this protocol was effective to produce mixtures with both fine and coarser-grained H$_2$O ice, we observed differences when using one or the other type of water ice. For example, the grains of the fine-grained water ice floated in liquid nitrogen forming aggregates, probably increasing the effective scatterer size of the water ice if they were not destroyed before solidification.

\begin{landscape}
\renewcommand{\tabularxcolumn}[1]{m{#1}}
\newcolumntype{s}{>{\hsize=.5\hsize}X}
\setlength\tabcolsep{9pt}
\renewcommand{\arraystretch}{2}
    \begin{table}[h]
        \tiny
        \begin{tabularx}{\hsize}{X s X X X X s X}
            \hline
            \textbf{\hspace{0pt}Mixing \newline Mode} & \textbf{CO$_2$}&   \textbf{Size of CO$_2$ ice} & \textbf{\hspace{0pt}Contaminant}  &  \textbf{Size of the \newline Contaminant} & \textbf{\hspace{0pt}Concentration of \newline contaminant}  & \textbf{T$_{Shroud}$ (K)} & \textbf{Duration of Measurements}\\
            \hline
            --- & Crushed & 400-800 \textmu m & --- & --- & --- & 128 K &  1st scan: 30 mins \newline Hyperspectral: 71 mins\\\hline
Frost & Slab & Wedge shaped. \newline Max dim: 8 x 5 x 2.8 cm  & Water frost & Unknown.  &  From \textmu m to \newline $\sim$mm-thick layer & 160 K  & 1st scan: 30 mins \newline Hyperspectral: 71 mins\\
            \hline
            \multirow{5}{\hsize}{\hspace{0pt}Intimate mixture} & \multirow{2}{*}{Frost} & \multirow{2}{\hsize}{Few \textmu m. Agglomerates of hundreds of \textmu m} & SPIPA-A  & 4.5 $\pm$ 2.5 \textmu m & \multirow{4}{*}{3, 5, 10 wt$\%$} & 148 K & \multirow{5}{\hsize}{1st scan: 30 mins \newline Hyperspectral: 71 mins}\\
            & & & SPIPA-B  & 69 $\pm$ 31 \textmu m & & 138 K& \\
            \cline{2-5}
            \cline{7-7}
            & \multirow{3}{*}{Crushed} & \multirow{3}{*}{400-800 \textmu m} & SPIPA-A  & 4.5 $\pm$ 2.5 \textmu m & & 140 K & \\
            & & & SPIPA-B & 69 $\pm$ 31 \textmu m & & 150 K & \\
            \cline{6-7}
            & & & JSC Mars-1 & {71.7$\% >$ 150\textmu m \newline 28.3$\% <$ 150\textmu m \citep{Allen:1997}} & 25, 50, 65, 90, 95 wt$\%$ & 160 K & \\
            \hline
            \multirow{2}{\hsize}{\hspace{0pt}Ternary \newline Intimate mixture} & \multirow{2}{*}{Frost} & \multirow{2}{\hsize}{Few \textmu m. Agglomerates of hundreds of \textmu m} & SPIPA-A \newline JSC Mars-1  & 4.5 $\pm$ 2.5 \textmu m \newline $<$100 \textmu m & 3 wt$\%$ \newline 0.2 wt$\%$ & \multirow{2}{*}{See Fig  \ref{Fig:cake_temperature}} & \multirow{2}{\hsize}{ Total time: 11scans * 32 mins}\\
          &  & & SPIPA-B \newline JSC Mars-1  & 69 $\pm$ 31 \textmu m \newline $<$100 \textmu m & 3 wt$\%$  \newline 0.2 wt$\%$   & & \\
          \hline
        \end{tabularx}
        \caption{Summary of the compositions, concentrations, mixture types and acquisition conditions of the experiments reported in this study.}
        \label{Table:mixtures}
    \end{table}
\end{landscape}

\subsection{Instrument description} \label{M:SCITEAS}
We have acquired the reflectance data presented in this study with the Simulation Chamber for Imaging Temporal Evolution of Analogous Samples (SCITEAS) at LOSSy, the Laboratory for Outflow Studies of Sublimating icy materials of the University of Bern. SCITEAS is a vacuum and cold chamber monitored with an imaging system that allows us to characterise ice-bearing samples. \citet{Pommerol:2015} first described SCITEAS in 2015. In the first paper of this series \citep{Yoldi:2020}, we have detailed improvements to the system. Further details, together with a list of the published studies carried out with SCITEAS can be found in \citet{Pommerol:2019}. Therefore, we only provide here the main characteristics of the instrument:
\begin{itemize}
    \item Acquisition of hyper/multispectral cubes between 0.38 and 2.5 \textmu m.  
    \item Samples can be kept at low temperature (down to 100~K) by flowing liquid nitrogen (LN$_2$) through a cold shroud that surrounds the sample holder. 
    \item The pressure can be reduced down to 10$^{-6}$ mbar.
    \item The spectral sampling is defined by the user. For hyperspectral acquisitions, we typically work with a sampling of 15 nm in the visible (VIS, 0.4-0.95 \textmu m) and of 6 nm in the near-infrared (NIR, 0.95- 2.500 \textmu m).
    \item The geometrical configuration of the instrument is set by default so that the incidence angle is around 20$^\circ$, the emission angle is 30$^\circ$ and the phase angles are 50$^\circ$ in the visible and 10$^\circ$ in the near-infrared \citep{Pommerol:2015}.
    \item The samples holder are made of aluminum and covered by a layer of black aluminum tape which has a constant reflectance of 0.05 over the entire spectral range investigated here. The holders are rectangular with LxWxH dimensions of 4x2x2cm.
\end{itemize}

At the time these experiments were conducted, the vacuum pumps of SCITEAS could not pump the CO$_2$ out of the chamber at a rate that would avoid building up pressure inside of it. The chamber was also not equipped with a cold plate below the sample. Both features were added later as a revised version of the chamber (SCITEAS-2) was built \citep{Cerubini:2022}. For safety reasons, we did not close the lid of the chamber completely when working with CO$_2$ ice to let the gas produced by sublimation escape from the chamber.

Performing the experiments under the laboratory atmosphere is of course different from the Martian conditions. There are two aspects to consider: First, the absolute pressure is higher by a factor of $\sim$ 100 and second the sublimation of CO$_2$ in the laboratory takes place in an inert atmosphere of nitrogen and oxygen with only traces of CO$_2$ while on Mars there is an equilibrium between a CO$_2$ ice surface and a CO$_2$ atmosphere. As a result of the absolute difference of pressure, sublimation in the lab will also occur at a higher temperature (195K vs. 140K), following the Clausius–Clapeyron relation. This will not affect the behaviour of water which is always far below its sublimation temperature in both cases. Higher pressure can also increase the vertical heat transport within the sample through gas conduction and convection, which in turn could have consequences for the metamorphism of the sample at various depths.
Sublimating CO$_2$ ice in an inert gas atmosphere rather than in equilibrium with its pure gas phase is also likely to change the outcome of the sublimation by affecting the sublimation speed and possible vertical transport and recondensation of CO$_2$ within the porosity of the sample \citep{Blackburn:2010}.
Future work will ideally attempt to perform these sublimation experiments in conditions closer to the Martian surface and find out if the possible differences discussed here are indeed significant.
Note that the influence of lab pressure is not only relevant for the sublimation but also for the growth. For example, \citep{kieffer:1968} suggest that for a process limited by diffusion, the size of the frost crystals is proportional to the relative concentration of the condensible gas. Consequently, we can expect the Martian crystals of frost to be smaller than the ones grown in the laboratory. The effect of pressure (and temperature) on the morphology of crystals grown was studied experimentally by \citet{Anya:2019}. 

\subsubsection{Acquisition time} \label{M:acquisition}
The acquisition of a hyperspectral cube under the current setting lasts 70 min \citep{Yoldi:2020}. However, CO$_2$ sublimates faster than that at the range of temperatures and pressures we work with. We have modified that spectral sampling to reduce the duration of the measurements: even though we keep the 6 nm sampling around absorption features in the NIR, we measure the reflectance only at specific wavelengths in the continuum. This acquisition mode lasts 30 min and will be referred to as \textit{quick scan}.

Most of our measurements consist of a first, quick scan followed by a complete one. The second scan provides us with the entire shape of the spectra but with a larger CO$_2$ grain size due to sintering and/or reduced concentration of the finest grains of CO$_2$ due to sublimation. To monitor the sublimation of ices in the ternary mixtures at high temporal resolution, we have used only quick scans.  

\subsubsection{Sublimation of ice from the samples} \label{M:sublimation}
At a constant pressure (i.e. 1 atm) and for a given grain size, temperature controls the sublimation of our ices. Table \ref{Table:mixtures} shows the mean temperature measured on the shroud for each experiment. To reach such low temperatures at 1 atm, we filled the cylinder containing the sample holder (see Fig 2 in \citet{Pommerol:2015}) with LN$_2$ at the beginning of the experiment. 

\begin{figure}[h]
    \centering\includegraphics[width=\textwidth]{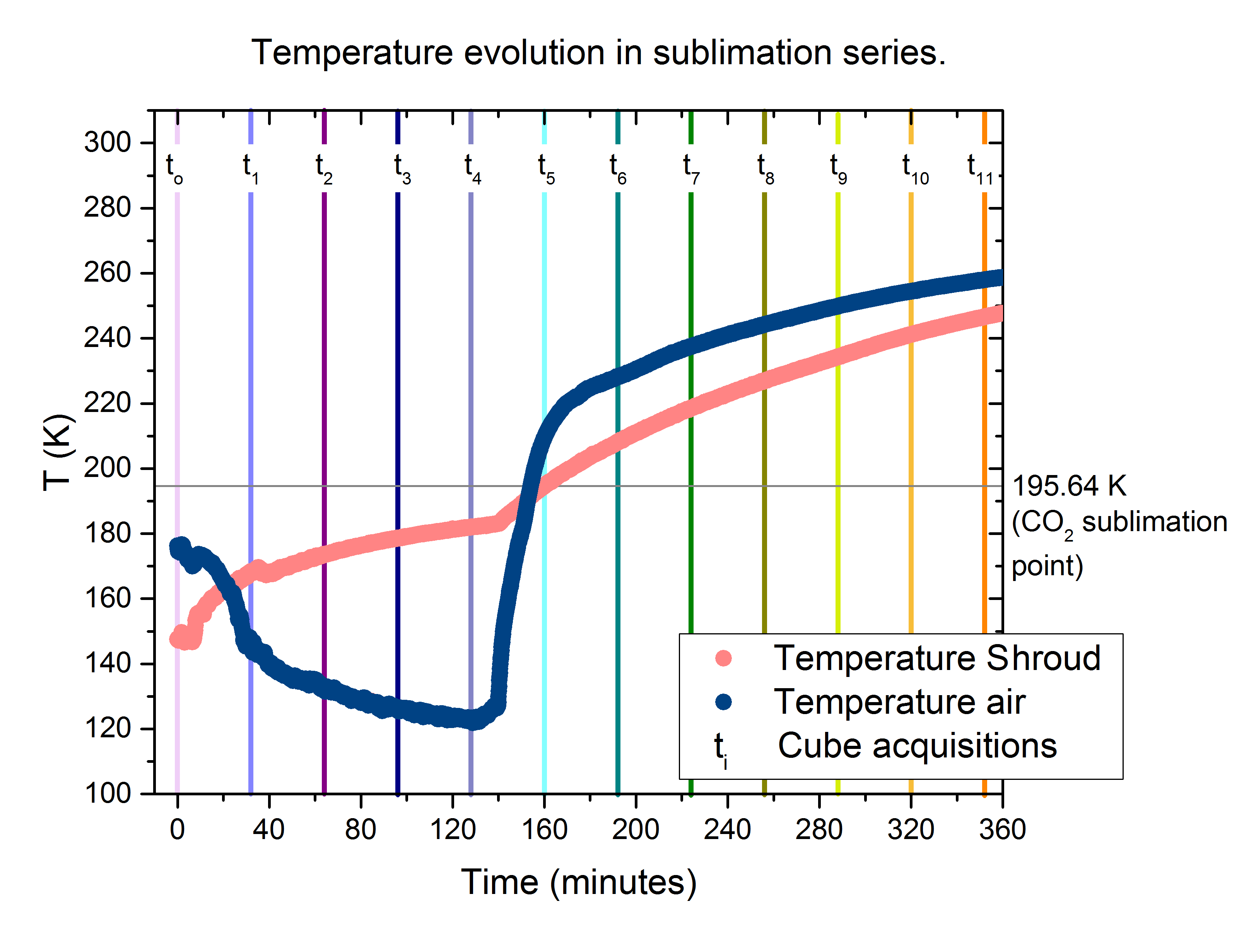}
    \caption[Temperature evolution in the ternary mixture sublimation series.]{Evolution of the temperature of the shroud and the air at the surface surrounding the samples. The vertical lines indicate the time of acquisition of each of the 12 quick scans performed during the experiment. The horizontal line indicates the sublimation temperature of CO$_2$ at atmospheric pressure.} 
    \label{Fig:cake_temperature}
\end{figure}

The evolution of temperature measured in the shroud and the air near the ternary mixtures is shown in Figure \ref{Fig:cake_temperature}, where we have also marked the times at which each measurement (12 in total) started. The steep increase of temperature observed at t=140 min was intentional to speed up sublimation. 

\subsubsection{Water frost condensation on the samples} \label{M:frost_deposition}
As mentioned before, we were not able to seal the cold chamber during the experiments. However, the constant and intense out-gassing of the LN$_2$ from the sample holder keeps the chamber very dry, as nitrogen pushes the air out of the compartment. This keeps water frost deposition to a minimum, but does not completely avoid it.

To assess the amount of water cold trapped onto our samples, we added a control sample of pure CO$_2$ ice in each set of experiments. The H$_2$O index at 1.5 \textmu m (see Section \ref{M:bands}) of the control samples allows us to assess the water frost trapped by the samples. While we do not have any way to directly assess the quantity of frost deposited, we can use empirical relations derived from previous work to estimate the frost thickness from the strength of the water absorption bands. In our previous work \citep{Yoldi:2020}, we had used the 2-\textmu m band of H$_2$O for this purpose and an empirical relation derived from the data published by \citet{clark1981}. Here, we cannot use the 2-\textmu m absorption band of water because of the strong overlapping absorption of CO$_2$ but the 1.5-\textmu m H$_2O$ band does not suffer from this problem. Digitizing the data from \citet{clark1981} for water frost grown over a Mauna Kea dust sample, we find a logarithmic relationship between the 1.5–\textmu m band depth and the frost thickness. Digitizing as well the two spectra from \citet{kieffer:1968} for H$_2$O frost grown over CO$_2$ (same as in our work) and assuming a typical porosity of 90 \% \citep{Li:2021}, we notice that these two points fall over the same trend. This trend allows us to roughly estimate the frost thicknesses mentioned in the following sections, which vary between 25 and 400 \textmu m depending on the sample and the duration of the exposure to moist air. While the consistency between the results of \citep{kieffer:1968} and \citet{clark1981} gives us confidence that these estimates are credible, an independent way to assess the thickness of the frost layer would be preferable and new developments are planned to allow such observations.

\subsection{Data processing}
\label{M:processing}
We convert the raw data into reflectance factor (REFF) units (as defined by \citet{Hapke1993}) through a calibration process detailed in \citet{Pommerol:2015}. We provide here the main steps of that process: 

\begin{enumerate}
    \item Before the measurement of the actual samples, we perform similar measurements (same cameras settings, spectral ranges and sampling, sample position) with a large, nearly-Lambertian surface (Spectralon\textsuperscript{TM} (Labsphere)) covering the field-of-view of the camera.
    \item Regularly during the measurements of the spectral cubes, we close a shutter installed at the monochromator and acquire dark images with the exact same settings as the actual images. 
    \item We interpolate temporally the dark signal and subtract it from the corresponding cube.
    \item We divide the dark-signal corrected data acquired from the actual sample by the dark-signal corrected data acquired with the Spectralon surface.
\end{enumerate}
The calibration procedure outputs a reflectance hyper/multispectral cube in which we define Regions of Interest (ROIs) to extract averaged reflectance spectra from the desired areas.

\subsubsection{Uncertainties}
\label{M:uncertainties}
 We detail in the first paper of this series \citep{Yoldi:2020}, the procedure that we follow to assess the uncertainties of our measurements. Therefore, we only present here the major points of the procedure.

\begin{itemize}
    \item Our sample holders contain an additional calibration target whose reflectance is known (i.e. Spectralon$^TM$ from Labsphere). By tracking the variation of its reflectance throughout the measurements, we assess their precision and accuracy.
    \item The average precision on our calibration target is $\pm$0.3$\%$ in the VIS and 2$\%$ in the NIR. This translates into signal-to-noise ratios (SNRs) of 333 and 50, respectively.
    \item Let us assume the scenario in which the uncertainty in the precision is entirely due to photon noise. In that case, the uncertainty varies with the square root of the signal, that is, of the reflectance of the sample. Following this, the SNR drops to 95 in VIS (relative error of 1$\%$) and 16 in NIR (relative error of 6$\%$) for a sample with a reflectance of 10$\%$. 
    \item We do not work with individual pixels, but with ROIs formed by hundreds to thousands of pixels. Consequently, the SNRs of our measurements increase in average to much higher values (several hundreds) limited by the readout noise of the cameras and dark noise. 
    \item We assess the accuracy of our measurements by comparing the theoretical reflectance of the calibration target with our measurements. This comparison shows a variability of 6$\%$ with the visible camera and 3$\%$ with the near-infrared.
\end{itemize}

Even though accuracy limitations control our uncertainties, the SNR and absolute accuracy have different implications on our spectral analyses. We favour the use of relative spectral criteria, where ratios between wavelengths of the same spectra are computed (i.e., slopes and band depths). These relative criteria are not affected by the absolute accuracy that affects the reflectance at all wavelengths in the same way and are only influenced by the SNR. 

\subsubsection{Spectral analysis}
\label{M:spectra}
We study the reflectance spectra of our samples through the reflectance of the continuum, the depth of the absorption bands, the spectral slopes and the CaSSIS colours. We have chosen these parameters because they have been repeatedly used in other spectrophotometric studies (e.g., \citep{Langevin:2007, Appere:2011, Brown:2014, Fornasier:2015}). We build on some of their choices and definition of parameters because we want our data to be comparable with existing data sets.

\paragraph{Reflectance in the continuum} The reflectance of the continuum allows us to distinguish bright from dark materials at a specific wavelength. It depends on various variables such as the single scattering albedo of the material or the size of grains. Both CO$_2$ and H$_2$O ices are bright in the continuum, but other materials, such as salts, are also bright in the continuum \citep{Cerubini:2022}. We have measured the reflectance of the continuum at 0.940 \textmu m. \citet{Langevin:2007} used the reflectance at 1.08 \textmu as a reference for the reflectance in the continuum, based on a compromise between OMEGA instrumental considerations and the scattering of aerosols in the Martian atmosphere \citep{Langevin:2007}. The difference in reflectance between these wavelengths is negligible for the CO$_2$ and JSC Mars-1 since they show flat spectra in this region. It is not the case for water ice, which presents a blue slope in the near-IR, but the difference of reflectance between these two positions lies within the uncertainties of the instrument. 

\paragraph{Evaluation of band depths} \label{M:bands}  The presence and position of absorption bands are specific to the materials (composition and state of the matter), and their depth to the length of the optical path inside the material (linked to the size and/or the amount of material present). We have analysed the absorption signatures of carbon dioxide at 1.435 and 2.281 \textmu m and of water at 1.5 \textmu m. For every band, we have used the evaluators proposed by \citet{Langevin:2007}. It is not straightforward to assess the strength of the CO$_2$ signature at 1.435 \textmu m, since it lies close to a CO$_2$ atmospheric absorption (around 1.444 \textmu m) and it overlaps with the short wavelength side of the water band at 1.5 \textmu m (which starts at 1.38 \textmu m \citep{Langevin:2007}). To minimise the impact of these two features, \citet{Langevin:2007} defined an evaluator (Eq  \eqref{EQ:evaluator}) from the reflectance (RF) of the spectra at 1.429, 1.385 and 1.443 \textmu m, which, adapted to the sampling of SCITEAS resulted in 1.432, 1.39 and 1.444 \textmu m respectively. 

\begin{equation} \label{EQ:evaluator}
R = \frac{RF(1.432 \mu m)}{RF(1.39 \mu m)^{0.5} \times RF(1.444 \mu m)^{0.5}}
\end{equation}

Then, \citet{Langevin:2007} found a relationship between R and the CO$_2$ strength computed from water-free spectra (Eq  \eqref{EQ:band14}).
\begin{equation} \label{EQ:band14}
CO_2(1.435 \mu m) = 1.16(1-R)^{0.92}
\end{equation}
As warned in \citet{Langevin:2007}, the spectral element at 1.444 \textmu m lies in the absorption signature of CO$_2$, therefore (1-R) underestimates the strength of the band.

We used Eq \eqref{EQ:band22} and \eqref{EQ:band15} below to evaluate the absorption features at 2.281 \textmu m (CO$_2$) and 1.5 \textmu m (H$_2$O) respectively. The values of the wavelengths are slightly changed from the original values to adapt SCITEAS sampling. 

\begin{equation} \label{EQ:band22}
CO_2(2.281 \mu m) = 1 - \frac{RF(2.29 \mu m)}{RF(2.226 \mu m)^{0.3} \times RF(2.314 \mu m)^{0.7}}
\end{equation}
\begin{equation} \label{EQ:band15}
H_2O(1.5 \mu m) = 1 - \frac{RF(1.498 \mu m)}{RF(1.390 \mu m)^{0.7} \times RF(1.774 \mu m)^{0.3}}
\end{equation}

As the band depth is defined as a ratio of two values of reflectance, the uncertainty on the band depth is only affected by the SNR. 

\paragraph{Spectral slopes} The spectral slopes indicate how reflectance depends on the wavelength. They give us an idea of the colour of the materials, and is useful to identify the way in which materials are mixed \citep{Yoldi:2020}. We have assessed the spectral slopes with Eq \eqref{EQ:slope}. $\lambda _1$ and $\lambda _2$ have been set to 0.445 and 0.745 \textmu m in the visible and 1.1 and 1.7 \textmu m in the near-infrared. RF$_{\lambda _1}$ and RF$_{\lambda _2}$ refer to the values of reflectance at $\lambda _1$ and $\lambda _2$ respectively.

\begin{equation} \label{EQ:slope}
S_r = \frac{RF_{\lambda _2}- RF_{\lambda _1}}{RF_{\lambda _1} (\lambda _2 - \lambda _1)} \times 10^4 \qquad (\% /100 nm)
\end{equation}

As for the band depth calculation, the relative uncertainties are therefore only affected by the SNR. 

Throughout this study, we use the terms red, blue and flat slopes. A spectrum (or a region of a spectrum) has a red slope when its reflectance increases towards longer - or red- wavelengths. Red slopes imply positive spectral slope values, measured in $\%$/100 nm. A spectrum (or a region of a spectrum) has a blue slope when its reflectance increases towards shorter - or blue- wavelengths. Blue slopes imply negative spectral slope values. A spectrum (or a region of a spectrum) is flat when it shows a constant reflectance, or spectral slope values $\sim 0 \ \%$/100 nm.

\paragraph{Colour composites} In the majority of cases, we present the spectral criteria averaged per ROI. In some cases, however, we also show the spatial variation of these criteria by showing the values per pixel. For that, we can either produce monochromatic images where the signal represents a particular spectral criterion, or we can combine three spectral criteria in RGB composites. When choosing the latter, we stretch each channel separately to emphasise the colours. We indicate the stretching applied to each channel with each RGB composite. 

\subsubsection{Colour analysis}
\label{M: colour}
The Colour and Stereo Surface Imaging System (CaSSIS, \citep{Thomas:2017}) is the visible imager of the Exomars Trace Gas orbiter (EM-TGO). It consists of a 4-mirrors telescope with a focal length of 880mm, a 2k x 2k pixels CMOS sensor and a rotation mechanism to align the telescope and sensor with respect to the ground track and obtain nearly-simultaneous stereo images of the surface. On top of the CMOS sensor, four broadband bandpass filters are mounted to image the surface in four colours. The respective filter names, effective central wavelengths and bandwidths (in nm) are: BLU (494.8, 133.6), PAN (678.2, 231.9), RED (836.0, 98.5), and NIR (939.3, 121.8). Because of the non Sun-synchronous orbit of TGO and the high sensitivity of the instrument, CaSSIS is able to observe the surface at variable local Solar times throughout the seasons and collects a unique dataset to study the diurnal and seasonal Martian volatiles cycles.

The simulated CaSSIS colours inform us on the possibilities and limitations of the instrument when it comes to detecting and distinguishing CO$_2$ and H$_2$O ices. It also provides us with prior knowledge to target the observations of the Martian surface. We have convolved the data measured with SCITEAS with the spectral response of the instrument \citep{Thomas:2017}. This methodology is automatised in the calibration software of SCITEAS and was detailed in \citet{Yoldi:2020}. Briefly, it is a process in which we take into account \begin{inparaenum}[(i)]
\item the illumination conditions at an average Sun-Mars distance and a simulated incidence angle, and
\item the specificities of CaSSIS (angular aperture of the telescope, area and reflectivity of the mirror, quantum efficiency, transmission of the filters, etc.),
\end{inparaenum} to simulate the reflectance that would have been measured by CaSSIS from samples like ours. 

To compute the parameters introduced so far (reflectance in the continuum, indices and spectral slopes), we need to know the reflectance of the samples at specific wavelengths. However, to simulate the CaSSIS colours, we need to know the reflectance of the samples in the spectral intervals that correspond to each filter (BLU, PAN, RED or NIR). The more measurements we have in an interval, the more accurate our simulations are. That is why the CaSSIS colours are the only spectral parameter that we derive from hyperspectral measurements and therefore, they represent the samples when they have already undergone sintering and sublimation.

Note that the bandpass filters of CaSSIS were designed to be very close to those of HiRISE with the infrared filter of HiRISE (IR) split into the RED and NIR filters of CaSSIS to provide additional sensitivity to iron-bearing mineralogy. Therefore, nearly all observations valid for the CaSSIS filters are also relevant for HiRISE.
\section{Results}
\label{S:Results}

\subsection{Pure CO$_2$ ice} \label{R:sizes}

\begin{figure}[h!]
\centering\includegraphics[width=\textwidth]{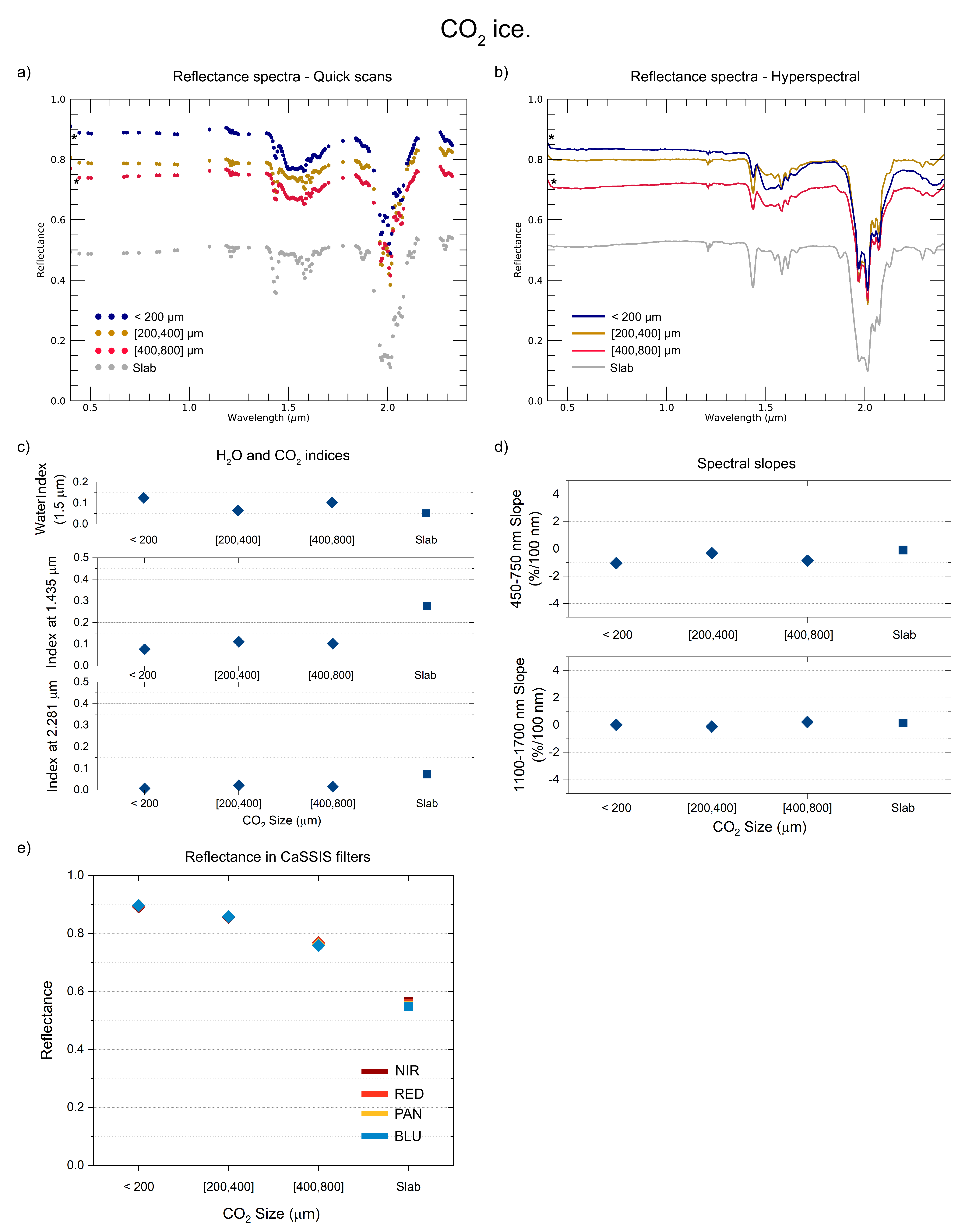}
\caption[Reflectance analysis of pure CO$_2$ ice of various grain sizes.]{a) Multispectral reflectance spectra for various CO$_2$ grain sizes. b) Hyperspectral reflectance spectra for various CO$_2$ grain sizes. c) Strength of various absorption features: the one from H$_2$O at 1.5 \textmu m (indicative of the cold-trapped water) and the ones at 1.435 and 2.281 \textmu m of CO$_2$ ice. d) VIS and NIR spectral slopes of the samples. e) Simulated reflectance measured in the CaSSIS filters. * Stars indicate calibration artifacts at the shortest wavelengths. All CO$_2$ samples appear contaminated by H$_2$O frost with estimated thicknesses of 100 \textmu m, 50 \textmu m,  75 \textmu m and 25 \textmu m (from the finest to the coarsest size fraction) based on the strength of the 1.5\textmu m feature.}
\label{Fig:sizes}
\end{figure}

Fig \ref{Fig:sizes} shows the reflectance spectra and their byproducts for various size distributions of CO$_2$ ice. Fig \ref{Fig:sizes}a and Fig \ref{Fig:sizes}b evince that the reflectance of the samples is inversely proportional to the size of their grains. The differences in reflectance between Fig \ref{Fig:sizes}a and Fig \ref{Fig:sizes}b are due to the sintering and/or sublimation of the finest particles of CO$_2$ ice during the experiments. Fine grains are more affected by this morphological evolution. The slab shows relatively low reflectance ($\sim$0.5) in the continuum, which suggests that part of the photons crossed the slab and reached the black sample holder. \citet{Kieffer:1970} reports that water frost causes reflectance to peak at 1.8 and 2.24 \textmu m; we do observe variations around those wavelengths between the multi- and hyperspectral measurements in, for instance, the slab.

The strength of the water index at 1.5 \textmu m (Fig \ref{Fig:sizes}c) indicates that, exposed to the same conditions of relative humidity and temperature, the smallest particles ({\textless 200} \textmu m) cold-trapped more water frost than larger size fractions. Granular ices present lower CO$_2$ indices than the slab as the path of the photons is longer in continuous media. Both CO$_2$ indices (at 1.435 and 2.281 \textmu m) remain relatively constant (around 0.1 and 0, respectively) for the particulate samples.

The VIS and NIR spectral slopes are relatively flat (i.e., around 0 $\%$/100 nm) for all the CO$_2$ samples (Fig \ref{Fig:sizes}d). The trapped water frost explains the small shift of particulate CO$_2$ ice towards blue slopes ( $ \sim -$1 $\%$/100 nm); it was shown in \citet{Yoldi:2020} that low amounts of water frost on samples suffice to lower their visible slopes to negative values. This particular spectral behaviour was also already modeled by \citep{Warren:1990} and \citep{Singh:2016} with similar results.

Finally, we show in Fig \ref{Fig:sizes}e the simulated CaSSIS colours. Because we obtain the CaSSIS colours from the hyperspectral measurements, the samples have already evolved. For this reason, the CaSSIS colours do not show a large difference in reflectance between the fine and the coarse ice grains. We see no significant difference of reflectance between measurements within the different filters of CaSSIS.

\subsection{CO$_2$ slab with water frost} \label{R:frost}

\begin{figure}[h!]
\centering\includegraphics[width=\textwidth]{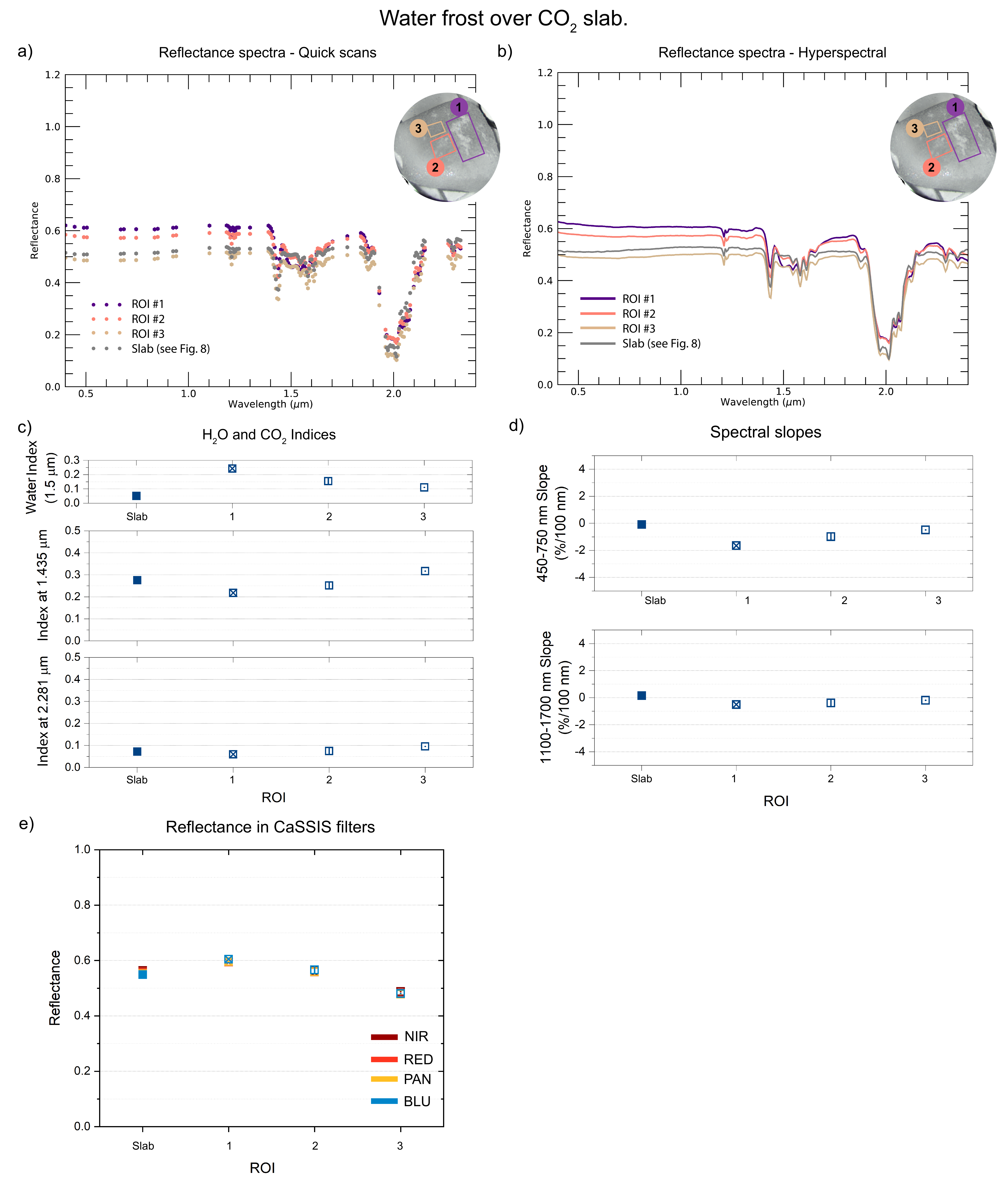}
\caption[Reflectance analysis of water frost condensed over a CO$_2$ ice slab.]{Water frost over CO$_2$ slab with different amount/thickness of frost within three ROIs identified on the RGB colour composite. a) Multispectral reflectance spectra of the ROIs. b) Hyperspectral reflectance spectra of the ROIs. c) Strength of various absorption features: the one from H$_2$O at 1.5 \textmu m and the ones at 1.435 and 2.281 \textmu m from CO$_2$ ice. d) VIS and NIR spectral slopes of the ROIs. e) Simulated reflectance measured in the CaSSIS filters. Based on the observed strengths of the 1.5 \textmu m band compared to earlier experimental work, we estimate the frost thicknesses in the three ROIs (\# 1, 2 and 3) to be 400, 150 and 75 \textmu m, respectively.}
\label{Fig:frost}
\end{figure}

Fig \ref{Fig:frost} shows the reflectance spectra and their byproducts for three ROIs drawn on the CO$_2$ slab covered with water frost (Fig \ref{Fig:frost}a). ROI \#1 is the one with most water frost ($\sim$ 400 \textmu m thickness), and ROI \#3 the one with the least ($\sim$ 75 \textmu m thickness). For comparison purposes, we have added the spectrum of the slab shown in Fig \ref{Fig:sizes}, on which we did not condense any water intentionally but is still slightly affected by water frost ($\sim$ 25 \textmu m). 

ROIs with more water frost condensed show high reflectance levels (Fig \ref{Fig:frost}a and Fig \ref{Fig:frost}b). Compared to the slab that had not been exposed to atmospheric water, ROI \#3 shows slightly lower reflectance. We attribute this small variation to a putative difference in the roughness of the ice surface caused by the outdoor exposure of the CO$_2$ slab. Fig \ref{Fig:frost}a and Fig \ref{Fig:frost}b do not show great differences in the absolute reflectance levels, as slabs do not evolve during the experiments as much as small particles do.

The H$_2$O indices at 1.5 \textmu m follow the amount of water frost in each ROI (Fig \ref{Fig:frost}c). Both CO$_2$ indices are negatively correlated with water frost thickness, as fewer photons reach the CO$_2$ slab due to scattering by the fine-grained H$_2$O frost at the surface. ROI \#3 shows slightly higher indices than the water frost-free slab, which can as well be explained by differences in the thickness of the slabs. 

We see in Fig \ref{Fig:frost}d that the VIS slope is slightly more sensitive to the amount of water frost condensed on the slabs than the NIR slope. The presence of water frost turns the visible spectral slope blue. 

Measured with CaSSIS, the ROIs with water frost would appear up to 20$\%$ brighter than the non-contaminated slab (Fig \ref{Fig:frost}e). Because of the blue slope induced by water frost, ROIs with more frost appear brighter in the BLU filter ($\sim$ 400 - 550 nm) than in the red ones (PAN: $\sim$ 550 - 800 nm; RED: $\sim$ 800 - 900 nm; RED: $\sim$ 900 - 1100 nm).

\subsection{CO$_2$ and H$_2$O ice intimate mixtures} \label{R:intimate}

\begin{figure}[h!]
\centering\includegraphics[width=\textwidth]{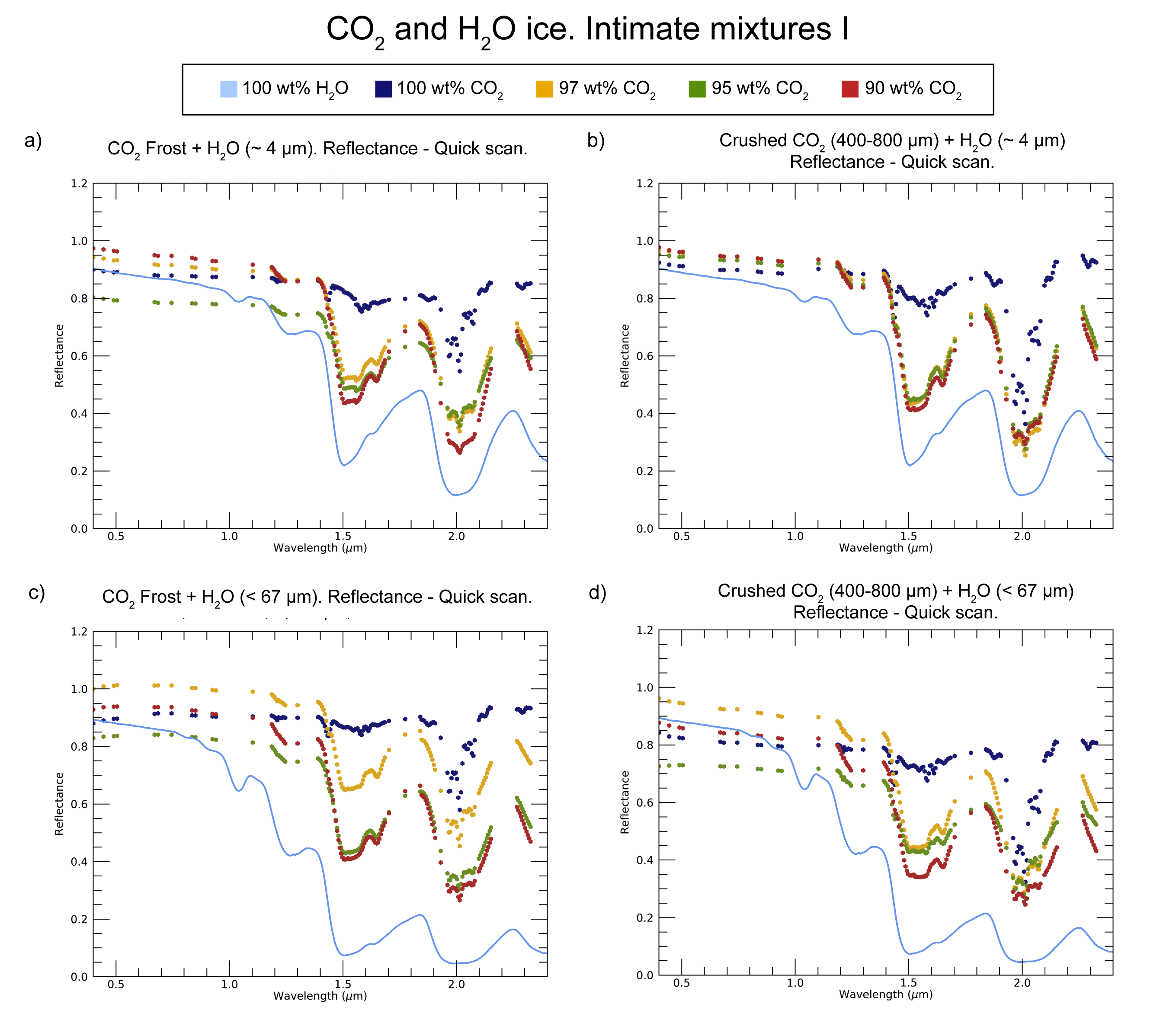}
\caption[Multispectral (quick scans) reflectance spectra of intimate mixtures of CO$_2$ and H$_2$O ice.]{Multispectral (quick scans) spectra of intimate mixtures of CO$_2$ and H$_2$O ices. The legend shows the CO$_2$ content of each sample. a) CO$_2$ frost with fine grained H$_2$O ice ($\sim$ 4 \textmu m). b) Crushed CO$_2$ with fine grained H$_2$O ice ($\sim$ 4 \textmu m). c) CO$_2$ frost with coarser grained H$_2$O ice (< 67 \textmu m). d) Crushed CO$_2$ with coarser grained H$_2$O ice (< 67 \textmu m).}
\label{Fig:spipa_multispectra}
\end{figure}

Fig \ref{Fig:spipa_multispectra} shows the reflectance spectra of the CO$_2$ (frost, crushed) and H$_2$O (fine, coarser) intimate mixtures. For all the different combinations, we see that the pure CO$_2$ sample is not the brightest. This is the result of the difficulties of mixing CO$_2$ ice with other components discussed previously. 

We see that water ice affects the reflectance spectra of the mixtures out of proportion to its concentration; water concentrations as low as 3 wt$\%$ are enough to dominate the spectrum. We see different behaviours in the continuum and the absorptive regions of the spectra. The reflectance of the continuum and other weak absorption features of water seems to be controlled by the texture (grain size, shape, surface roughness, density...) of the samples, which are not homogeneous due to the formation of agglomerates (see Fig \ref{Fig:asso}a). The reflectance in the spectral regions of the absorption bands of water (centred at 1.5 and 2.0 \textmu m), however, is more sensitive to the composition of the samples.

\begin{figure}[h!]
\centering\includegraphics[width=\textwidth]{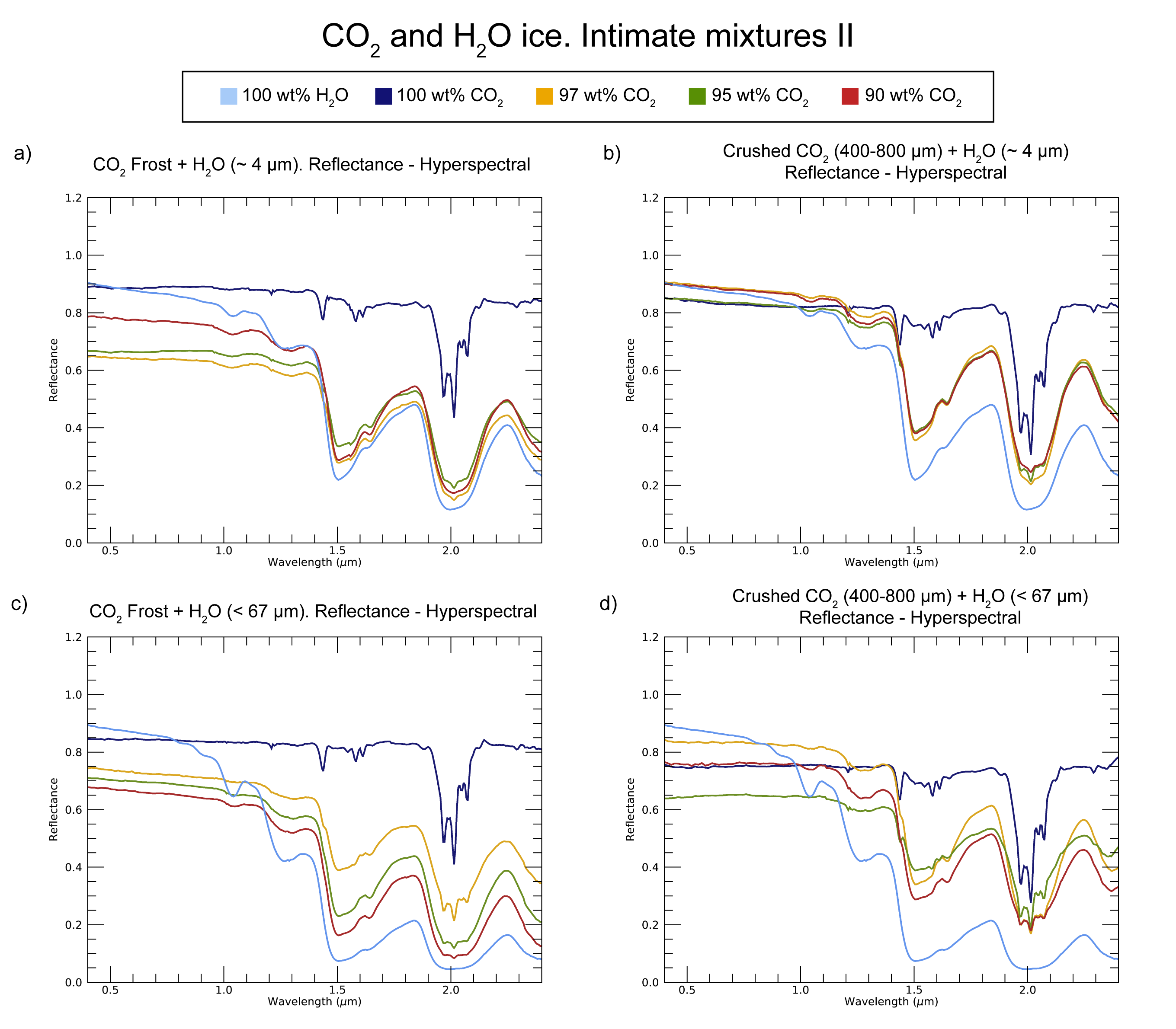}
\caption[Hyperspectral reflectance spectra of intimate mixtures of CO$_2$ and H$_2$O ice.]{Hyperspectral spectra of intimate mixtures of CO$_2$ and H$_2$O ices. The legend shows the CO$_2$ content of each sample. a) CO$_2$ frost (10-100 \textmu m) with fine grained H$_2$O ice ($\sim$ 4 \textmu m). b) Crushed CO$_2$ with fine grained H$_2$O ice ($\sim$ 4 \textmu m). c) CO$_2$ frost with coarser grained H$_2$O ice ($\sim$ 67 \textmu m). d) Crushed CO$_2$ with coarser grained H$_2$O ice ($\sim$ 67 \textmu m).}
\label{Fig:spipa_hyperspectra}
\end{figure}

Fig \ref{Fig:spipa_hyperspectra} shows the reflectance spectra of the CO$_2$ (frost, crushed) and H$_2$O (fine, coarser) intimate mixtures. We have added the pure spectra of the fine and coarser grained H$_2$O ices for comparison. 

Overall, we observe that the samples with CO$_2$ frost (10-100 \textmu m) sublimate faster than those with crushed CO$_2$. By the shape of the CO$_2$ absorption bands within that of water ice at 2.0 \textmu m, we see that CO$_2$ ice resists sublimation better when mixed with coarser H$_2$O ice particles than with finer ones.

Comparing Fig \ref{Fig:spipa_multispectra} and Fig \ref{Fig:spipa_hyperspectra}, we see that samples with more water show a greater drop in the reflectance from the multi- to the hyperspectral figures. As CO$_2$ ice sublimates, the CO$_2$ - to - H$_2$O decreases and water ice forms a lag that dominates the spectrum. 

\begin{figure}[h!]
\centering\includegraphics[width = \textwidth]{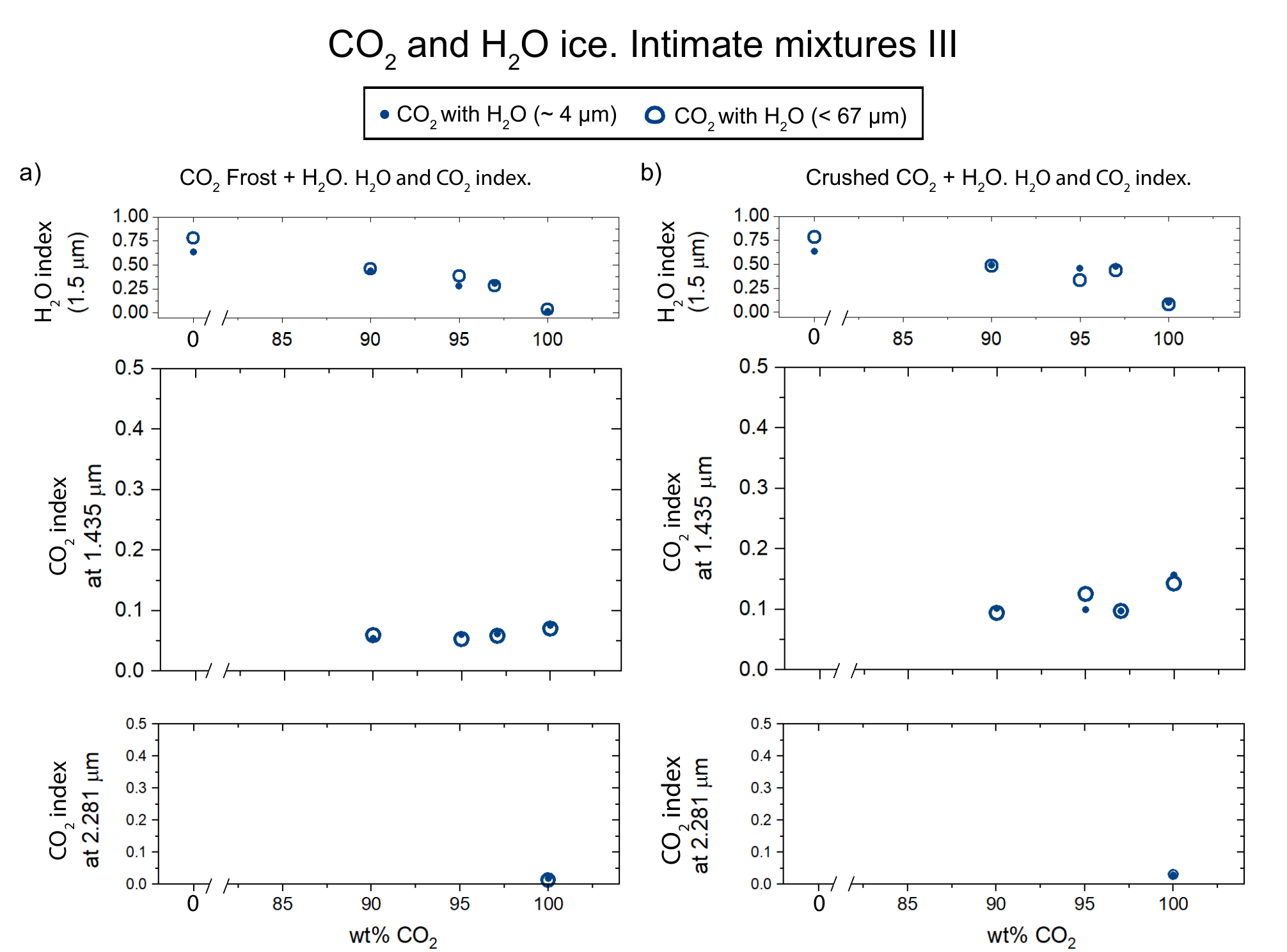}
\caption[Analysis of spectral criteria for intimate mixtures of CO$_2$ and H$_2$O ice.]{Analysis of different spectral criteria for intimate mixtures of CO$_2$ and H$_2$O ices in the form of H$_2$O ice particles of either $\sim$ 4 \textmu m or $\sim$ 67 \textmu m (full vs. open symbols), a) Using CO$_2$ frost (10-100 \textmu m). b) Using crushed CO$_2$ (400-800 \textmu m)}
\label{Fig:spipa_criteria1}
\end{figure}

 Fig \ref{Fig:spipa_criteria1} shows the CO$_2$ and H$_2$O indices for intimate mixtures with CO$_2$ frost (Fig \ref{Fig:spipa_criteria1}a) and crushed CO$_2$ (Fig \ref{Fig:spipa_criteria1}b). The H$_2$O indices of mixtures with 5 wt.$\%$ of water ice are already half of that of the same mixtures with 100 wt$\%$ of water ice.
 
The CO$_2$ indices decrease with the concentration of CO$_2$. A reduction of 10 wt$\%$ of CO$_2$ ice within the mixtures reduces the index at 1.4 \textmu m by $\sim$38$\%$ for the samples with CO$_2$ frost and by $\sim$33$\%$ for the samples with crushed CO$_2$. The index at 2.281 \textmu m is so weak that it cannot be measured in the majority of the cases.  

\begin{figure}[h!]
\centering\includegraphics[width = \textwidth]{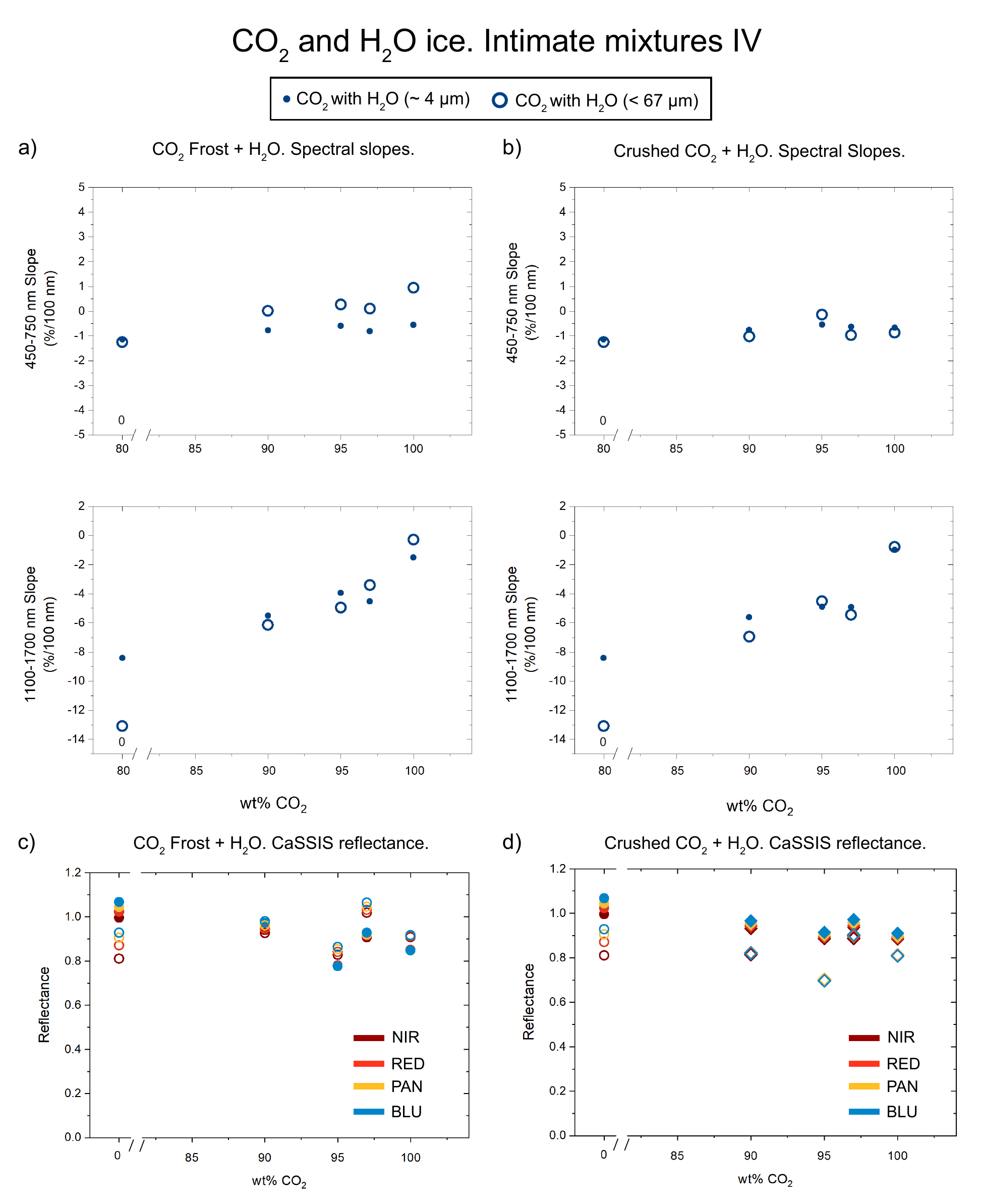}
\caption[Analysis of spectral criteria for intimate mixtures of CO$_2$ and H$_2$O ice, second part.]{Analysis of spectral criteria for intimate mixtures of H$_2$O ice and CO$_2$ ice in the form of left) frost (10-100 \textmu m) or right) crushed particles between 400 and 800 \textmu m. a) and b) Visible (top) and near-IR (bottom) spectral slopes. c) and d) Simulated reflectance with the CaSSIS filters.} 
\label{Fig:spipa_criteria2}
\end{figure}

We see that, in general, the VIS and NIR spectral slopes are bluer with increasing concentration of H$_2$O ice in the mixture (Fig \ref{Fig:spipa_criteria2}a and \ref{Fig:spipa_criteria2}b).

As seen in Fig \ref{Fig:spipa_multispectra} and Fig \ref{Fig:spipa_hyperspectra}, the visible part of the spectra is dominated by the texture of the samples rather than by their composition. Consequently, the reflectance measured with CaSSIS (Fig \ref{Fig:spipa_criteria2}c and \ref{Fig:spipa_criteria2}e) does not overall follow the composition of the samples. As CO$_2$ ice covers water ice and the spectral slope becomes neutral, the different filters measure similar values of reflectance. 

\subsection{Crushed CO$_2$ ice and JSC Mars-1 intimate mixtures} \label{R:JSC Mars-1}

\begin{figure}[h!]
\centering\includegraphics[width=\textwidth]{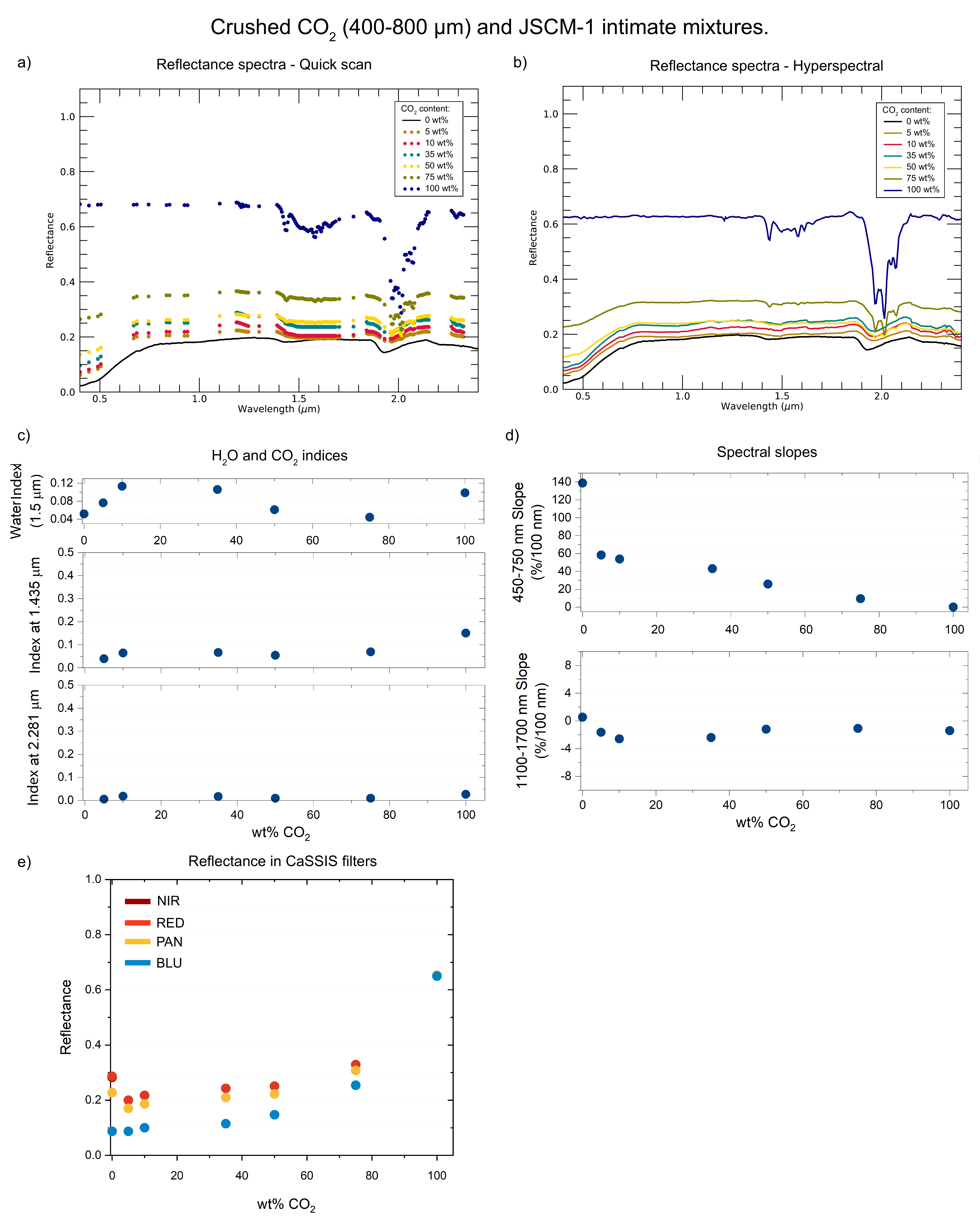}
\caption[Reflectance analysis of crushed CO$_2$ ice and JSC Mars-1 intimate mixtures.]{Reflectance analysis of crushed CO$_2$ ice and JSC Mars-1 intimate mixtures. a) Multispectral (quick scans) reflectance spectra. b) Hyperspectral reflectance spectra c) Strength of various absorption features: the one from H$_2$O at 1.5 \textmu m (indicative of the cold-trapped water) and the ones at 1.435 and 2.281 \textmu m from CO$_2$ ice. d) VIS and NIR spectral slopes of the samples. e) Simulated reflectance measured in the CaSSIS filters.} 
\label{Fig:JSCM1}
\end{figure}

Fig \ref{Fig:JSCM1}a and Fig \ref{Fig:JSCM1}b show that the reflectance of the intimate mixtures of JSC Mars-1 and crushed CO$_2$ drops as the amount of ice decreases. Compared to the spectra of crushed ice of same size distribution (Fig \ref{Fig:sizes}a,Fig \ref{Fig:sizes} b), the pure CO$_2$ ice shown here is about 20$\%$ darker. This can be due to the broad size distribution that we are using (400-800 \textmu m); if the histogram of grain size distribution peaks towards 800 \textmu m, then the reflectance is expected to be weaker than if it peaks towards 400 \textmu m. We also see that samples which initially contained up to 35 wt$\%$ of CO$_2$ ice only show a weak CO$_2$ absorption around 2.29 \textmu m. In the VIS, we can identify the red slope characteristic of JSC Mars-1 when the weight of dust is only a quarter of that of the sample.

We analyse in Fig \ref{Fig:JSCM1}c the CO$_2$ and H$_2$O signatures of the samples. The spectrum of JSC Mars-1 shown here was not measured together with the rest of the samples, and therefore its H$_2$O index reflects its degree of hydration. The water index shows no correlation between the cold-trapped water frost and the CO$_2$ wt$\%$. Instead, we attribute the variation in cold-trapped water frost to the process of filling of the sample holders, which was longer than usual as there were more samples to fill. Both CO$_2$ signatures tend to increase with the weight percentage of CO$_2$; as observed previously, the signature at 1.435 \textmu m is stronger than the one at 2.281 \textmu m.

Fig \ref{Fig:JSCM1}d shows that the VIS slope of the sample with 5 wt$\%$ of CO$_2$ ice is reduced by a 50$\%$ compared to JSC Mars-1, and it continues decreasing as the percentage of carbon dioxide increases. In the NIR, the spectral slope remains flat-to-bluish, oscillating from 0 to -4 $\%$/100 nm following the amount of water frost condensed at the surface of the samples. 

Measured with CaSSIS (Fig \ref{Fig:JSCM1}e), the presence of JSC Mars-1 results in reflectance measurements in the BLU filter around 30$\%$ weaker than in other filters. This gap between the reflectance measured in the different filters narrows as the percentage of ice increases until it disappears entirely for the pure CO$_2$ sample. With 75wt$\%$ of CO$_2$ ice, the BLU filter still measures 20$\%$ lower reflectance than the other ones. This value is similar to the ones measured in \citep{Yoldi:2020} for water ice and JSC Mars-1 intimate mixtures. 

\subsection{Sublimation series of the ternary mixtures} \label{R:Cake}

\begin{figure}[h!]
    \centering\includegraphics[width=\textwidth]{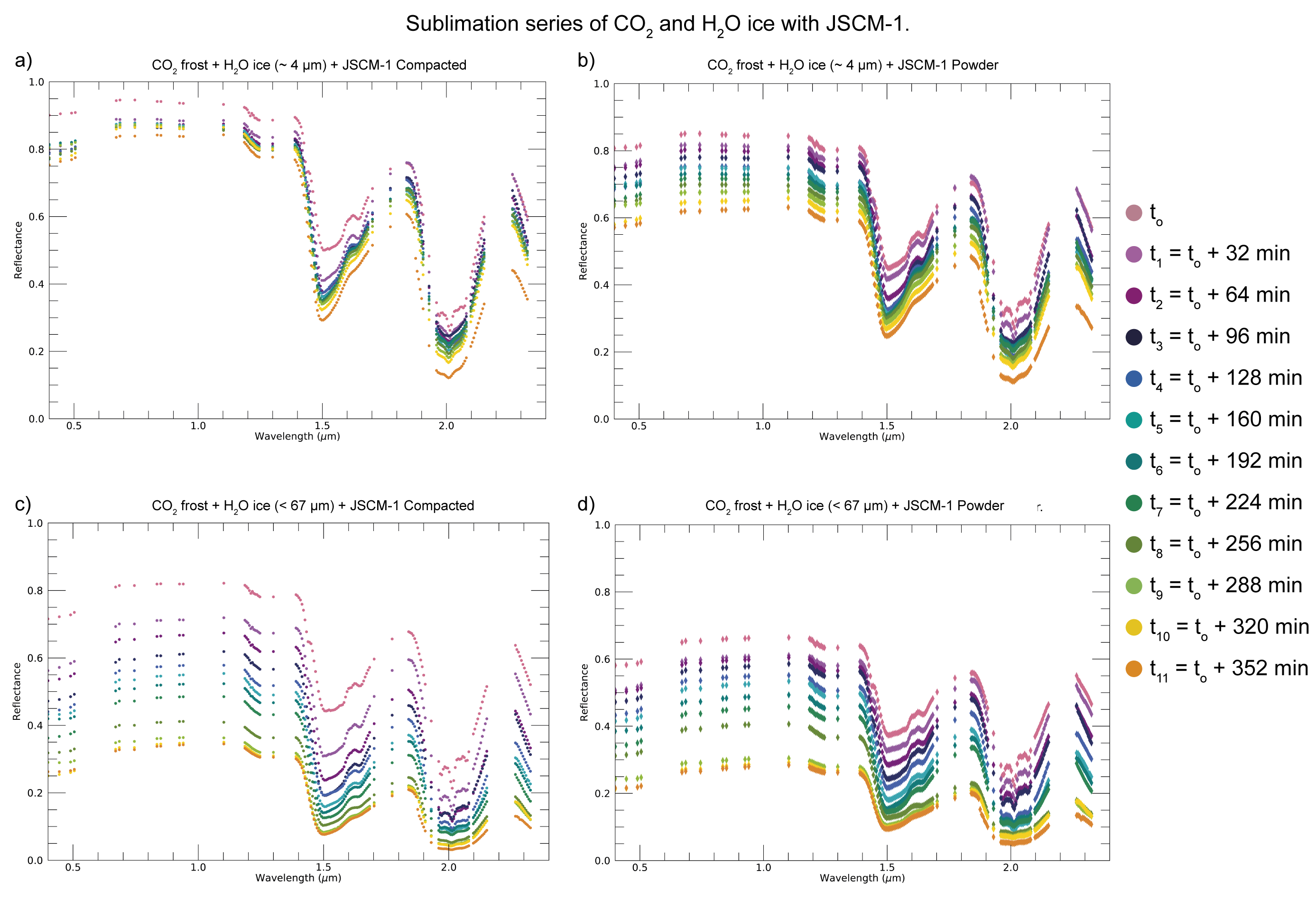}
    \caption[Reflectance spectra of ternary intimate mixtures (CO$_2$ and H$_2$O ice with JSC Mars-1).]{Reflectance spectra of ternary intimate mixtures (CO$_2$ and H$_2$O ice with JSC Mars-1). a) Compacted mixture of CO$_2$ frost (10-100 \textmu m), fine grained H$_2$O ice ($\sim$ 4 \textmu m) and JSC Mars-1. b) Powdered mixture of CO$_2$ frost, fine grained H$_2$O ice and JSC Mars-1. c) Compacted mixture of CO$_2$ frost, coarser grained H$_2$O ice ($\sim$ 67 \textmu m) and JSC Mars-1. d) Powdered mixture of CO$_2$ frost, coarser grained H$_2$O ice and JSC Mars-1.} 
    \label{Fig:cake_spectra}
\end{figure}

Fig \ref{Fig:cake_spectra} is a 2x2 matrix of plots showing the evolution of the reflectance spectra of the ternary mixtures as the CO$_2$ and H$_2$O ices sublimated. The rows correspond to the size of water ice (fine or coarser) and the columns to the texture of the samples (compact or powder). The four samples initially had the same composition: 96.8 wt$\%$ of CO$_2$ frost, 3 wt$\%$ of water ice and 0.2 wt$\%$ of JSC Mars-1 ( particles smaller than 100 \textmu m).

We observe similarities with the binary mixtures, namely: \begin{inparaenum}[1)]
    \item A shape of the spectra dominated by water ice, even if present at low concentrations.
    \item A first, relatively abrupt, darkening of the sample (between t$_0$ and t$_1$), explained by the quick sublimation of cold-trapped water frost.
    \item A progressive change in the shape of the water bands towards broader and less complex bands, indicating the warming up of the samples.
    \item The VIS (400-950nm) shows greater variation in these plots, since that region is more sensitive to changes in the texture. 
\end{inparaenum} 

The compacted samples initially show a higher reflectance than the powdered ones, partly explained by the trapped water frost.

We observe a non-intuitive behaviour of the absorption band of CO$_2$ around 2 \textmu m in the samples with fine H$_2$O ice, which does not seem to occur with coarser H$_2$O ice. The band decreases to a minimum at t$_3$ to later on grow again and be noticeable by the end of the experiment, when the temperature of the air above the sample was of 250 K (55 K above the sublimation point of CO$_2$ at atmospheric pressure (Fig \ref{Fig:cake_temperature})).

\begin{figure}[h!]
    \centering\includegraphics[width=0.76\textwidth]{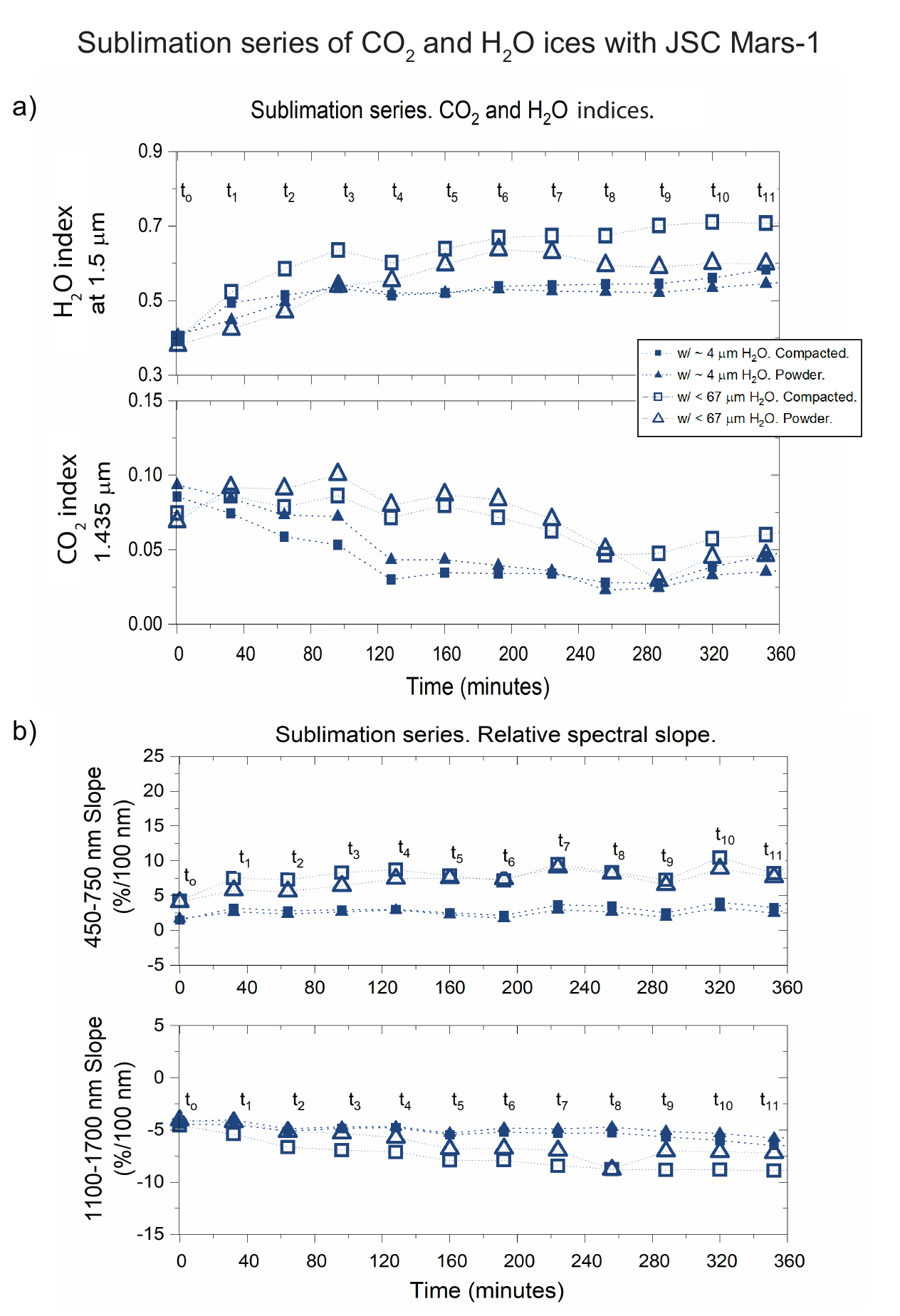}
    \caption[Analysis of spectral criteria of ternary mixtures of CO$_2$/H$_2$O ice with JSC Mars-1.]{Analysis of spectral criteria of ternary mixtures of CO$_2$/H$_2$O ice with JSC Mars-1. a) Analysis of the water (1.5 \textmu m) and carbon dioxide (1.435 and 2.281 \textmu m) indices. b)  VIS (top) and NIR (bottom) spectral slopes.} 
    \label{Fig:cake_criteria}
\end{figure}

The evolution of carbon dioxide and water indices (Fig \ref{Fig:cake_criteria}a) shows a growing H$_2$O index at 1.5 \textmu m as CO$_2$ sublimates. For all the samples, there is an initial, steep increase of the water index, that slows down after t$_3$. As observed in Fig \ref{Fig:cake_spectra}, the evolution of the CO$_2$ index depends on the size of the water ice. The CO$_2$ indices in samples with fine grained water ice decreases by around half from t$_0$ to t$_4$ and stay constant around 0.025 until t$_9$. The CO$_2$ indices in samples with coarser grained water ice oscillate around 0.075 from t$_0$ to t$_4$ and then decrease down to 0.05 until t$_9$. After that, the CO$_2$ indices of every sample start increasing until the end of the measurement. We do not detect any absorption at 2.281 \textmu m during the duration of the experiment.

Fig \ref{Fig:cake_criteria}b shows the evolution of the visible and near-infrared spectral slopes of the ternary mixtures. The ternary mixtures are a combination of components with different spectral slopes; \begin{inparaenum}[(i)]
    \item CO$_2$ ice, with no slope in both the VIS and NIR ranges,
    \item H$_2$O ice, with blue slopes in the visible ($\sim$-1$\%/$100 nm) and near-infrared (SPIPA-A: -8.4$\%/$100 nm, SPIPA-B: -13$\%/$100 nm) spectral ranges, and
    \item JSC Mars-1, red in the visible ($\sim$140$\%/$100 nm) and flat in the near-infrared range.
\end{inparaenum} Because of the neutral slopes of CO$_2$ ice, this component only impacts the reflectance slopes by how much it covers or uncovers the other components.

In the visible range of the spectrum, the slopes are expected to become redder as water ice sublimates and the relative abundance of JSC Mars-1 increases. The ternary mixtures with coarser grained water ice show such a behaviour, as a dust lag formed with the sublimating ice. The spectral slopes of ternary mixtures with fine grained H$_2$O ice, however, remained constant around 2.5 $\%$/100 nm during the six hours of the experiment; the scattering by small ice particles dominated the reflectance. We can assume that we would have seen the same dust lag after some additional hours of sublimation. In the same way, the NIR spectral slopes of these samples are dominated by water ice. Because there is no water absorption associated with the transient decrease in the spectral slope of the powdered sample with coarser grained water ice (see Fig \ref{Fig:cake_spectra}a) at t$_8$, we assume this point to possibly be an instrumental artefact.

\subsection{Comparison of criteria. Addition of water ice and dust experiments} \label{R:comparison}

\begin{figure}[h!]
    \centering\includegraphics[width=0.8\textwidth]{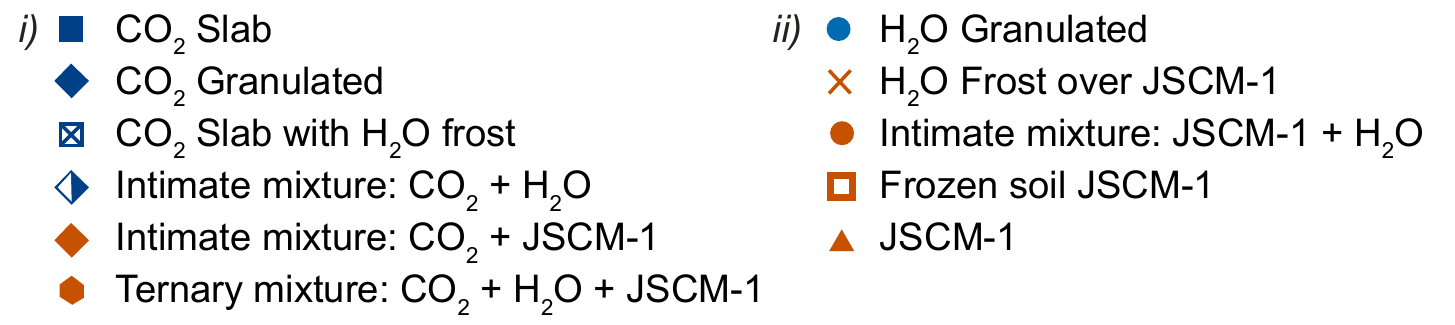}
    \caption[Legend for comparison criteria figures.]{Legends applicable to Figs. \ref{Fig:comparison_criteria_1} and \ref{Fig:comparison_criteria_2}. i) Legend applicable to the CO$_2$ experiments. ii) Legend applicable to the H$_2$O experiments, imported from \cite{Yoldi:2020}}
    \label{Fig:comparison_criteria_legend}
        
\end{figure}

\begin{figure}[h!]
    \centering\includegraphics[width=\textwidth]{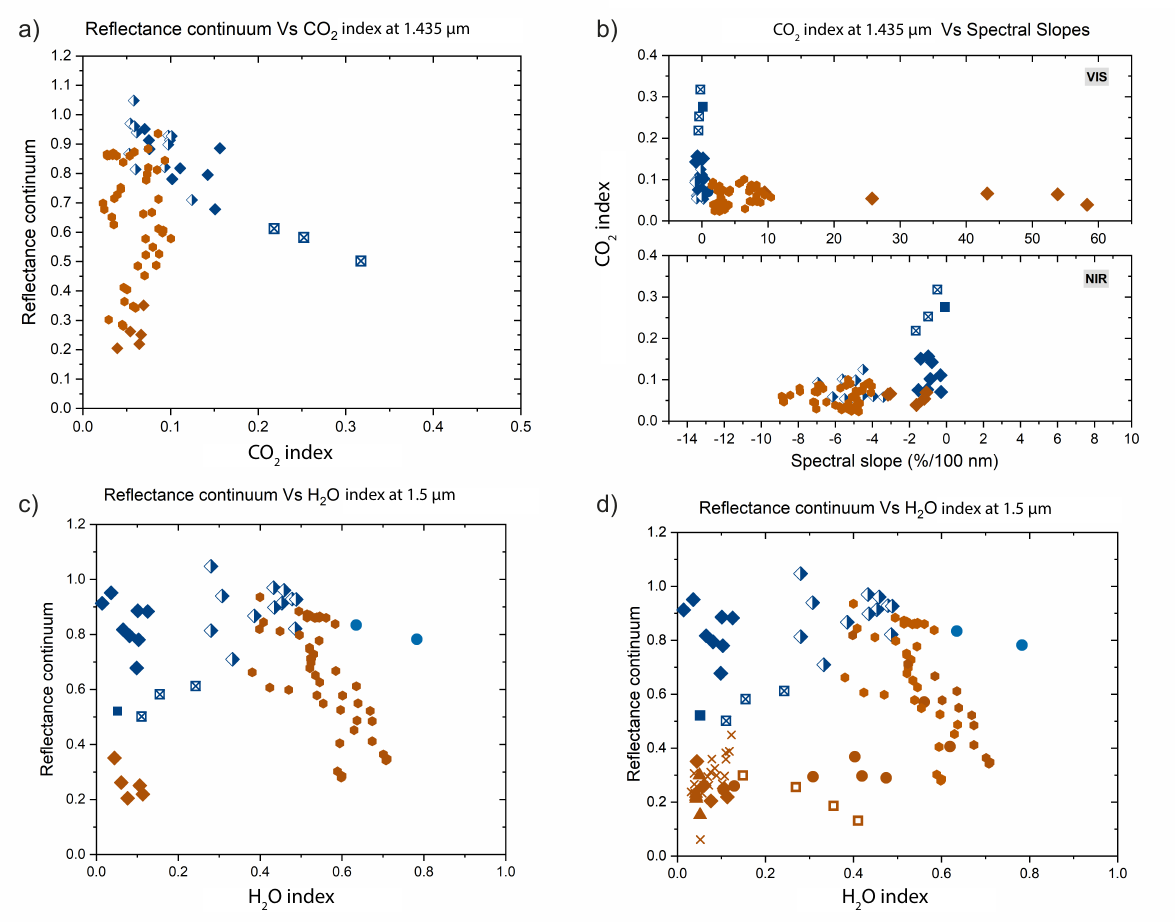}
    \caption[Comparison of spectral criteria for every experiment. Addition of water ice and dust experiments.]{Comparison of spectral criteria between experiments. Addition of H$_2$O/dust experiments from \cite{Yoldi:2020}. The legend for these plots is provided in Figure \ref{Fig:comparison_criteria_legend}. Tip: the colours inform about the composition of the samples, and the symbols about their mixing mode/texture  a) Reflectance of the continuum v. CO$_2$ index (1.435 \textmu m). b) CO$_2$ index (1.435 \textmu m) VS the VIS (top) and NIR (bottom) spectral slopes. c) Reflectance of the continuum VS the H$_2$O index (1.5 \textmu m). d) Same as c) with water ice/dust experiments.} 
    \label{Fig:comparison_criteria_1}
\end{figure}

We compare now the criteria defined in Section \ref{S:Methods} for all the experiments. We aim to find trends among the samples to eventually identify them in terms of composition and mixing mode. We used the same strategy in the first part of this study, where we focused on water ice and dust associations \citep{Yoldi:2020}. Hence, we have added to some of the plots the experiments with water ice and JSC Mars-1 conducted in \citet{Yoldi:2020}. We provide here a summary of the water ice experiments to ease reading of the figure. The corresponding legend is given in Fig \ref{Fig:comparison_criteria_legend}\textit{ii)}.

\begin{enumerate}
    \item \textit{JSCM-1 + Water frost samples}: experiments in which water frost was condensed onto various size fractions of JSC Mars-1. By sporadically exposing the samples to the atmosphere, we studied the reflectance of different thickness of water frost on the regolith.
    \item \textit{Intimate mixtures}: JSC Mars-1 and SPIPA-A or SPIPA-B mixtures. We mixed the original size distribution of JSC Mars-1 with 10, 35, 50 and 75 wt$\%$ of SPIPA-A and SPIPA-B separately. 
    \item \textit{Frozen soils}: experiments in which we thoroughly wet various size fractions of JSC Mars-1 and froze them. The result was a matrix of water ice surrounding the JSC Mars-1 grains. We measured the reflectance of the samples as ice sublimated. 
\end{enumerate}

Fig \ref{Fig:comparison_criteria_1}a compares the reflectance of the continuum (0.940 \textmu m) to the CO$_2$ indices of the samples. For clarity, the X-axis only shows values up to 0.5. The pure ice samples (blue symbols) occupy reflectance values from $\sim$ 1.0 to 0.5 and CO$_2$ indices between 0.1 and 0.3. JSC Mars-1 and crushed CO$_2$ intimate mixtures group on the bottom left of the graph, with reflectance values and CO$_2$ indices lower than 0.4 and 0.1 respectively. The ternary mixtures cover a large range of reflectance values and are grouped on the left side of the graph with CO$_2$ indices lower than $\sim$0.1. 

We compare in Fig \ref{Fig:comparison_criteria_1}b the CO$_2$ index and the VIS (top graph) and NIR (bottom) spectral slopes. The graph at the top is L-shaped. There is a vertical column around 0-values of the visible spectral slopes formed by pure ice samples. Samples with JSC Mars-1 fill the horizontal section of the graph with CO$_2$ index values of 0.1 and a wide range of spectral slopes depending on their content of JSC Mars-1. For reference, the visible spectral slope of pure JSC Mars-1 exceeds 100$\%$/100 nm. The graph on the bottom shows a shrunk, flipped L-shape. This plot differentiates intimate mixtures of CO$_2$ and H$_2$O mixtures (around -3 and -7 $\%$/100 nm) from pure CO$_2$ samples that trapped water frost (around 0 and -2 $\%$/100 nm). Ternary mixtures show the bluest slope of all samples, ranging from -4 to almost -10 $\%$/100 nm. Intimate mixtures of crushed CO$_2$ and JSC Mars-1 fall together with the pure CO$_2$ samples, except for the points that correspond to samples with more JSC Mars-1.

In Fig \ref{Fig:comparison_criteria_1}c, we compare the reflectance in the continuum and the H$_2$O index at 1.5 \textmu m. The experiments with CO$_2$ and hydrated minerals (JSC Mars-1), the granular CO$_2$ that trapped atmospheric water and the experiments where water frost had condensed on a CO$_2$ slab cluster to the left, below H$_2$O indices of 0.2. The exception is the ROI on the slab that was completely covered by water frost, whose water index is $\sim$0.25. As seen previously, this group of experiments presents a wide range of reflectance values, from $\sim$1.0 for the finest CO$_2$ grains to 0.2 for the samples containing high amounts of JSC Mars-1. Intimate mixtures of CO$_2$ and H$_2$O ice form a group above reflectance values of 0.7 and between H$_2$O index of 0.2 and $\sim$0.5. These points get progressively closer to the values of pure water ice. They do not, however, follow a linear path since -as seen previously- the reflectance spectra in the visible is more dependent on the texture of the sample than on the CO$_2$-to-H$_2$O ratio. Finally, ternary mixtures are seen at high values of H$_2$O indices (0.4-0.8) and spread reflectance values (0.25-1.0). 

In Fig \ref{Fig:comparison_criteria_1}d, we incorporate the water ice and JSC Mars-1 samples from \citet{Yoldi:2020}. CO$_2$ and JSC Mars-1 intimate mixtures cluster together with the JSC Mars-1 samples on which water frost condensed, which indicates that intimate mixtures trapped atmospheric water. Fig \ref{Fig:comparison_criteria_1}d also evinces that CO$_2$ ice (non-absorptive material) intimately mixed with only a 3 wt$\%$ of water ice shows water indices as strong as JSC Mars-1 (absorptive material) intimately mixed with 35 wt$\%$ of water ice. We also note that as the loose ternary mixtures sublimate, they join the points representing the intimate mixtures of water ice and JSC Mars-1.

\begin{figure}[ht!]
    \centering\includegraphics[width=\textwidth]{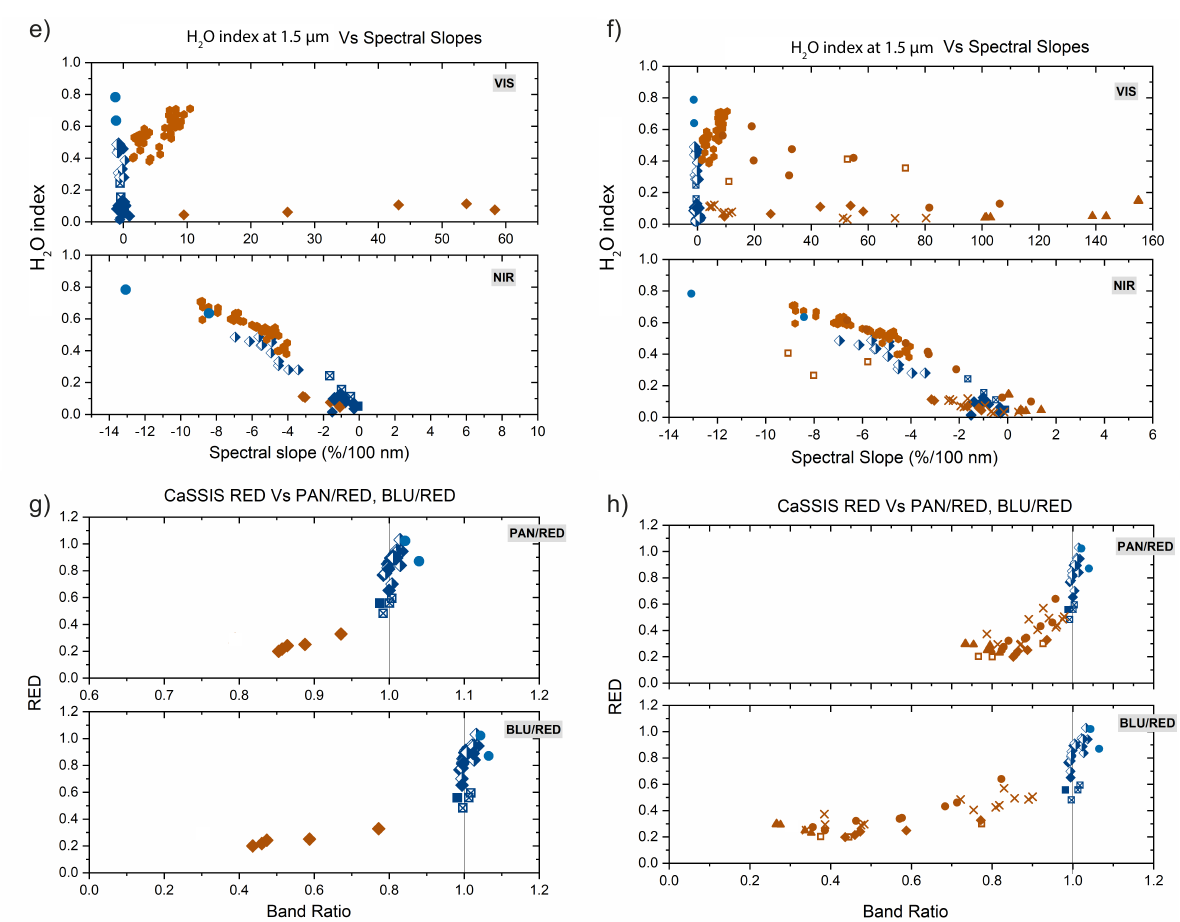}
    \caption[Comparison of spectral criteria for every experiment. Addition of water ice and dust experiments.]{Continuation of Figure \ref{Fig:comparison_criteria_1}. Comparison of spectral criteria between experiments. Addition of H$_2$O/dust experiments from \cite{Yoldi:2020}. The legend for these plots is shown in Figure \ref{Fig:comparison_criteria_legend} e) Water index at 1.5 \textmu m versus the VIS (top) and NIR (bottom) spectral slopes. f)  Same as \textit{e)} with water ice/dust experiments. g) Reflectance simulated in the RED filter of CaSSIS versus the PAN/RED (top) and BLU/RED (bottom) ratios. h) Same as \textit{g)} with the water ice /dust experiments.} 
    \label{Fig:comparison_criteria_2}
\end{figure}

In the top graph of Fig \ref{Fig:comparison_criteria_2}e, we compare the water index at 1.5 \textmu m with the VIS spectral slope. The distribution of the points is similar to that observed in Fig \ref{Fig:comparison_criteria_1}b. Here, however, the ternary mixtures are easily distinguishable from the JSC Mars-1 and CO$_2$ ice intimate mixtures, since they show high H$_2$O indices (0.4 to 0.7 compared to a maximum of 0.2 for the mixtures without water ice). This graph also distinguishes between CO$_2$ samples that trapped small quantities of atmospheric water and intimate mixtures of CO$_2$ and H$_2$O ices. The graph on the bottom compares the water index to the NIR spectral slope: all the samples follow a similar path from flat slopes towards the slope values of pure water ice. 

When we add the water experiments (Fig \ref{Fig:comparison_criteria_2}f), we see that, for the VIS, most of the water experiments fill in the gap left by the L-shaped graph drawn by the CO$_2$ samples. The JSCM-1/water frost experiments overlap again the CO$_2$ and JSC Mars-1 intimate mixtures. Looking at the NIR spectral slopes, the JSC Mars-1 and water frozen soils fall outside the trend drawn by the CO$_2$ experiments. 

Finally, we compare in Fig \ref{Fig:comparison_criteria_2}g the reflectance measured in the RED filter of CaSSIS to the PAN/RED and BLU/RED filter ratios. We have zoomed into the X-axis of the PAN/RED band ratio. We do not include in these graphs the ternary mixtures since we only performed quick scans with those samples, which does not allow us to perform CaSSIS colour simulations (see Section \ref{S:Methods}). Both graphs allow us to separate the samples bearing JSC Mars-1 from those of pure ice. All the pure ice samples gather around band ratios of 1, samples with pure water being redder than the ones with CO$_2$. When we add the water ice samples to the plot (Fig \ref{Fig:comparison_criteria_2}h), CO$_2$ and JSC Mars-1 intimate mixtures do not align with the water ice and JSC Mars-1 intimate mixtures. Instead, intimate mixtures of CO$_2$ and JSC Mars-1 fall in between the intimate mixtures of water ice and JSC Mars-1 and the frozen soils. This is because CO$_2$ ice has a redder spectral slope than water ice. Hence, the intimate mixtures with CO$_2$ and JSC Mars-1 plot below and parallel to the intimate mixtures with H$_2$O ice and JSC Mars-1.
\section{Discussion} \label{S:Discussion}

\subsection{Spectro-photometry of pure CO$_2$}
We have measured the spectra of various size fractions of CO$_2$ ice in the spectral range 0.4-2.4 \textmu m. According to \citet{Singh:2016}, in a semi-infinite layer, the size of the CO$_2$ ice grains affects less the reflectance than the size of H$_2$O ice. In our experiments however, the samples were only 2-cm thick. While the particles of the CO$_2$ frost (10-100 \textmu m, see Fig. \ref{Fig:co2}) are small enough that this particular sample is optically thick (see the discussion of this topic by \citet{Hapke1993}), the dark neutral substrate of the sample holder (constant reflectance of 0.05) certainly affects the continuum level of the slab and crushed slab samples. Indeed, because of the low absorptivity of CO$_2$ in the continuum, photons were able to reach the bottom of the sample holder. The fact that the samples are not optically thick probably explains the dependence of continuum reflectance on particle size. But with the exception of the slab sample (reflectance reduced to 0.5), the effects of the substrate are relatively minor, with a maximum decrease of reflectance of 0.2 only between the coarsest fraction (400-800 \textmu m) and the optically thick snow. This is less that what would be expected from models which predict a very high transparency of samples made of such large CO$_2$ particles \citep{Singh:2016}. The fine-scale texture at the surface of the grains as well as the presence of interfaces, pores or defects in the interiors of the grains probably increases the scattering and decreases the penetration compared to idealised smooth and compact particles. The situation is different at wavelengths where the absorption is not negligible and the variations of the absorption bands intensity between the different particle size fractions are essentially the results of the change of particle size. However, spectral criteria such as band depth which consist essentially in a contrast of reflectance between the band and the continuum will still be affected. Therefore, care must be taken when comparing directly these spectra of pure CO$_2$ ice with Martian spectra but as the reflectance of the substrate and the thickness of the samples are known, these spectra remain nevertheless useful to compare to the outputs of models and simulation.

Our spectra of crushed CO$_2$ ice show a strength of the 2 \textmu m band similar to the one of the "fractured layer" shown by \citep{kieffer:1968}. We see much more structure in this band, however, probably because of the higher spectral resolution of our measurements. Similarly we see strong but narrow absorption at 1.44 and 1.6 \textmu m, which are not visible in the spectrum of \citep{kieffer:1968}. Our spectrum of CO$_2$ frost (10-100 \textmu m) appears very similar to the one labeled as "medium-grained" in \citep{kieffer:1968} despite the same differences as observed with crushed CO$_2$, again probably because of the different spectral resolution. While their is no estimate for how large this "medium-grained" frost is, it is larger than the "fine-grained" CO$_2$ frost which displays surface texture at the 10-20 \textmu m scale. This is consistent with the microscope observations of our CO$_2$ frost which show large particles of a few tens of \textmu m in diameter.

Note that on planetary surfaces as in the laboratory, different grain sizes can lead to different capacities of atmospheric water frost trapping, which also indirectly affect the reflectance of the samples. 

Regardless of the grain size, CO$_2$ is overall poorly absorptive in the studied spectral range, with some deep but narrow absorption bands longward of 1.1 \textmu m. In our laboratory spectra, CO$_2$ presents flat visible spectra; \citet{Singh:2016} modelled, however, a red slope at wavelengths shorter than 0.5 \textmu m, which we have not been able to reproduce, possibly due to the contamination of the surface of the samples by H$_2$O frost. 

Further comparison between modeling results \citep{Singh:2016} and our experimental spectra show other differences. In particular, the strength of the CO$_2$ absorption bands in our spectra are consistent with modeled spectra of much smaller particle size. For instance, our spectrum for the (400-800) \textmu m size fraction displays a band depth of about 0.5, similar to the modeled spectrum with a particle size of 100 \textmu m in the work of \citet{Singh:2016}. Also surprisingly, the spectrum of our compact slab shows band strengths much smaller than the modeled spectrum for a particle diameter of 1.5mm and seems more consistent with a modeled grain size of 500 \textmu m only. These discrepancies between experimental and modeled results probably originate from the non-ideal particles produced for the experiments that display irregularities in shape, fine-scales surface texture, internal defects, pores and joints between grains. All these features of complex real particles combine to scatter the light at scales much smaller than the actual diameter of spherical particles assumed in models.

\subsection{CO$_2$ as a component of a mixture}
As part of a mixture, and because of the flat spectrum of CO$_2$ ice, CO$_2$ acts as a reddening material when mixed with a bluer one (e.g., water), and as a bluing one when mixed with a redder one (e.g., JSC Mars-1). In this regard, CO$_2$ ice has similar effects on the reflectance spectra as the early stages of water frost deposition. Small amounts of water frost condensed on some samples increase the reflectance and flatten the spectral slopes of the materials underneath, without producing any significant increase in the H$_2$O indexes \citep{Yoldi:2020}. Unfortunately, these similarities complicate the differentiation of CO$_2$ and H$_2$O frosts from analysis of visible colours only. 

In the first paper of this series, we showed how small amounts of JSC Mars-1 were enough to mask water ice when both materials were intimately mixed. Here, we observe how small amounts of water ice intimately mixed with carbon dioxide ice can mask the CO$_2$. This is not a new result as it was already shown experimentally by \citet{Kieffer:1970} (from 0.8 \textmu m) and predicted theoretically \citep{Warren:1990}.

We explained in \citet{Yoldi:2020} that the reflectance of water ice and dust intimate mixtures does not increase linearly with the amount of water ice. Instead, the spectra of intimate mixtures are controlled by the most absorptive component. JSC Mars-1 is the absorptive material when mixed with H$_2$O ice, but when we mix H$_2$O and CO$_2$ ices, water ice is the most absorptive material. Consequently, the reflectance spectra of intimate mixtures of water and carbon dioxide ices are dominated by water ice. This situation reverses at wavelengths where water becomes less absorptive than CO$_2$. Parameters such as grain size or association mode between the materials can balance this behaviour of the reflectance spectra \citep{Yoldi:2020, Warren:1990, Singh:2016}. It is still important to remember that, when looking at planetary surfaces in the 0.4-2.4 \textmu m spectral range, a surface showing a water-like spectrum is not necessarily free of CO$_2$, in the same way as a surface showing a dust spectrum does not exclude the presence of ice \citep{Yoldi:2015}. 

\subsection{Ternary mixtures of CO$_2$, H$_2$O and JSC Mars-1}
\begin{figure}[!hbt]
    \centering\includegraphics[width=0.8\textwidth]{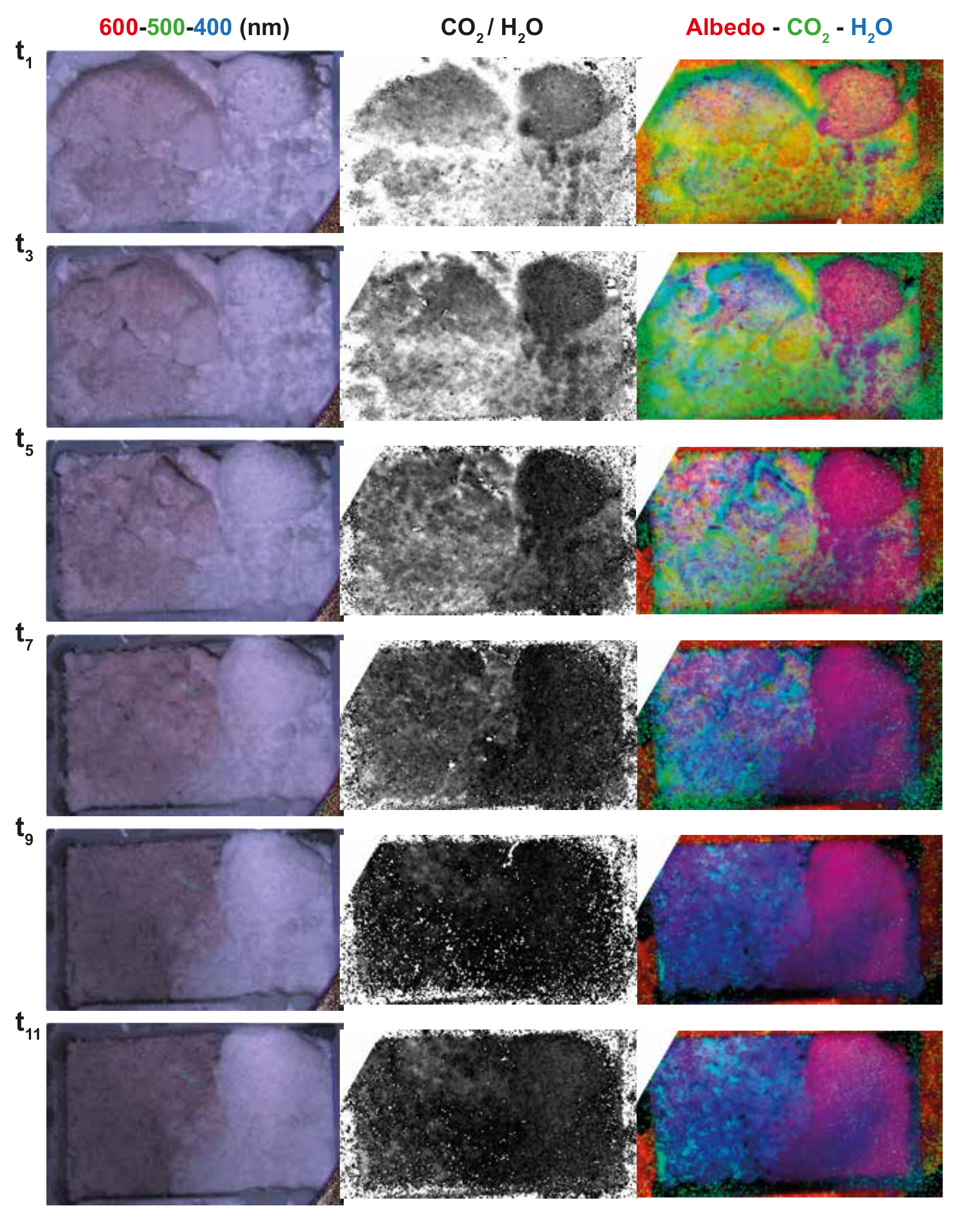}
    \caption{Sublimation series of the ternary mixtures of CO$_2$ frost (10-100 \textmu m), water ice (fine and coarser grained, respectively) and JSC Mars-1. In the left column, we show RGB composites with the signal at 600 nm in the red channel, at 500 nm in the green channel and 400 nm in the blue channel. The column in the middle is a monochromatic image containing the CO$_2$/H$_2$O ratio. The images in this column are stretched from 0.02 to 0.3. The right column shows RGB composites with the continuum reflectance in the red channel, the CO$_2$ index in the green channel and the H$_2$O index in the blue channel. Stretching: R(0.318-0.520), G(0.214-0.306), B(0.339-0.575). The rows correspond to the samples at odd times between t$_{1}$ and t$ _{11} $.}
    \label{Fig:Evol_cake}
\end{figure}
Fig \ref{Fig:Evol_cake} shows half of the steps of the sublimation process of ternary mixtures. The complete sequence can be found, in video format, in the supplementary material. We have produced RGB colour composites with three wavelengths (600, 500 and 400 nm) and three spectral parameters (reflectance in the continuum, CO$_2$ signature and H$_2$O signature), as well as monochromatic images representing the CO$_2$-to-H$_2$O ratio.  

The images on the left column of Fig \ref{Fig:Evol_cake} provide a good approximation to the natural colours of the samples and reveal a different evolution of the mixture with fine grained ($\sim$ 4.5 \textmu m, right) and coarser grained ($\sim$ 67 \textmu m, left) H$_2$O ice. The mixture with fine H$_2$O ice particles looks brighter than the one with coarser H$_2$O ice particles since the scattering by the small particles dominates the reflectance throughout the entire sequence. Besides the brightness, we observed a distinct evolution of the texture and morphology of the samples. In the mixtures with coarser H$_2$O ice particles, the compacted sample (top of the image, see Fig \ref{M:ternary_mixtures}) broke apart as the volatile components sublimated. At the same time, JSC Mars-1 deposited on top of the mixtures, conferring to the sample its characteristic reddish colour. Thus, the sample with coarser H$_2$O ice appears in t$_{11}$ as an almost flat deposit of JSC Mars-1 mixed with the remaining water ice. The sample with finer H$_2$O ice, however, seemed to have undergone a process of expansion (this is best appreciated on the video): while in t$_1$ we can still identify the flat surface of the compacted sample, from t$_5$ the sample resembles a porous snowball because of the ice filaments it has grown on the surface. The powdered sample (below the compacted one) went through the same process, which progressively softened its surface. 

The images on the column in the middle show the evolution of the CO$_2$/H$_2$O ratio as both sublimated. The pictures have been stretched so that black corresponds to CO$_2$/H$_2$O = 0.02 and white to CO$_2$/H$_2$O = 0.3. t$_1$ already presents the samples bearing fine ice slightly darker than the ones with coarse ice. This contrast increases with time so that at t5 the signatures of water already dominate the spectrum of the samples where H$_2$O ice is fine grained (4.5 \textmu m). The images at t$_9$ and t$_{11}$, although noisy, show a little brightening of the samples, related to the increase in the CO$_2$ signal we measured in Figure \ref{Fig:cake_criteria}a. 
Finally, the RGB composites of spectral criteria give an overview of the evolution of the albedo and the CO$_2$ and H$_2$O signatures, represented by the red, green and blue colours, respectively. The colours reveal that, from t$_5$, a layer of water covers the fine-grained mixtures, while green tones are still visible in the coarse-grained ones. In the pictures from t$_7$ and t$_9$, we observe that CO$_2$ remains preferentially in the creases of the material, and in t$_{11}$, the strengthening of the signal of CO$_2$ appears as sparse green dots. 

\begin{figure}[!hbt]
    \centering\includegraphics[width=\textwidth]{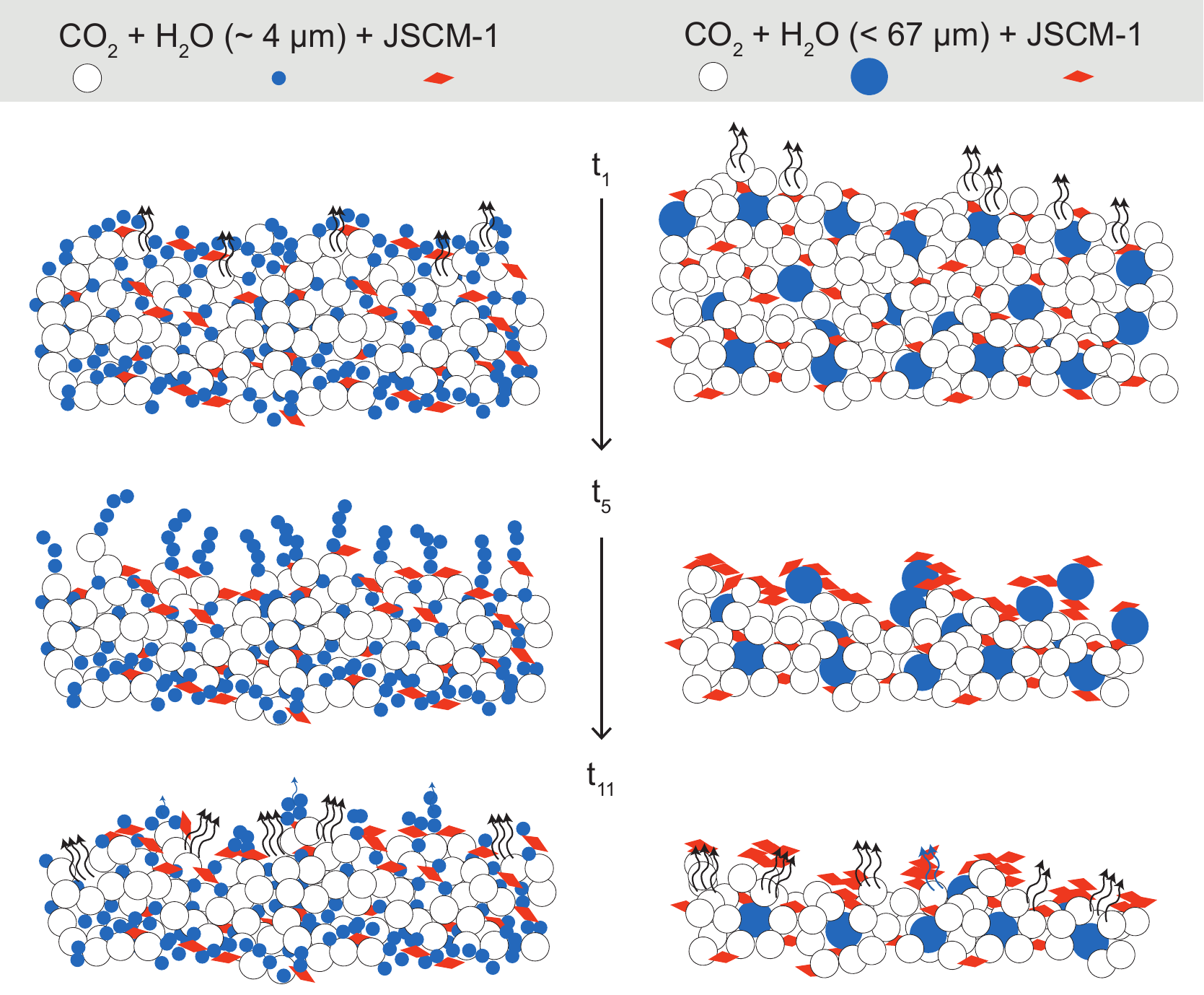}
    \caption{Schematic view of the sublimation processes for ternary mixtures of CO$_2$ frost (10-100 \textmu m), H$_2$O ice (two different particle sizes) and a finer fraction ($<$ 100 \textmu m) of the JSC Mars-1 regolith simulant. To the left, mixtures with fine H$_2$O ice particles ($\sim$ 4.5 \textmu m) and to the right with coarser($\sim$ 67 \textmu m) particles. The three rows represent different times in the sublimation process, increasing from top to bottom and corresponding approximately to t$_{1}$, t$_5$ and t$_{11}$ on Fig \ref{Fig:Evol_cake}. As CO$_2$ ice (blue particles) sublimates, the JSC Mars-1 dust analogue (red) accumulates at the surface in both cases. The process of sublimation seems to differ depending on the size of the water particles. In the case of coarse ($\sim$ 67 \textmu m) H$_2$O particles, they simply accumulate at the surface and eventually sublimate. The fine ($\sim$ 4.5 \textmu m) H$_2$O ice particles display a different behaviour as they clump together, forming long and thin filaments of water ice which progressively cover most of the surface but are occasionally ejected by the flux of CO$_2$ gas produced by the sublimation of the ice.}
    \label{Fig:Illustration_cake}
\end{figure}

From the information that these images provide, we have reconstructed the process of sublimation of the ternary mixtures (Fig \ref{Fig:Illustration_cake}). In the beginning (t$_1 \sim$ 160 K), the main component subject to sublimation is CO$_2$. As it sublimates in the mixture with coarser H$_2$O ice particles, the surface concentrations of water ice and JSC Mars-1 increase. Around t$_5$ ($\sim$ 195 K, i.e., the sublimation temperature of CO$_2$ ice at 1 atm), all the CO$_2$ has sublimated from the surface and the water ice and JSC Mars-1 form a mantle that protects the CO$_2$ ice underneath from rapid sublimation. In the mixtures with fine H$_2$O ice, however, when the grains of CO$_2$ sublimate, they carry along the fine grains of water ice. As the CO$_2$ gas expands, it cannot hold the water ice grains any longer, which fall again onto the sample. The grains of fine H$_2$O ice rearrange then in energetically suitable structures, in this case, filaments (t$_5$). This step explains the change of texture seen in these mixtures, which seemed to grow fluffy. These filaments are effective light-scatters; their signal dominates the reflectance and masks the CO$_2$ ice underneath.  

As the temperature in our experiment kept increasing, the sublimation accelerated. At some point (in our case t$_9$ or t$_{11}$), the pressure of the sublimating CO$_2$ gas was strong enough to clear its way out of the sample, piercing the sublimation lag and/or breaking the water ice structures. Some of the inner CO$_2$ grains then became visible at the places where this process had just occurred. 

This sequence of differential sublimation explains why the CO$_2$ index does not drop linearly with sublimation (see Fig \ref{Fig:cake_criteria}). Instead, the signal first drops, levels out and then increases for mixtures with fine grained H$_2$O ice and oscillates in the beginning to then drop and again rise for mixtures with coarser H$_2$O ice. It also explains why the H$_2$O signal does not increase linearly, but initially increases to eventually level out (fine H$_2$O) or slow down (coarser H$_2$O). The growth of the CO$_2$ signal once it seemed to be gone could also be explained by the release of previously adsorbed CO$_2$ on JSC Mars-1 microparticles or possibly by the decomposition of CO$_2$ clathrate hydrates  (Vincent Chevrier, personal communication). Presumably, how this sublimation sequence develops depends on factors such as the CO$_2$-to-H$_2$O ratio, the grain size of the refractory material or the association mode of the materials (as seen in \citet{Poch:2016}). 

Finally, these experiments provide experimental support for the hypothesis formulated in \citet{Appere:2011} and \citet{Pommerol:2013} regarding the disappearance and subsequent return of the spectral signal of CO$_2$. A thin mantle of fine water frost masks the signature of CO$_2$, which reappears as the water frost layer disrupts with the increase of temperature. 

\subsection{Use of spectral parameters to identify mixtures}
To identify CO$_2$ ice, the first thing to look for is its absorption feature at 1.435 \textmu m. As we have repeatedly noted, the detection of a CO$_2$ signature evinces the presence of that component, but the lack of a signature does not guarantee the absence of CO$_2$ since it could be mixed intimately with other components (e.g., water ice) and masked. We can also get a hint about the presence of CO$_2$ ice by looking at the continuum reflectance against the water index at 1.5 \textmu m (Fig \ref{Fig:comparison_criteria_1}d); CO$_2$-bearing experiments stand out with high reflectance at weak water indices.

\paragraph{CO$_2$ Vs. CO$_2$ + dust} To identify dust in the mixtures, looking at the CaSSIS colours provide a direct answer. Indeed, every sample containing JSC Mars-1 falls under BLU/RED ratios of 0.9, while pure ice samples fall around BLU/RED ratios of 1.0. Alternatively, comparing the visible spectral slope to either the water or the carbon dioxide indexes is a good strategy to identify most of the JSC Mars-1 - bearing samples.

\paragraph{CO$_2$ Vs. CO$_2$ + H$_2$O} In order to identify water ice, we look at the H$_2$O index; for example, by comparing the visible spectral slope against the water index. Comparing first the water index to the visible spectral slope and then to the NIR spectral slope allows us first to identify pure ice samples, and then to identify whether the samples are made of pure water ice, intimate mixtures of carbon dioxide and water ices or pure carbon dioxide. Looking at the CO$_2$ index versus the NIR spectral slope also allows us to differentiate between intimate mixtures of water and carbon dioxide ices, pure CO$_2$ samples and slabs.

\paragraph{Atmospheric frost}
In our previous publication, we pointed out the comparison of the H$_2$O index and the visible slope as being useful to identify water frost \citep{Yoldi:2020}. When we incorporate the CO$_2$ experiments, this identification becomes more difficult, since CO$_2$ and small condensates of water frost share the same reflectance characteristics: they are bright, they tend to flatten the spectral slope of the substrate they are mixed with, and they do not alter the absorption bands of water. Therefore, it is hard to tell whether the intimate mixtures with JSC Mars-1 have trapped water frost by looking at Fig \ref{Fig:comparison_criteria_2}f (VIS). We have to look as well at the near-infrared, where water and carbon dioxide behave differently, water being bluer than carbon dioxide. In Fig \ref{Fig:comparison_criteria_2}f (NIR) we observe that two of the symbols for CO$_2$ and JSC Mars-1 intimate mixtures in the graph at the bottom follow the trend drawn by the JSCM-1/water frost experiments instead of remaining with the pure CO$_2$ or pure JSC Mars-1 samples. The corresponding experiments have indeed trapped water frost. 

\subsection{Limitations of the measurements.}
As mentioned in Section \ref{S:Methods}, it is challenging to achieve homogeneous intimate mixtures in which the end members are present in extreme ratios (i.e., 10 wt$\%$ - 90 wt$\%$ instead of 50 wt$\%$- 50 wt$\%$). The task becomes even more challenging when both components have the same colour since we cannot assess visually how well they mix. The measured reflectance showed that we succeeded in obtaining different CO$_2$-to-H$_2$O ratios. However, due to the formation of agglomerates, the samples were not as homogeneous as desired (see for example Fig \ref{Fig:asso}a). \citet{Kieffer:1970} already noted the difficulty of linking the reflectance spectra to the composition of his samples, due to textural effects. In the same way, we are not able to identify the mixtures displaying more water ice from their reflectance in the visible. In the near-infrared, however, the presence of the absorption bands of water allows us to notice the increasing amount of water ice within the samples. 

Because of the fast sintering of CO$_2$ ice grains, fine granular surfaces are hard to achieve \citep{Eluszkiewicz:2005}. Further laboratory work is needed to achieve CO$_2$ samples free from agglomerates, to first study their reflectance with smooth surfaces, and then understand the nature of agglomerates as well as their effect on the reflectance. We need to understand, for instance, whether we are creating these agglomerates by manipulating the samples, whether they only form at atmospheric pressures or whether small granular sliding at the surface of ice is enough to create them. 
\section{Conclusions} \label{S:conclusions}
We have reported on the reflectance measurements carried out at LOSSy on analogues for the surface of Mars containing H$_2$O and CO$_2$ ice (at atmospheric pressure). We have measured the reflectance of pure CO$_2$ frost, ice grains and slabs, and we have mixed CO$_2$ with water ice (intimate mixture and as frost contamination) and/or JSC Mars-1. We have monitored the reflectance of a ternary mixture of CO$_2$, H$_2$O and JSC Mars-1 as the volatile components sublimated. We summarise below the general trends we've observed:    

\begin{itemize}
    \item The size of particles of CO$_2$ plays a key role in the reflectance in various ways. \begin{inparaenum}[i)] \item Large particles transport photons deep in the medium, while small particles scatter light more efficiently, masking the material underneath the ice \item Small particle sizes trap more of atmospheric water. \end{inparaenum}
    \item The reflectance in the visible of the mixtures with granulated CO$_2$ is dominated by effects of the texture rather than the composition itself. In the near-infrared, however, the composition drives the reflectance. 
    \item As a part of a mixture, CO$_2$ is easily masked by more absorbent components due to its weak absorptivity. The materials tested here (water ice and JSC Mars-1) being much more absorptive than CO$_2$, masked the characteristic absorption features of CO$_2$ out of proportion to their presence.  
    \item The evolution of the reflectance of our ternary mixtures as temperature increased was controlled by the size of the particles of water ice, even if present only at 3 wt$\%$. The sublimation of CO$_2$ caused the JSC Mars-1 and coarse-grained water ice to settle down and cover the ice underneath. The sublimation of CO$_2$ also caused the fine-grained water ice to form highly-scattering H$_2$O frost filaments over the CO$_2$ ice underneath, compatible with the observations of the northern, seasonal, polar cap of Mars by \citet{Appere:2011} and \citet{Pommerol:2013}.
    \item We have compared many spectral parameters in the visible and near-infrared spectral ranges to make our data useful for a wide range of instruments measuring the reflectance of the Martian surface. 
    \item If we want to know about the grain size of the CO$_2$ ice, it is useful to compare the reflectance in the continuum versus the CO$_2$ index. To study the colour of the sample is useful to spot dust within the ice samples: CaSSIS colours or comparing the visible spectral slope to the water or carbon dioxide indexes will reveal the presence of dust on a sample. If we compare the water index first to the visible and then to the NIR spectral slope, we can identify if pure ice samples are made only of water ice or carbon dioxide ice or if they are a mix of both. 
    \item It is hard to identify water frost over CO$_2$ ice by looking at the reflectance spectra of the sample. However, the comparison of the water index versus the visible and NIR spectral slopes is a good indicator of the presence of water frost over surfaces. 
    \item Comparing visible and NIR slopes can be a way of identifying CO$_2$ ice for instruments that cannot resolve absorption features. 
\end{itemize}

The methods that we have proposed in these two papers (see \citet{Yoldi:2020}) provide a useful tool to fully exploit reflectance data. Because of the variety of spectral criteria and ranges used, these tools can be used with the data of many instruments that measure or have measured the reflectance of the surface of Mars (i.e., OMEGA, CRISM, HRSC, HiRISE and CaSSIS). They are also useful to calibrate reflectance models. The data here presented are available to the scientific community since these laboratory experiences are critical for the calibration of reflectance models and the understanding of processes occurring on planetary surfaces.

\vspace{50px}
\textbf{Data availability}
All the reflectance spectra presented here are available on the \href{https://wiki.sshade.eu/sshade/databases/bypass}{Bern icY Planetary Analogues Solid Spectroscopy (BYPASS)}  database, hosted on the \href{https://wiki.sshade.eu/start}{Solid Spectroscopy Hosting Architecture of Databases and Expertise (SSHADE)} .\textit{ Direct DOI link to the data will be provided here upon acceptance.}

\vspace{50px}

\textbf{Acknowledgements}
This work has been developed in the framework of the National Center for Competence in Research PlanetS funded by the Swiss National Science Foundation (SNSF). The authors want to thank Vincent Chevrier, Aditya Khuller and an anonymous reviewer for their valuable and constructive comments on this article.





\bibliographystyle{model1-num-names}
\bibliography{biblio.bib}

\begin{thebibliography}{73}
\expandafter\ifx\csname natexlab\endcsname\relax\def\natexlab#1{#1}\fi
\providecommand{\bibinfo}[2]{#2}
\ifx\xfnm\relax \def\xfnm[#1]{\unskip,\space#1}\fi
\bibitem[{Filacchione et~al.(2016)Filacchione, Raponi, Capaccioni, Ciarniello,
  Tosi, Capria, De~Sanctis, Migliorini, Piccioni, Cerroni, Barucci, Fornasier,
  Schmitt, Quirico, Erard, Bockelee-Morvan, Leyrat, Arnold, Mennella,
  Ammannito, Bellucci, Benkhoff, Bibring, Blanco, Blecka, Carlson, Carsenty,
  Colangeli, Combes, Combi, Crovisier, Drossart, Encrenaz, Federico, Fink,
  Fonti, Fulchignoni, Ip, Irwin, Jaumann, Kuehrt, Langevin, Magni, McCord,
  Moroz, Mottola, Palomba, Schade, Stephan, Taylor, Tiphene, Tozzi, Beck,
  Biver, Bonal, Combe, Despan, Flamini, Formisano, Frigeri, Grassi, Gudipati,
  Kappel, Longobardo, Mancarella, Markus, Merlin, Orosei, Rinaldi, Cartacci,
  Cicchetti, Hello, Henry, Jacquinod, Reess, Noschese, Politi, and
  Peter}]{Filacchione:2016}
\bibinfo{author}{G.~Filacchione}, \bibinfo{author}{A.~Raponi},
  \bibinfo{author}{F.~Capaccioni}, \bibinfo{author}{M.~Ciarniello},
  \bibinfo{author}{F.~Tosi}, \bibinfo{author}{M.~T. Capria},
  \bibinfo{author}{M.~C. De~Sanctis}, \bibinfo{author}{A.~Migliorini},
  \bibinfo{author}{G.~Piccioni}, \bibinfo{author}{P.~Cerroni},
  \bibinfo{author}{M.~A. Barucci}, \bibinfo{author}{S.~Fornasier},
  \bibinfo{author}{B.~Schmitt}, \bibinfo{author}{E.~Quirico},
  \bibinfo{author}{S.~Erard}, \bibinfo{author}{D.~Bockelee-Morvan},
  \bibinfo{author}{C.~Leyrat}, \bibinfo{author}{G.~Arnold},
  \bibinfo{author}{V.~Mennella}, \bibinfo{author}{E.~Ammannito},
  \bibinfo{author}{G.~Bellucci}, \bibinfo{author}{J.~Benkhoff},
  \bibinfo{author}{J.~P. Bibring}, \bibinfo{author}{A.~Blanco},
  \bibinfo{author}{M.~I. Blecka}, \bibinfo{author}{R.~Carlson},
  \bibinfo{author}{U.~Carsenty}, \bibinfo{author}{L.~Colangeli},
  \bibinfo{author}{M.~Combes}, \bibinfo{author}{M.~Combi},
  \bibinfo{author}{J.~Crovisier}, \bibinfo{author}{P.~Drossart},
  \bibinfo{author}{T.~Encrenaz}, \bibinfo{author}{C.~Federico},
  \bibinfo{author}{U.~Fink}, \bibinfo{author}{S.~Fonti},
  \bibinfo{author}{M.~Fulchignoni}, \bibinfo{author}{W.-H. Ip},
  \bibinfo{author}{P.~Irwin}, \bibinfo{author}{R.~Jaumann},
  \bibinfo{author}{E.~Kuehrt}, \bibinfo{author}{Y.~Langevin},
  \bibinfo{author}{G.~Magni}, \bibinfo{author}{T.~McCord},
  \bibinfo{author}{L.~Moroz}, \bibinfo{author}{S.~Mottola},
  \bibinfo{author}{E.~Palomba}, \bibinfo{author}{U.~Schade},
  \bibinfo{author}{K.~Stephan}, \bibinfo{author}{F.~Taylor},
  \bibinfo{author}{D.~Tiphene}, \bibinfo{author}{G.~P. Tozzi},
  \bibinfo{author}{P.~Beck}, \bibinfo{author}{N.~Biver},
  \bibinfo{author}{L.~Bonal}, \bibinfo{author}{J.-P. Combe},
  \bibinfo{author}{D.~Despan}, \bibinfo{author}{E.~Flamini},
  \bibinfo{author}{M.~Formisano}, \bibinfo{author}{A.~Frigeri},
  \bibinfo{author}{D.~Grassi}, \bibinfo{author}{M.~S. Gudipati},
  \bibinfo{author}{D.~Kappel}, \bibinfo{author}{A.~Longobardo},
  \bibinfo{author}{F.~Mancarella}, \bibinfo{author}{K.~Markus},
  \bibinfo{author}{F.~Merlin}, \bibinfo{author}{R.~Orosei},
  \bibinfo{author}{G.~Rinaldi}, \bibinfo{author}{M.~Cartacci},
  \bibinfo{author}{A.~Cicchetti}, \bibinfo{author}{Y.~Hello},
  \bibinfo{author}{F.~Henry}, \bibinfo{author}{S.~Jacquinod},
  \bibinfo{author}{J.~M. Reess}, \bibinfo{author}{R.~Noschese},
  \bibinfo{author}{R.~Politi}, \bibinfo{author}{G.~Peter},
\newblock \bibinfo{title}{Seasonal exposure of carbon dioxide ice on the
  nucleus of comet 67p/churyumov-gerasimenko},
\newblock \bibinfo{journal}{Science} \bibinfo{volume}{354}
  (\bibinfo{year}{2016}) \bibinfo{pages}{1563--1566}.
\bibitem[{Cruikshank et~al.(1993)Cruikshank, T.~L., T.~C., T.~R., C., B.,
  R.~H., and M.~J.}]{Cruikshank:2016}
\bibinfo{author}{D.~P. Cruikshank}, \bibinfo{author}{R.~T.~L.},
  \bibinfo{author}{O.~T.~C.}, \bibinfo{author}{G.~T.~R.},
  \bibinfo{author}{d.~B. C.}, \bibinfo{author}{S.~B.},
  \bibinfo{author}{B.~R.~H.}, \bibinfo{author}{B.~M.~J.},
\newblock \bibinfo{title}{Ices on the surface of triton},
\newblock \bibinfo{journal}{Science} \bibinfo{volume}{261}
  (\bibinfo{year}{1993}) \bibinfo{pages}{742–745}.
\bibitem[{Mahaffy et~al.(2013)Mahaffy, Webster, Atreya, Franz, Wong, Conrad,
  Harpold, Jones, Leshin, Manning, Owen, Pepin, Squyres, and
  Trainer}]{Mahaffy:2013}
\bibinfo{author}{P.~R. Mahaffy}, \bibinfo{author}{C.~R. Webster},
  \bibinfo{author}{S.~K. Atreya}, \bibinfo{author}{H.~Franz},
  \bibinfo{author}{M.~Wong}, \bibinfo{author}{P.~G. Conrad},
  \bibinfo{author}{D.~Harpold}, \bibinfo{author}{J.~J. Jones},
  \bibinfo{author}{L.~A. Leshin}, \bibinfo{author}{H.~Manning},
  \bibinfo{author}{T.~Owen}, \bibinfo{author}{R.~O. Pepin},
  \bibinfo{author}{S.~Squyres}, \bibinfo{author}{M.~Trainer},
\newblock \bibinfo{title}{Abundance and isotopic composition of gases in the
  martian atmosphere from the curiosity rover},
\newblock \bibinfo{journal}{Science} \bibinfo{volume}{341}
  (\bibinfo{year}{2013}) \bibinfo{pages}{263--266}.
\bibitem[{Pollack et~al.(1990)Pollack, Haberle, Schaeffer, and
  Lee}]{Pollack:1990}
\bibinfo{author}{J.~B. Pollack}, \bibinfo{author}{R.~M. Haberle},
  \bibinfo{author}{J.~Schaeffer}, \bibinfo{author}{H.~Lee},
\newblock \bibinfo{title}{Simulations of the general circulation of the martian
  atmosphere: 1. polar processes},
\newblock \bibinfo{journal}{Journal of Geophysical Research: Solid Earth}
  \bibinfo{volume}{95} (\bibinfo{year}{1990}) \bibinfo{pages}{1447--1473}.
\bibitem[{Hourdin et~al.(1995)Hourdin, Forget, and Talagrand}]{Hourdin:1995}
\bibinfo{author}{F.~Hourdin}, \bibinfo{author}{F.~Forget},
  \bibinfo{author}{O.~Talagrand},
\newblock \bibinfo{title}{The sensitivity of the martian surface pressure and
  atmospheric mass budget to various parameters: A comparison between numerical
  simulations and viking observations},
\newblock \bibinfo{journal}{Journal of Geophysical Research: Planets}
  \bibinfo{volume}{100} (\bibinfo{year}{1995}) \bibinfo{pages}{5501--5523}.
\bibitem[{James et~al.(1992)James, Kieffer, and Paige}]{James:1992}
\bibinfo{author}{P.~B. James}, \bibinfo{author}{H.~H. Kieffer},
  \bibinfo{author}{D.~A. Paige},
\newblock \bibinfo{title}{The seasonal cycle of carbon dioxide on mars},
\newblock \bibinfo{journal}{Mars}  (\bibinfo{year}{1992})
  \bibinfo{pages}{934--968}.
\bibitem[{Jakosky(1985)}]{Jakosky:1985}
\bibinfo{author}{B.~M. Jakosky},
\newblock \bibinfo{title}{The seasonal cycle of water on mars},
\newblock \bibinfo{journal}{Space science reviews} \bibinfo{volume}{41}
  (\bibinfo{year}{1985}) \bibinfo{pages}{131--200}.
\bibitem[{Kahn et~al.(1990)Kahn, Lee, Martin, and Zurek}]{Kahn:1990}
\bibinfo{author}{R.~Kahn}, \bibinfo{author}{S.~Lee},
  \bibinfo{author}{T.~Martin}, \bibinfo{author}{R.~Zurek},
\newblock \bibinfo{title}{The martian dust cycle},
\newblock in: \bibinfo{booktitle}{Bulletin of the American Astronomical
  Society}, volume~\bibinfo{volume}{22}, p. \bibinfo{pages}{1076}.
\bibitem[{Piqueux et~al.(2016)Piqueux, Kleinböhl, Hayne, Heavens, Kass,
  McCleese, Schofield, and Shirley}]{Piqueux:2016}
\bibinfo{author}{S.~Piqueux}, \bibinfo{author}{A.~Kleinböhl},
  \bibinfo{author}{P.~O. Hayne}, \bibinfo{author}{N.~G. Heavens},
  \bibinfo{author}{D.~M. Kass}, \bibinfo{author}{D.~J. McCleese},
  \bibinfo{author}{J.~T. Schofield}, \bibinfo{author}{J.~H. Shirley},
\newblock \bibinfo{title}{Discovery of a widespread low-latitude diurnal co2
  frost cycle on mars},
\newblock \bibinfo{journal}{Journal of Geophysical Research: Planets}
  \bibinfo{volume}{121} (\bibinfo{year}{2016}) \bibinfo{pages}{1174--1189}.
\bibitem[{Neugebauer et~al.(1971)Neugebauer, Münch, Kieffer, Chase, and
  Miner}]{Neugebauer:1971}
\bibinfo{author}{G.~Neugebauer}, \bibinfo{author}{G.~Münch},
  \bibinfo{author}{H.~Kieffer}, \bibinfo{author}{S.~C.~J. Chase},
  \bibinfo{author}{E.~Miner},
\newblock \bibinfo{title}{Mariner 1969 infrared radiometer results:
  Temperatures and thermal properties of the martian surface},
\newblock \bibinfo{journal}{Astronomical Journal} \bibinfo{volume}{76}
  (\bibinfo{year}{1971}) \bibinfo{pages}{719}.
\bibitem[{Leighton and Murray(1966)}]{Leighton:1966}
\bibinfo{author}{R.~B. Leighton}, \bibinfo{author}{B.~C. Murray},
\newblock \bibinfo{title}{Behavior of carbon dioxide and other volatiles on
  mars},
\newblock \bibinfo{journal}{Science} \bibinfo{volume}{153}
  (\bibinfo{year}{1966}) \bibinfo{pages}{136--144}.
\bibitem[{{Warren}(1984)}]{Warren:1984}
\bibinfo{author}{S.~G. {Warren}},
\newblock \bibinfo{title}{{Optical constants of ice from the ultraviolet to the
  microwave}} \bibinfo{volume}{23} (\bibinfo{year}{1984})
  \bibinfo{pages}{1206--1225}.
\bibitem[{{Warren}(1986)}]{Warren:1986}
\bibinfo{author}{S.~G. {Warren}},
\newblock \bibinfo{title}{{Optical constants of carbon dioxide ice}}
  \bibinfo{volume}{25} (\bibinfo{year}{1986}) \bibinfo{pages}{2650--2674}.
\bibitem[{Hansen(1997)}]{Hansen:1997}
\bibinfo{author}{Hansen},
\newblock \bibinfo{title}{The infrared absorption spectrum of carbon dioxide
  ice from 1.8 to 333 \textmu m},
\newblock \bibinfo{journal}{Journal of Geophysical Research: Planets}
  \bibinfo{volume}{102} (\bibinfo{year}{1997}) \bibinfo{pages}{21569--21587}.
\bibitem[{Hansen(2005)}]{Hansen:2005}
\bibinfo{author}{Hansen},
\newblock \bibinfo{title}{Ultraviolet to near‐infrared absorption spectrum of
  carbon dioxide ice from 0.174 to 1.8 \textmu m},
\newblock \bibinfo{journal}{Journal of Geophysical Research: Planets}
  \bibinfo{volume}{110} (\bibinfo{year}{2005}).
\bibitem[{{Bohren} and {Barkstrom}(1974)}]{Bohren:1974}
\bibinfo{author}{C.~F. {Bohren}}, \bibinfo{author}{B.~R. {Barkstrom}},
\newblock \bibinfo{title}{{Theory of the optical properties of snow}},
\newblock \bibinfo{journal}{Journal of Geophysical Research}
  \bibinfo{volume}{79} (\bibinfo{year}{1974}) \bibinfo{pages}{4527--4535}.
\bibitem[{{Wiscombe} and {Warren}(1980)}]{Wiscombe:1980}
\bibinfo{author}{W.~J. {Wiscombe}}, \bibinfo{author}{S.~G. {Warren}},
\newblock \bibinfo{title}{{A Model for the Spectral Albedo of Snow. I: Pure
  Snow.}},
\newblock \bibinfo{journal}{Journal of Atmospheric Sciences}
  \bibinfo{volume}{37} (\bibinfo{year}{1980}) \bibinfo{pages}{2712--2733}.
\bibitem[{{Hapke}(1993)}]{Hapke1993}
\bibinfo{author}{B.~{Hapke}}, \bibinfo{title}{{Theory of reflectance and
  emittance spectroscopy}}, \bibinfo{year}{1993}.
\bibitem[{Singh and Flanner(2016)}]{Singh:2016}
\bibinfo{author}{D.~Singh}, \bibinfo{author}{M.~G. Flanner},
\newblock \bibinfo{title}{An improved carbon dioxide snow spectral albedo
  model: Application to martian conditions},
\newblock \bibinfo{journal}{Journal of Geophysical Research: Planets}
  \bibinfo{volume}{121} (\bibinfo{year}{2016}) \bibinfo{pages}{2037--2054}.
  \bibinfo{note}{2016JE005040}.
\bibitem[{{Khuller} et~al.(2021){Khuller}, {Christensen}, and
  {Warren}}]{khuller:2021}
\bibinfo{author}{A.~R. {Khuller}}, \bibinfo{author}{P.~R. {Christensen}},
  \bibinfo{author}{S.~G. {Warren}},
\newblock \bibinfo{title}{{Spectral Albedo of Dusty Martian H$_{2}$O Snow and
  Ice}},
\newblock \bibinfo{journal}{Journal of Geophysical Research (Planets)}
  \bibinfo{volume}{126} (\bibinfo{year}{2021}) \bibinfo{pages}{e06910}.
\bibitem[{Yoldi et~al.(2015)Yoldi, Pommerol, Jost, Poch, Gouman, and
  Thomas}]{Yoldi:2015}
\bibinfo{author}{Z.~Yoldi}, \bibinfo{author}{A.~Pommerol},
  \bibinfo{author}{B.~Jost}, \bibinfo{author}{O.~Poch},
  \bibinfo{author}{J.~Gouman}, \bibinfo{author}{N.~Thomas},
\newblock \bibinfo{title}{Vis-nir reflectance of water ice/regolith analogue
  mixtures and implications for the detectability of ice mixed within planetary
  regoliths},
\newblock \bibinfo{journal}{Geophysical Research Letters} \bibinfo{volume}{42}
  (\bibinfo{year}{2015}) \bibinfo{pages}{6205--6212}.
  \bibinfo{note}{2015GL064780}.
\bibitem[{{Thomas} et~al.(2005){Thomas}, {Malin}, {James}, {Cantor},
  {Williams}, and {Gierasch}}]{Thomas:2005}
\bibinfo{author}{P.~C. {Thomas}}, \bibinfo{author}{M.~C. {Malin}},
  \bibinfo{author}{P.~B. {James}}, \bibinfo{author}{B.~A. {Cantor}},
  \bibinfo{author}{R.~M.~E. {Williams}}, \bibinfo{author}{P.~{Gierasch}},
\newblock \bibinfo{title}{{South polar residual cap of Mars: Features,
  stratigraphy, and changes}} \bibinfo{volume}{174} (\bibinfo{year}{2005})
  \bibinfo{pages}{535--559}.
\bibitem[{Titus et~al.(2003)Titus, Kieffer, and Christensen}]{Titus:2003}
\bibinfo{author}{T.~N. Titus}, \bibinfo{author}{H.~H. Kieffer},
  \bibinfo{author}{P.~R. Christensen},
\newblock \bibinfo{title}{Exposed water ice discovered near the south pole of
  mars},
\newblock \bibinfo{journal}{Science} \bibinfo{volume}{299}
  (\bibinfo{year}{2003}) \bibinfo{pages}{1048--1051}.
\bibitem[{{Plaut} et~al.(2007){Plaut}, {Picardi}, {Safaeinili}, {Ivanov},
  {Milkovich}, {Cicchetti}, {Kofman}, {Mouginot}, {Farrell}, {Phillips},
  {Clifford}, {Frigeri}, {Orosei}, {Federico}, {Williams}, {Gurnett},
  {Nielsen}, {Hagfors}, {Heggy}, {Stofan}, {Plettemeier}, {Watters},
  {Leuschen}, and {Edenhofer}}]{Plaut:2007}
\bibinfo{author}{J.~J. {Plaut}}, \bibinfo{author}{G.~{Picardi}},
  \bibinfo{author}{A.~{Safaeinili}}, \bibinfo{author}{A.~B. {Ivanov}},
  \bibinfo{author}{S.~M. {Milkovich}}, \bibinfo{author}{A.~{Cicchetti}},
  \bibinfo{author}{W.~{Kofman}}, \bibinfo{author}{J.~{Mouginot}},
  \bibinfo{author}{W.~M. {Farrell}}, \bibinfo{author}{R.~J. {Phillips}},
  \bibinfo{author}{S.~M. {Clifford}}, \bibinfo{author}{A.~{Frigeri}},
  \bibinfo{author}{R.~{Orosei}}, \bibinfo{author}{C.~{Federico}},
  \bibinfo{author}{I.~P. {Williams}}, \bibinfo{author}{D.~A. {Gurnett}},
  \bibinfo{author}{E.~{Nielsen}}, \bibinfo{author}{T.~{Hagfors}},
  \bibinfo{author}{E.~{Heggy}}, \bibinfo{author}{E.~R. {Stofan}},
  \bibinfo{author}{D.~{Plettemeier}}, \bibinfo{author}{T.~R. {Watters}},
  \bibinfo{author}{C.~J. {Leuschen}}, \bibinfo{author}{P.~{Edenhofer}},
\newblock \bibinfo{title}{{Subsurface Radar Sounding of the South Polar Layered
  Deposits of Mars}},
\newblock \bibinfo{journal}{Science} \bibinfo{volume}{316}
  (\bibinfo{year}{2007}) \bibinfo{pages}{92}.
\bibitem[{{Bierson} et~al.(2016){Bierson}, {Phillips}, {Smith}, {Wood},
  {Putzig}, {Nunes}, and {Byrne}}]{Bierson:2016}
\bibinfo{author}{C.~J. {Bierson}}, \bibinfo{author}{R.~J. {Phillips}},
  \bibinfo{author}{I.~B. {Smith}}, \bibinfo{author}{S.~E. {Wood}},
  \bibinfo{author}{N.~E. {Putzig}}, \bibinfo{author}{D.~{Nunes}},
  \bibinfo{author}{S.~{Byrne}},
\newblock \bibinfo{title}{{Stratigraphy and evolution of the buried CO$_{2}$
  deposit in the Martian south polar cap}},
\newblock \bibinfo{journal}{Geophysical Research Letters} \bibinfo{volume}{43}
  (\bibinfo{year}{2016}) \bibinfo{pages}{4172--4179}.
\bibitem[{{Bibring} et~al.(2004){Bibring}, {Langevin}, {Poulet}, {Gendrin},
  {Gondet}, {Berth{\'e}}, {Soufflot}, {Drossart}, {Combes}, {Bellucci},
  {Moroz}, {Mangold}, {Schmitt}, {OMEGA Team}, {Erard}, {Forni}, {Manaud},
  {Poulleau}, {Encrenaz}, {Fouchet}, {Melchiorri}, {Altieri}, {Formisano},
  {Bonello}, {Fonti}, {Capaccioni}, {Cerroni}, {Coradini}, {Kottsov},
  {Ignatiev}, {Titov}, {Zasova}, {Pinet}, {Sotin}, {Hauber}, {Hoffman},
  {Jaumann}, {Keller}, {Arvidson}, {Mustard}, {Duxbury}, and
  {Forget}}]{Bibring:2004}
\bibinfo{author}{J.-P. {Bibring}}, \bibinfo{author}{Y.~{Langevin}},
  \bibinfo{author}{F.~{Poulet}}, \bibinfo{author}{A.~{Gendrin}},
  \bibinfo{author}{B.~{Gondet}}, \bibinfo{author}{M.~{Berth{\'e}}},
  \bibinfo{author}{A.~{Soufflot}}, \bibinfo{author}{P.~{Drossart}},
  \bibinfo{author}{M.~{Combes}}, \bibinfo{author}{G.~{Bellucci}},
  \bibinfo{author}{V.~{Moroz}}, \bibinfo{author}{N.~{Mangold}},
  \bibinfo{author}{B.~{Schmitt}}, \bibinfo{author}{{OMEGA Team}},
  \bibinfo{author}{S.~{Erard}}, \bibinfo{author}{O.~{Forni}},
  \bibinfo{author}{N.~{Manaud}}, \bibinfo{author}{G.~{Poulleau}},
  \bibinfo{author}{T.~{Encrenaz}}, \bibinfo{author}{T.~{Fouchet}},
  \bibinfo{author}{R.~{Melchiorri}}, \bibinfo{author}{F.~{Altieri}},
  \bibinfo{author}{V.~{Formisano}}, \bibinfo{author}{G.~{Bonello}},
  \bibinfo{author}{S.~{Fonti}}, \bibinfo{author}{F.~{Capaccioni}},
  \bibinfo{author}{P.~{Cerroni}}, \bibinfo{author}{A.~{Coradini}},
  \bibinfo{author}{V.~{Kottsov}}, \bibinfo{author}{N.~{Ignatiev}},
  \bibinfo{author}{D.~{Titov}}, \bibinfo{author}{L.~{Zasova}},
  \bibinfo{author}{P.~{Pinet}}, \bibinfo{author}{C.~{Sotin}},
  \bibinfo{author}{E.~{Hauber}}, \bibinfo{author}{H.~{Hoffman}},
  \bibinfo{author}{R.~{Jaumann}}, \bibinfo{author}{U.~{Keller}},
  \bibinfo{author}{R.~{Arvidson}}, \bibinfo{author}{J.~{Mustard}},
  \bibinfo{author}{T.~{Duxbury}}, \bibinfo{author}{F.~{Forget}},
\newblock \bibinfo{title}{{Perennial water ice identified in the south polar
  cap of Mars}},
\newblock \bibinfo{journal}{Nature} \bibinfo{volume}{428}
  (\bibinfo{year}{2004}) \bibinfo{pages}{627--630}.
\bibitem[{Calvin and Martin(1994)}]{Calvin:1994}
\bibinfo{author}{W.~M. Calvin}, \bibinfo{author}{T.~Z. Martin},
\newblock \bibinfo{title}{Spatial variability in the seasonal south polar cap
  of mars},
\newblock \bibinfo{journal}{Journal of Geophysical Research: Planets}
  \bibinfo{volume}{99} (\bibinfo{year}{1994}) \bibinfo{pages}{21143--21152}.
\bibitem[{Kieffer et~al.(2000)Kieffer, Titus, Mullins, and
  Christensen}]{Kieffer:2000}
\bibinfo{author}{H.~H. Kieffer}, \bibinfo{author}{T.~N. Titus},
  \bibinfo{author}{K.~F. Mullins}, \bibinfo{author}{P.~R. Christensen},
\newblock \bibinfo{title}{Mars south polar spring and summer behavior observed
  by tes: Seasonal cap evolution controlled by frost grain size},
\newblock \bibinfo{journal}{Journal of Geophysical Research: Planets}
  \bibinfo{volume}{105} (\bibinfo{year}{2000}) \bibinfo{pages}{9653--9699}.
\bibitem[{{Kieffer}(2007)}]{Kieffer:2007}
\bibinfo{author}{H.~H. {Kieffer}},
\newblock \bibinfo{title}{{Cold jets in the Martian polar caps}},
\newblock \bibinfo{journal}{Journal of Geophysical Research (Planets)}
  \bibinfo{volume}{112} (\bibinfo{year}{2007}) \bibinfo{pages}{E08005}.
\bibitem[{Langevin et~al.(2007)Langevin, Bibring, Montmessin, Forget,
  Vincendon, Dout{\'e}, Poulet, and Gondet}]{Langevin:2007}
\bibinfo{author}{Y.~Langevin}, \bibinfo{author}{J.-P. Bibring},
  \bibinfo{author}{F.~Montmessin}, \bibinfo{author}{F.~Forget},
  \bibinfo{author}{M.~Vincendon}, \bibinfo{author}{S.~Dout{\'e}},
  \bibinfo{author}{F.~Poulet}, \bibinfo{author}{B.~Gondet},
\newblock \bibinfo{title}{Observations of the south seasonal cap of mars during
  recession in 2004–2006 by the omega visible/near-infrared imaging
  spectrometer on board mars express},
\newblock \bibinfo{journal}{Journal of Geophysical Research: Planets}
  \bibinfo{volume}{112} (\bibinfo{year}{2007}) \bibinfo{pages}{n/a--n/a}.
  \bibinfo{note}{E08S12}.
\bibitem[{{Murchie} et~al.(2007){Murchie}, {Arvidson}, {Bedini}, {Beisser},
  {Bibring}, {Bishop}, {Boldt}, {Cavender}, {Choo}, {Clancy}, {Darlington},
  {Des Marais}, {Espiritu}, {Fort}, {Green}, {Guinness}, {Hayes}, {Hash},
  {Heffernan}, {Hemmler}, {Heyler}, {Humm}, {Hutcheson}, {Izenberg}, {Lee},
  {Lees}, {Lohr}, {Malaret}, {Martin}, {McGovern}, {McGuire}, {Morris},
  {Mustard}, {Pelkey}, {Rhodes}, {Robinson}, {Roush}, {Schaefer}, {Seagrave},
  {Seelos}, {Silverglate}, {Slavney}, {Smith}, {Shyong}, {Strohbehn}, {Taylor},
  {Thompson}, {Tossman}, {Wirzburger}, and {Wolff}}]{murchie:2007}
\bibinfo{author}{S.~{Murchie}}, \bibinfo{author}{R.~{Arvidson}},
  \bibinfo{author}{P.~{Bedini}}, \bibinfo{author}{K.~{Beisser}},
  \bibinfo{author}{J.~P. {Bibring}}, \bibinfo{author}{J.~{Bishop}},
  \bibinfo{author}{J.~{Boldt}}, \bibinfo{author}{P.~{Cavender}},
  \bibinfo{author}{T.~{Choo}}, \bibinfo{author}{R.~T. {Clancy}},
  \bibinfo{author}{E.~H. {Darlington}}, \bibinfo{author}{D.~{Des Marais}},
  \bibinfo{author}{R.~{Espiritu}}, \bibinfo{author}{D.~{Fort}},
  \bibinfo{author}{R.~{Green}}, \bibinfo{author}{E.~{Guinness}},
  \bibinfo{author}{J.~{Hayes}}, \bibinfo{author}{C.~{Hash}},
  \bibinfo{author}{K.~{Heffernan}}, \bibinfo{author}{J.~{Hemmler}},
  \bibinfo{author}{G.~{Heyler}}, \bibinfo{author}{D.~{Humm}},
  \bibinfo{author}{J.~{Hutcheson}}, \bibinfo{author}{N.~{Izenberg}},
  \bibinfo{author}{R.~{Lee}}, \bibinfo{author}{J.~{Lees}},
  \bibinfo{author}{D.~{Lohr}}, \bibinfo{author}{E.~{Malaret}},
  \bibinfo{author}{T.~{Martin}}, \bibinfo{author}{J.~A. {McGovern}},
  \bibinfo{author}{P.~{McGuire}}, \bibinfo{author}{R.~{Morris}},
  \bibinfo{author}{J.~{Mustard}}, \bibinfo{author}{S.~{Pelkey}},
  \bibinfo{author}{E.~{Rhodes}}, \bibinfo{author}{M.~{Robinson}},
  \bibinfo{author}{T.~{Roush}}, \bibinfo{author}{E.~{Schaefer}},
  \bibinfo{author}{G.~{Seagrave}}, \bibinfo{author}{F.~{Seelos}},
  \bibinfo{author}{P.~{Silverglate}}, \bibinfo{author}{S.~{Slavney}},
  \bibinfo{author}{M.~{Smith}}, \bibinfo{author}{W.~J. {Shyong}},
  \bibinfo{author}{K.~{Strohbehn}}, \bibinfo{author}{H.~{Taylor}},
  \bibinfo{author}{P.~{Thompson}}, \bibinfo{author}{B.~{Tossman}},
  \bibinfo{author}{M.~{Wirzburger}}, \bibinfo{author}{M.~{Wolff}},
\newblock \bibinfo{title}{{Compact Reconnaissance Imaging Spectrometer for Mars
  (CRISM) on Mars Reconnaissance Orbiter (MRO)}},
\newblock \bibinfo{journal}{Journal of Geophysical Research (Planets)}
  \bibinfo{volume}{112} (\bibinfo{year}{2007}) \bibinfo{pages}{E05S03}.
\bibitem[{{Bell} et~al.(2009){Bell}, {Wolff}, {Malin}, {Calvin}, {Cantor},
  {Caplinger}, {Clancy}, {Edgett}, {Edwards}, {Fahle}, {Ghaemi}, {Haberle},
  {Hale}, {James}, {Lee}, {McConnochie}, {Noe Dobrea}, {Ravine}, {Schaeffer},
  {Supulver}, and {Thomas}}]{Bell:2009}
\bibinfo{author}{J.~F. {Bell}}, \bibinfo{author}{M.~J. {Wolff}},
  \bibinfo{author}{M.~C. {Malin}}, \bibinfo{author}{W.~M. {Calvin}},
  \bibinfo{author}{B.~A. {Cantor}}, \bibinfo{author}{M.~A. {Caplinger}},
  \bibinfo{author}{R.~T. {Clancy}}, \bibinfo{author}{K.~S. {Edgett}},
  \bibinfo{author}{L.~J. {Edwards}}, \bibinfo{author}{J.~{Fahle}},
  \bibinfo{author}{F.~{Ghaemi}}, \bibinfo{author}{R.~M. {Haberle}},
  \bibinfo{author}{A.~{Hale}}, \bibinfo{author}{P.~B. {James}},
  \bibinfo{author}{S.~W. {Lee}}, \bibinfo{author}{T.~{McConnochie}},
  \bibinfo{author}{E.~{Noe Dobrea}}, \bibinfo{author}{M.~A. {Ravine}},
  \bibinfo{author}{D.~{Schaeffer}}, \bibinfo{author}{K.~D. {Supulver}},
  \bibinfo{author}{P.~C. {Thomas}},
\newblock \bibinfo{title}{{Mars Reconnaissance Orbiter Mars Color Imager
  (MARCI): Instrument description, calibration, and performance}},
\newblock \bibinfo{journal}{Journal of Geophysical Research (Planets)}
  \bibinfo{volume}{114} (\bibinfo{year}{2009}) \bibinfo{pages}{E08S92}.
\bibitem[{Brown et~al.(2010)Brown, Calvin, McGuire, and Murchie}]{Brown:2010}
\bibinfo{author}{A.~J. Brown}, \bibinfo{author}{W.~M. Calvin},
  \bibinfo{author}{P.~C. McGuire}, \bibinfo{author}{S.~L. Murchie},
\newblock \bibinfo{title}{Compact reconnaissance imaging spectrometer for mars
  (crism) south polar mapping: First mars year of observations},
\newblock \bibinfo{journal}{Journal of Geophysical Research: Planets}
  \bibinfo{volume}{115} (\bibinfo{year}{2010}).
\bibitem[{McEwen et~al.(????)McEwen, Eliason, Bergstrom, Bridges, Hansen,
  Delamere, Grant, Gulick, Herkenhoff, Keszthelyi, Kirk, Mellon, Squyres,
  Thomas, and Weitz}]{McEwen:2007}
\bibinfo{author}{A.~S. McEwen}, \bibinfo{author}{E.~M. Eliason},
  \bibinfo{author}{J.~W. Bergstrom}, \bibinfo{author}{N.~T. Bridges},
  \bibinfo{author}{C.~J. Hansen}, \bibinfo{author}{W.~A. Delamere},
  \bibinfo{author}{J.~A. Grant}, \bibinfo{author}{V.~C. Gulick},
  \bibinfo{author}{K.~E. Herkenhoff}, \bibinfo{author}{L.~Keszthelyi},
  \bibinfo{author}{R.~L. Kirk}, \bibinfo{author}{M.~T. Mellon},
  \bibinfo{author}{S.~W. Squyres}, \bibinfo{author}{N.~Thomas},
  \bibinfo{author}{C.~M. Weitz},
\newblock \bibinfo{title}{Mars reconnaissance orbiter's high resolution imaging
  science experiment (hirise)},
\newblock \bibinfo{journal}{Journal of Geophysical Research: Planets}
  \bibinfo{volume}{112} (????).
\bibitem[{Pommerol et~al.(2011)Pommerol, Portyankina, Thomas, Aye, Hansen,
  Vincendon, and Langevin}]{Pommerol:2011}
\bibinfo{author}{A.~Pommerol}, \bibinfo{author}{G.~Portyankina},
  \bibinfo{author}{N.~Thomas}, \bibinfo{author}{K.-M. Aye},
  \bibinfo{author}{C.~J. Hansen}, \bibinfo{author}{M.~Vincendon},
  \bibinfo{author}{Y.~Langevin},
\newblock \bibinfo{title}{Evolution of south seasonal cap during martian
  spring: Insights from high-resolution observations by hirise and crism on
  mars reconnaissance orbiter},
\newblock \bibinfo{journal}{Journal of Geophysical Research: Planets}
  \bibinfo{volume}{116} (\bibinfo{year}{2011}) \bibinfo{pages}{n/a--n/a}.
  \bibinfo{note}{E08007}.
\bibitem[{Brown et~al.(2014)Brown, Piqueux, and Titus}]{Brown:2014}
\bibinfo{author}{A.~J. Brown}, \bibinfo{author}{S.~Piqueux},
  \bibinfo{author}{T.~N. Titus},
\newblock \bibinfo{title}{Interannual observations and quantification of
  summertime h$_2$o ice deposition on the martian co$_2$ ice south polar cap},
\newblock \bibinfo{journal}{Earth and Planetary Science Letters}
  \bibinfo{volume}{406} (\bibinfo{year}{2014}) \bibinfo{pages}{102--109}.
\bibitem[{{Christensen} et~al.(2004){Christensen}, {Jakosky}, {Kieffer},
  {Malin}, {McSween}, {Nealson}, {Mehall}, {Silverman}, {Ferry}, {Caplinger},
  and {Ravine}}]{Christensen:2004}
\bibinfo{author}{P.~R. {Christensen}}, \bibinfo{author}{B.~M. {Jakosky}},
  \bibinfo{author}{H.~H. {Kieffer}}, \bibinfo{author}{M.~C. {Malin}},
  \bibinfo{author}{J.~{McSween}, Harry~Y.}, \bibinfo{author}{K.~{Nealson}},
  \bibinfo{author}{G.~L. {Mehall}}, \bibinfo{author}{S.~H. {Silverman}},
  \bibinfo{author}{S.~{Ferry}}, \bibinfo{author}{M.~{Caplinger}},
  \bibinfo{author}{M.~{Ravine}},
\newblock \bibinfo{title}{{The Thermal Emission Imaging System (THEMIS) for the
  Mars 2001 Odyssey Mission}},
\newblock \bibinfo{journal}{Space Science Reviews} \bibinfo{volume}{110}
  (\bibinfo{year}{2004}) \bibinfo{pages}{85--130}.
\bibitem[{{Christensen} et~al.(2001){Christensen}, {Bandfield}, {Hamilton},
  {Ruff}, {Kieffer}, {Titus}, {Malin}, {Morris}, {Lane}, {Clark}, {Jakosky},
  {Mellon}, {Pearl}, {Conrath}, {Smith}, {Clancy}, {Kuzmin}, {Roush}, {Mehall},
  {Gorelick}, {Bender}, {Murray}, {Dason}, {Greene}, {Silverman}, and
  {Greenfield}}]{Christensen:2001}
\bibinfo{author}{P.~R. {Christensen}}, \bibinfo{author}{J.~L. {Bandfield}},
  \bibinfo{author}{V.~E. {Hamilton}}, \bibinfo{author}{S.~W. {Ruff}},
  \bibinfo{author}{H.~H. {Kieffer}}, \bibinfo{author}{T.~N. {Titus}},
  \bibinfo{author}{M.~C. {Malin}}, \bibinfo{author}{R.~V. {Morris}},
  \bibinfo{author}{M.~D. {Lane}}, \bibinfo{author}{R.~L. {Clark}},
  \bibinfo{author}{B.~M. {Jakosky}}, \bibinfo{author}{M.~T. {Mellon}},
  \bibinfo{author}{J.~C. {Pearl}}, \bibinfo{author}{B.~J. {Conrath}},
  \bibinfo{author}{M.~D. {Smith}}, \bibinfo{author}{R.~T. {Clancy}},
  \bibinfo{author}{R.~O. {Kuzmin}}, \bibinfo{author}{T.~{Roush}},
  \bibinfo{author}{G.~L. {Mehall}}, \bibinfo{author}{N.~{Gorelick}},
  \bibinfo{author}{K.~{Bender}}, \bibinfo{author}{K.~{Murray}},
  \bibinfo{author}{S.~{Dason}}, \bibinfo{author}{E.~{Greene}},
  \bibinfo{author}{S.~{Silverman}}, \bibinfo{author}{M.~{Greenfield}},
\newblock \bibinfo{title}{{Mars Global Surveyor Thermal Emission Spectrometer
  experiment: Investigation description and surface science results}},
\newblock \bibinfo{journal}{Journal of Geophysical Research}
  \bibinfo{volume}{106} (\bibinfo{year}{2001}) \bibinfo{pages}{23823--23872}.
\bibitem[{{Wagstaff} et~al.(2008){Wagstaff}, {Titus}, {Ivanov}, {Casta{\~n}o},
  and {Bandfield}}]{Wagstaff:2008}
\bibinfo{author}{K.~L. {Wagstaff}}, \bibinfo{author}{T.~N. {Titus}},
  \bibinfo{author}{A.~B. {Ivanov}}, \bibinfo{author}{R.~{Casta{\~n}o}},
  \bibinfo{author}{J.~L. {Bandfield}},
\newblock \bibinfo{title}{{Observations of the north polar water ice annulus on
  Mars using THEMIS and TES}},
\newblock \bibinfo{journal}{Planetary and Space Science} \bibinfo{volume}{56}
  (\bibinfo{year}{2008}) \bibinfo{pages}{256--265}.
\bibitem[{{Houben} et~al.(1997){Houben}, {Haberle}, {Young}, and
  {Zent}}]{Houben:1997}
\bibinfo{author}{H.~{Houben}}, \bibinfo{author}{R.~M. {Haberle}},
  \bibinfo{author}{R.~E. {Young}}, \bibinfo{author}{A.~P. {Zent}},
\newblock \bibinfo{title}{{Modeling the Martian seasonal water cycle}},
\newblock \bibinfo{journal}{Journal of Geophysical Research}
  \bibinfo{volume}{102} (\bibinfo{year}{1997}) \bibinfo{pages}{9069--9084}.
\bibitem[{Langevin et~al.(2005)Langevin, Poulet, Bibring, Schmitt, Dout{\'e},
  and Gondet}]{Langevin:2005}
\bibinfo{author}{Y.~Langevin}, \bibinfo{author}{F.~Poulet},
  \bibinfo{author}{J.-P. Bibring}, \bibinfo{author}{B.~Schmitt},
  \bibinfo{author}{S.~Dout{\'e}}, \bibinfo{author}{B.~Gondet},
\newblock \bibinfo{title}{Summer evolution of the north polar cap of mars as
  observed by omega/mars express},
\newblock \bibinfo{journal}{Science} \bibinfo{volume}{307}
  (\bibinfo{year}{2005}) \bibinfo{pages}{1581--1584}.
\bibitem[{App{\'e}r{\'e} et~al.(2011)App{\'e}r{\'e}, Schmitt, Langevin,
  Dout{\'e}, Pommerol, Forget, Spiga, Gondet, and Bibring}]{Appere:2011}
\bibinfo{author}{T.~App{\'e}r{\'e}}, \bibinfo{author}{B.~Schmitt},
  \bibinfo{author}{Y.~Langevin}, \bibinfo{author}{S.~Dout{\'e}},
  \bibinfo{author}{A.~Pommerol}, \bibinfo{author}{F.~Forget},
  \bibinfo{author}{A.~Spiga}, \bibinfo{author}{B.~Gondet},
  \bibinfo{author}{J.-P. Bibring},
\newblock \bibinfo{title}{Winter and spring evolution of northern seasonal
  deposits on mars from omega on mars express},
\newblock \bibinfo{journal}{Journal of Geophysical Research: Planets}
  \bibinfo{volume}{116} (\bibinfo{year}{2011}) \bibinfo{pages}{n/a--n/a}.
  \bibinfo{note}{E05001}.
\bibitem[{Hansen et~al.(2013)Hansen, Byrne, Portyankina, Bourke, Dundas,
  McEwen, Mellon, Pommerol, and Thomas}]{Hansen:2013}
\bibinfo{author}{C.~Hansen}, \bibinfo{author}{S.~Byrne},
  \bibinfo{author}{G.~Portyankina}, \bibinfo{author}{M.~Bourke},
  \bibinfo{author}{C.~Dundas}, \bibinfo{author}{A.~McEwen},
  \bibinfo{author}{M.~Mellon}, \bibinfo{author}{A.~Pommerol},
  \bibinfo{author}{N.~Thomas},
\newblock \bibinfo{title}{Observations of the northern seasonal polar cap on
  mars: I. spring sublimation activity and processes},
\newblock \bibinfo{journal}{Icarus} \bibinfo{volume}{225}
  (\bibinfo{year}{2013}) \bibinfo{pages}{881 -- 897}. \bibinfo{note}{Mars Polar
  Science V}.
\bibitem[{Portyankina et~al.(2013)Portyankina, Pommerol, Aye, Hansen, and
  Thomas}]{Portyankina:2013}
\bibinfo{author}{G.~Portyankina}, \bibinfo{author}{A.~Pommerol},
  \bibinfo{author}{K.-M. Aye}, \bibinfo{author}{C.~J. Hansen},
  \bibinfo{author}{N.~Thomas},
\newblock \bibinfo{title}{Observations of the northern seasonal polar cap on
  mars ii: Hirise photometric analysis of evolution of northern polar dunes in
  spring},
\newblock \bibinfo{journal}{Icarus} \bibinfo{volume}{225}
  (\bibinfo{year}{2013}) \bibinfo{pages}{898 -- 910}. \bibinfo{note}{Mars Polar
  Science V}.
\bibitem[{Pommerol et~al.(2013)Pommerol, Appéré, Portyankina, Aye, Thomas,
  and Hansen}]{Pommerol:2013}
\bibinfo{author}{A.~Pommerol}, \bibinfo{author}{T.~Appéré},
  \bibinfo{author}{G.~Portyankina}, \bibinfo{author}{K.-M. Aye},
  \bibinfo{author}{N.~Thomas}, \bibinfo{author}{C.~Hansen},
\newblock \bibinfo{title}{Observations of the northern seasonal polar cap on
  mars iii: Crism/hirise observations of spring sublimation},
\newblock \bibinfo{journal}{Icarus} \bibinfo{volume}{225}
  (\bibinfo{year}{2013}) \bibinfo{pages}{911 -- 922}. \bibinfo{note}{Mars Polar
  Science V}.
\bibitem[{Quirico and Schmitt(1997)}]{Quirico:1997}
\bibinfo{author}{E.~Quirico}, \bibinfo{author}{B.~Schmitt},
\newblock \bibinfo{title}{Near-infrared spectroscopy of simple hydrocarbons and
  carbon oxides diluted in solid n$_2$ and as pure ices: Implications for
  triton and pluto},
\newblock \bibinfo{journal}{Icarus} \bibinfo{volume}{127}
  (\bibinfo{year}{1997}) \bibinfo{pages}{354 -- 378}.
\bibitem[{{{\"O}berg} et~al.(2007){{\"O}berg}, {Fraser}, {Boogert}, {Bisschop},
  {Fuchs}, {van Dishoeck}, and {Linnartz}}]{Oeberg:2007}
\bibinfo{author}{K.~I. {{\"O}berg}}, \bibinfo{author}{H.~J. {Fraser}},
  \bibinfo{author}{A.~C.~A. {Boogert}}, \bibinfo{author}{S.~E. {Bisschop}},
  \bibinfo{author}{G.~W. {Fuchs}}, \bibinfo{author}{E.~F. {van Dishoeck}},
  \bibinfo{author}{H.~{Linnartz}},
\newblock \bibinfo{title}{{Effects of CO$_{2}$ on H\_2O band profiles and band
  strengths in mixed H\_2O:CO$_{2}$ ices}},
\newblock \bibinfo{journal}{Astronomy and Astrophysics} \bibinfo{volume}{462}
  (\bibinfo{year}{2007}) \bibinfo{pages}{1187--1198}.
\bibitem[{{Kieffer}(1968)}]{kieffer:1968}
\bibinfo{author}{H.~H. {Kieffer}}, \bibinfo{title}{{Near infrared spectral
  reflectance of simulated Martian frosts}}, Ph.D. thesis, California Institute
  of Technology, United States, \bibinfo{year}{1968}.
\bibitem[{Kieffer(1970)}]{Kieffer:1970}
\bibinfo{author}{H.~Kieffer},
\newblock \bibinfo{title}{Spectral reflectance of co$_2$-h$_2$o frosts},
\newblock \bibinfo{journal}{Journal of Geophysical Research}
  \bibinfo{volume}{75} (\bibinfo{year}{1970}) \bibinfo{pages}{501--509}.
\bibitem[{Bernstein et~al.(2005)Bernstein, Cruikshank, and
  Sandford}]{Bernstein:2005}
\bibinfo{author}{M.~P. Bernstein}, \bibinfo{author}{D.~P. Cruikshank},
  \bibinfo{author}{S.~A. Sandford},
\newblock \bibinfo{title}{Near-infrared laboratory spectra of solid
  h$_2$o/co$_2$ and ch$_3$oh/co$_2$ ice mixtures},
\newblock \bibinfo{journal}{Icarus} \bibinfo{volume}{179}
  (\bibinfo{year}{2005}) \bibinfo{pages}{527--534}.
\bibitem[{{Oancea} et~al.(2012){Oancea}, {Grasset}, {Le Menn}, {Bollengier},
  {Bezacier}, {Le Mou{\'e}lic}, and {Tobie}}]{Oancea2012}
\bibinfo{author}{A.~{Oancea}}, \bibinfo{author}{O.~{Grasset}},
  \bibinfo{author}{E.~{Le Menn}}, \bibinfo{author}{O.~{Bollengier}},
  \bibinfo{author}{L.~{Bezacier}}, \bibinfo{author}{S.~{Le Mou{\'e}lic}},
  \bibinfo{author}{G.~{Tobie}},
\newblock \bibinfo{title}{{Laboratory infrared reflection spectrum of carbon
  dioxide clathrate hydrates for astrophysical remote sensing applications}},
\newblock \bibinfo{journal}{Icarus} \bibinfo{volume}{221}
  (\bibinfo{year}{2012}) \bibinfo{pages}{900--910}.
\bibitem[{Grisolle et~al.(2011)Grisolle, Appéré, Schmitt, Beck, Brissaud, and
  Douté}]{Grisolle:2011}
\bibinfo{author}{F.~Grisolle}, \bibinfo{author}{T.~Appéré},
  \bibinfo{author}{B.~Schmitt}, \bibinfo{author}{P.~Beck},
  \bibinfo{author}{O.~Brissaud}, \bibinfo{author}{S.~Douté},
\newblock \bibinfo{title}{Influence of condensing water frost on the near-ir
  spectrum of co$_2$ snow}  (\bibinfo{year}{2011}).
\bibitem[{Philippe et~al.(2015)Philippe, Schmitt, Beck, and
  Brissaud}]{Sylvain:2015}
\bibinfo{author}{S.~Philippe}, \bibinfo{author}{B.~Schmitt},
  \bibinfo{author}{P.~Beck}, \bibinfo{author}{O.~Brissaud},
\newblock \bibinfo{title}{Sublimation of co$_2$ ice with h$_2$o ice
  contamination: analogy with the sublimation of mars seasonal caps}
  (\bibinfo{year}{2015}).
\bibitem[{Schmitt et~al.(2020)Schmitt, Philippe, Beck, and
  Brissaud}]{Bernard:2020}
\bibinfo{author}{B.~Schmitt}, \bibinfo{author}{S.~Philippe},
  \bibinfo{author}{P.~Beck}, \bibinfo{author}{O.~Brissaud},
\newblock \bibinfo{title}{Sublimation at grain boundaries of polycristalline
  co$_2$ slab ice: the clue to the strong spring albedo increase of the martian
  seasonal polar caps}  (\bibinfo{year}{2020}).
\bibitem[{Portyankina et~al.(2019)Portyankina, Merrison, Iversen, Yoldi,
  Hansen, Aye, Pommerol, and Thomas}]{Anya:2019}
\bibinfo{author}{G.~Portyankina}, \bibinfo{author}{J.~Merrison},
  \bibinfo{author}{J.~Iversen}, \bibinfo{author}{Z.~Yoldi},
  \bibinfo{author}{C.~Hansen}, \bibinfo{author}{K.-M. Aye},
  \bibinfo{author}{A.~Pommerol}, \bibinfo{author}{N.~Thomas},
\newblock \bibinfo{title}{Laboratory investigations of the physical state of
  co2 ice in a simulated martian environment},
\newblock \bibinfo{journal}{Icarus} \bibinfo{volume}{322}
  (\bibinfo{year}{2019}) \bibinfo{pages}{210 -- 220}.
\bibitem[{Yoldi et~al.(2020)Yoldi, Pommerol, Poch, and Thomas}]{Yoldi:2020}
\bibinfo{author}{Z.~Yoldi}, \bibinfo{author}{A.~Pommerol},
  \bibinfo{author}{O.~Poch}, \bibinfo{author}{N.~Thomas},
\newblock \bibinfo{title}{Reflectance study of ice and mars soil simulant
  associations – i. h$_2$o ice},
\newblock \bibinfo{journal}{Icarus}  (\bibinfo{year}{2020})
  \bibinfo{pages}{114169}.
\bibitem[{Pommerol et~al.(2019)Pommerol, Jost, Poch, Yoldi, Brouet,
  Gracia-Bern{\'a}, Cerubini, Galli, Wurz, Gundlach, Blum, Carrasco, Szopa, and
  Thomas}]{Pommerol:2019}
\bibinfo{author}{A.~Pommerol}, \bibinfo{author}{B.~Jost},
  \bibinfo{author}{O.~Poch}, \bibinfo{author}{Z.~Yoldi},
  \bibinfo{author}{Y.~Brouet}, \bibinfo{author}{A.~Gracia-Bern{\'a}},
  \bibinfo{author}{R.~Cerubini}, \bibinfo{author}{A.~Galli},
  \bibinfo{author}{P.~Wurz}, \bibinfo{author}{B.~Gundlach},
  \bibinfo{author}{J.~Blum}, \bibinfo{author}{N.~Carrasco},
  \bibinfo{author}{C.~Szopa}, \bibinfo{author}{N.~Thomas},
\newblock \bibinfo{title}{Experimenting with mixtures of water ice and dust as
  analogues for icy planetary material},
\newblock \bibinfo{journal}{Space Science Reviews} \bibinfo{volume}{215}
  (\bibinfo{year}{2019}) \bibinfo{pages}{37}.
\bibitem[{Forget et~al.(1995)Forget, Hansen, and Pollack}]{Forget:1995}
\bibinfo{author}{F.~Forget}, \bibinfo{author}{G.~B. Hansen},
  \bibinfo{author}{J.~B. Pollack},
\newblock \bibinfo{title}{Low brightness temperatures of martian polar caps:
  Co$_2$ clouds or low surface emissivity?},
\newblock \bibinfo{journal}{Journal of Geophysical Research: Planets}
  \bibinfo{volume}{100} (\bibinfo{year}{1995}) \bibinfo{pages}{21219--21234}.
\bibitem[{Kieffer et~al.(2006)Kieffer, Christensen, and Titus}]{Kieffer:2006}
\bibinfo{author}{H.~H. Kieffer}, \bibinfo{author}{P.~R. Christensen},
  \bibinfo{author}{T.~N. Titus},
\newblock \bibinfo{title}{Co$_2$ jets formed by sublimation beneath translucent
  slab ice in mars' seasonal south polar ice cap},
\newblock \bibinfo{journal}{Nature} \bibinfo{volume}{442}
  (\bibinfo{year}{2006}).
\bibitem[{Brouet(2013)}]{Yann:thesis}
\bibinfo{author}{Y.~Brouet}, \bibinfo{title}{Contribution r{\`a} la
  d{\'e}termination de la permittivit{\'e} des noyaux com{\'e}taires et des
  ast{\'e}roides}, \bibinfo{year}{2013}.
\bibitem[{Allen et~al.(1997)Allen, Morris, Lindstrom, Lindstrom, and
  Lockwood}]{Allen:1997}
\bibinfo{author}{C.~C. Allen}, \bibinfo{author}{R.~V. Morris},
  \bibinfo{author}{D.~J. Lindstrom}, \bibinfo{author}{M.~M. Lindstrom},
  \bibinfo{author}{J.~P. Lockwood},
\newblock \bibinfo{title}{Jsc mars-1: Martian regolith simulant},
\newblock \bibinfo{journal}{Lunar and Planetary Science}
  \bibinfo{volume}{XXVIII} (\bibinfo{year}{1997}).
\bibitem[{Wolff et~al.(2009)Wolff, Smith, Clancy, Arvidson, Kahre, Seelos~IV,
  Murchie, and Savijärvi}]{Wolff:2009}
\bibinfo{author}{M.~J. Wolff}, \bibinfo{author}{M.~D. Smith},
  \bibinfo{author}{R.~T. Clancy}, \bibinfo{author}{R.~Arvidson},
  \bibinfo{author}{M.~Kahre}, \bibinfo{author}{F.~Seelos~IV},
  \bibinfo{author}{S.~Murchie}, \bibinfo{author}{H.~Savijärvi},
\newblock \bibinfo{title}{Wavelength dependence of dust aerosol single
  scattering albedo as observed by the compact reconnaissance imaging
  spectrometer},
\newblock \bibinfo{journal}{Journal of Geophysical Research: Planets}
  \bibinfo{volume}{114} (\bibinfo{year}{2009}).
\bibitem[{Goetz et~al.(2010)Goetz, Pike, Hviid, Madsen, Morris, Hecht, Staufer,
  Leer, Sykulska, Hemmig, Marshall, Morookian, Parrat, Vijendran, Bos,
  El~Maarry, Keller, Kramm, Markiewicz, Drube, Blaney, Arvidson, Bell~III,
  Reynolds, Smith, Woida, Woida, and Tanner}]{Goetz:2010}
\bibinfo{author}{W.~Goetz}, \bibinfo{author}{W.~T. Pike},
  \bibinfo{author}{S.~F. Hviid}, \bibinfo{author}{M.~B. Madsen},
  \bibinfo{author}{R.~V. Morris}, \bibinfo{author}{M.~H. Hecht},
  \bibinfo{author}{U.~Staufer}, \bibinfo{author}{K.~Leer},
  \bibinfo{author}{H.~Sykulska}, \bibinfo{author}{E.~Hemmig},
  \bibinfo{author}{J.~Marshall}, \bibinfo{author}{J.~M. Morookian},
  \bibinfo{author}{D.~Parrat}, \bibinfo{author}{S.~Vijendran},
  \bibinfo{author}{B.~J. Bos}, \bibinfo{author}{M.~R. El~Maarry},
  \bibinfo{author}{H.~U. Keller}, \bibinfo{author}{R.~Kramm},
  \bibinfo{author}{W.~J. Markiewicz}, \bibinfo{author}{L.~Drube},
  \bibinfo{author}{D.~Blaney}, \bibinfo{author}{R.~E. Arvidson},
  \bibinfo{author}{J.~F. Bell~III}, \bibinfo{author}{R.~Reynolds},
  \bibinfo{author}{P.~H. Smith}, \bibinfo{author}{P.~Woida},
  \bibinfo{author}{R.~Woida}, \bibinfo{author}{R.~Tanner},
\newblock \bibinfo{title}{Microscopy analysis of soils at the phoenix landing
  site, mars: Classification of soil particles and description of their optical
  and magnetic properties},
\newblock \bibinfo{journal}{Journal of Geophysical Research: Planets}
  \bibinfo{volume}{115} (\bibinfo{year}{2010}).
\bibitem[{Pommerol et~al.(2015)Pommerol, Jost, Poch, El-Maarry, Vuitel, and
  Thomas}]{Pommerol:2015}
\bibinfo{author}{A.~Pommerol}, \bibinfo{author}{B.~Jost},
  \bibinfo{author}{O.~Poch}, \bibinfo{author}{M.~El-Maarry},
  \bibinfo{author}{B.~Vuitel}, \bibinfo{author}{N.~Thomas},
\newblock \bibinfo{title}{The sciteas experiment: Optical characterizations of
  sublimating icy planetary analogues},
\newblock \bibinfo{journal}{Planetary and Space Science}
  \bibinfo{volume}{109–110} (\bibinfo{year}{2015}) \bibinfo{pages}{106 --
  122}.
\bibitem[{Cerubini et~al.(2022)Cerubini, Pommerol, Yoldi, and
  Thomas}]{Cerubini:2022}
\bibinfo{author}{R.~Cerubini}, \bibinfo{author}{A.~Pommerol},
  \bibinfo{author}{Z.~Yoldi}, \bibinfo{author}{N.~Thomas},
\newblock \bibinfo{title}{Near-infrared reflectance spectroscopy of sublimating
  salty ice analogues. implications for icy moons},
\newblock \bibinfo{journal}{Planetary and Space Science} \bibinfo{volume}{211}
  (\bibinfo{year}{2022}) \bibinfo{pages}{105391}.
\bibitem[{Blackburn et~al.(2010)Blackburn, Bryson, Chevrier, Roe, and
  White}]{Blackburn:2010}
\bibinfo{author}{D.~G. Blackburn}, \bibinfo{author}{K.~L. Bryson},
  \bibinfo{author}{V.~F. Chevrier}, \bibinfo{author}{L.~A. Roe},
  \bibinfo{author}{K.~F. White},
\newblock \bibinfo{title}{Sublimation kinetics of co2 ice on mars},
\newblock \bibinfo{journal}{Planetary and Space Science} \bibinfo{volume}{58}
  (\bibinfo{year}{2010}) \bibinfo{pages}{780--791}.
\bibitem[{{Clark} and {Lucey}(1981)}]{clark1981}
\bibinfo{author}{R.~N. {Clark}}, \bibinfo{author}{P.~G. {Lucey}},
\newblock \bibinfo{title}{{Spectral Properties of Ice Mineral Mixtures:
  Implications on the Composition on the Galilean Satellites and Other Icy
  Bodies.}},
\newblock in: \bibinfo{booktitle}{Bulletin of the American Astronomical
  Society}, volume~\bibinfo{volume}{13}, p. \bibinfo{pages}{700}.
\bibitem[{Li et~al.(2021)Li, Yang, and Wang}]{Li:2021}
\bibinfo{author}{S.-M. Li}, \bibinfo{author}{K.-S. Yang},
  \bibinfo{author}{C.-C. Wang},
\newblock \bibinfo{title}{A semi-empirical model for predicting frost
  properties},
\newblock \bibinfo{journal}{Processes} \bibinfo{volume}{9}
  (\bibinfo{year}{2021}).
\bibitem[{{Fornasier, S.} et~al.(2015){Fornasier, S.}, {Hasselmann, P. H.},
  {Barucci, M. A.}, {Feller, C.}, {Besse, S.}, {Leyrat, C.}, {Lara, L.},
  {Gutierrez, P. J.}, {Oklay, N.}, {Tubiana, C.}, {Scholten, F.}, {Sierks, H.},
  {Barbieri, C.}, {Lamy, P. L.}, {Rodrigo, R.}, {Koschny, D.}, {Rickman, H.},
  {Keller, H. U.}, {Agarwal, J.}, {A\'{}Hearn, M. F.}, {Bertaux, J.-L.},
  {Bertini, I.}, {Cremonese, G.}, {Da Deppo, V.}, {Davidsson, B.}, {Debei, S.},
  {De Cecco, M.}, {Fulle, M.}, {Groussin, O.}, {G\"uttler, C.}, {Hviid, S. F.},
  {Ip, W.}, {Jorda, L.}, {Knollenberg, J.}, {Kovacs, G.}, {Kramm, R.},
  {K\"uhrt, E.}, {K\"uppers, M.}, {La Forgia, F.}, {Lazzarin, M.}, {Lopez
  Moreno, J. J.}, {Marzari, F.}, {Matz, K.-D.}, {Michalik, H.}, {Moreno, F.},
  {Mottola, S.}, {Naletto, G.}, {Pajola, M.}, {Pommerol, A.}, {Preusker, F.},
  {Shi, X.}, {Snodgrass, C.}, {Thomas, N.}, and {Vincent,
  J.-B.}}]{Fornasier:2015}
\bibinfo{author}{{Fornasier, S.}}, \bibinfo{author}{{Hasselmann, P. H.}},
  \bibinfo{author}{{Barucci, M. A.}}, \bibinfo{author}{{Feller, C.}},
  \bibinfo{author}{{Besse, S.}}, \bibinfo{author}{{Leyrat, C.}},
  \bibinfo{author}{{Lara, L.}}, \bibinfo{author}{{Gutierrez, P. J.}},
  \bibinfo{author}{{Oklay, N.}}, \bibinfo{author}{{Tubiana, C.}},
  \bibinfo{author}{{Scholten, F.}}, \bibinfo{author}{{Sierks, H.}},
  \bibinfo{author}{{Barbieri, C.}}, \bibinfo{author}{{Lamy, P. L.}},
  \bibinfo{author}{{Rodrigo, R.}}, \bibinfo{author}{{Koschny, D.}},
  \bibinfo{author}{{Rickman, H.}}, \bibinfo{author}{{Keller, H. U.}},
  \bibinfo{author}{{Agarwal, J.}}, \bibinfo{author}{{A\'{}Hearn, M. F.}},
  \bibinfo{author}{{Bertaux, J.-L.}}, \bibinfo{author}{{Bertini, I.}},
  \bibinfo{author}{{Cremonese, G.}}, \bibinfo{author}{{Da Deppo, V.}},
  \bibinfo{author}{{Davidsson, B.}}, \bibinfo{author}{{Debei, S.}},
  \bibinfo{author}{{De Cecco, M.}}, \bibinfo{author}{{Fulle, M.}},
  \bibinfo{author}{{Groussin, O.}}, \bibinfo{author}{{G\"uttler, C.}},
  \bibinfo{author}{{Hviid, S. F.}}, \bibinfo{author}{{Ip, W.}},
  \bibinfo{author}{{Jorda, L.}}, \bibinfo{author}{{Knollenberg, J.}},
  \bibinfo{author}{{Kovacs, G.}}, \bibinfo{author}{{Kramm, R.}},
  \bibinfo{author}{{K\"uhrt, E.}}, \bibinfo{author}{{K\"uppers, M.}},
  \bibinfo{author}{{La Forgia, F.}}, \bibinfo{author}{{Lazzarin, M.}},
  \bibinfo{author}{{Lopez Moreno, J. J.}}, \bibinfo{author}{{Marzari, F.}},
  \bibinfo{author}{{Matz, K.-D.}}, \bibinfo{author}{{Michalik, H.}},
  \bibinfo{author}{{Moreno, F.}}, \bibinfo{author}{{Mottola, S.}},
  \bibinfo{author}{{Naletto, G.}}, \bibinfo{author}{{Pajola, M.}},
  \bibinfo{author}{{Pommerol, A.}}, \bibinfo{author}{{Preusker, F.}},
  \bibinfo{author}{{Shi, X.}}, \bibinfo{author}{{Snodgrass, C.}},
  \bibinfo{author}{{Thomas, N.}}, \bibinfo{author}{{Vincent, J.-B.}},
\newblock \bibinfo{title}{Spectrophotometric properties of the nucleus of comet
  67p/churyumov-gerasimenko from the osiris instrument onboard the rosetta
  spacecraft},
\newblock \bibinfo{journal}{A\&A} \bibinfo{volume}{583} (\bibinfo{year}{2015})
  \bibinfo{pages}{A30}.
\bibitem[{Thomas et~al.(2017)Thomas, Cremonese, Ziethe, Gerber, Br{\"a}ndli,
  Bruno, Erismann, Gambicorti, Gerber, Ghose, Gruber, Gubler, Mischler, Jost,
  Piazza, Pommerol, Rieder, Roloff, Servonet, Trottmann, Uthaicharoenpong,
  Zimmermann, Vernani, Johnson, Pel{\`o}, Weigel, Viertl, De~Roux, Lochmatter,
  Sutter, Casciello, Hausner, Ficai~Veltroni, Da~Deppo, Orleanski, Nowosielski,
  Zawistowski, Szalai, Sodor, Tulyakov, Troznai, Banaskiewicz, Bridges, Byrne,
  Debei, El-Maarry, Hauber, Hansen, Ivanov, Keszthelyi, Kirk, Kuzmin, Mangold,
  Marinangeli, Markiewicz, Massironi, McEwen, Okubo, Tornabene, Wajer, and
  Wray}]{Thomas:2017}
\bibinfo{author}{N.~Thomas}, \bibinfo{author}{G.~Cremonese},
  \bibinfo{author}{R.~Ziethe}, \bibinfo{author}{M.~Gerber},
  \bibinfo{author}{M.~Br{\"a}ndli}, \bibinfo{author}{G.~Bruno},
  \bibinfo{author}{M.~Erismann}, \bibinfo{author}{L.~Gambicorti},
  \bibinfo{author}{T.~Gerber}, \bibinfo{author}{K.~Ghose},
  \bibinfo{author}{M.~Gruber}, \bibinfo{author}{P.~Gubler},
  \bibinfo{author}{H.~Mischler}, \bibinfo{author}{J.~Jost},
  \bibinfo{author}{D.~Piazza}, \bibinfo{author}{A.~Pommerol},
  \bibinfo{author}{M.~Rieder}, \bibinfo{author}{V.~Roloff},
  \bibinfo{author}{A.~Servonet}, \bibinfo{author}{W.~Trottmann},
  \bibinfo{author}{T.~Uthaicharoenpong}, \bibinfo{author}{C.~Zimmermann},
  \bibinfo{author}{D.~Vernani}, \bibinfo{author}{M.~Johnson},
  \bibinfo{author}{E.~Pel{\`o}}, \bibinfo{author}{T.~Weigel},
  \bibinfo{author}{J.~Viertl}, \bibinfo{author}{N.~De~Roux},
  \bibinfo{author}{P.~Lochmatter}, \bibinfo{author}{G.~Sutter},
  \bibinfo{author}{A.~Casciello}, \bibinfo{author}{T.~Hausner},
  \bibinfo{author}{I.~Ficai~Veltroni}, \bibinfo{author}{V.~Da~Deppo},
  \bibinfo{author}{P.~Orleanski}, \bibinfo{author}{W.~Nowosielski},
  \bibinfo{author}{T.~Zawistowski}, \bibinfo{author}{S.~Szalai},
  \bibinfo{author}{B.~Sodor}, \bibinfo{author}{S.~Tulyakov},
  \bibinfo{author}{G.~Troznai}, \bibinfo{author}{M.~Banaskiewicz},
  \bibinfo{author}{J.~C. Bridges}, \bibinfo{author}{S.~Byrne},
  \bibinfo{author}{S.~Debei}, \bibinfo{author}{M.~R. El-Maarry},
  \bibinfo{author}{E.~Hauber}, \bibinfo{author}{C.~J. Hansen},
  \bibinfo{author}{A.~Ivanov}, \bibinfo{author}{L.~Keszthelyi},
  \bibinfo{author}{R.~Kirk}, \bibinfo{author}{R.~Kuzmin},
  \bibinfo{author}{N.~Mangold}, \bibinfo{author}{L.~Marinangeli},
  \bibinfo{author}{W.~J. Markiewicz}, \bibinfo{author}{M.~Massironi},
  \bibinfo{author}{A.~S. McEwen}, \bibinfo{author}{C.~Okubo},
  \bibinfo{author}{L.~L. Tornabene}, \bibinfo{author}{P.~Wajer},
  \bibinfo{author}{J.~J. Wray},
\newblock \bibinfo{title}{The colour and stereo surface imaging system (cassis)
  for the exomars trace gas orbiter},
\newblock \bibinfo{journal}{Space Science Reviews} \bibinfo{volume}{212}
  (\bibinfo{year}{2017}) \bibinfo{pages}{1897--1944}.
\bibitem[{{Warren} et~al.(1990){Warren}, {Wiscombe}, and
  {Firestone}}]{Warren:1990}
\bibinfo{author}{S.~G. {Warren}}, \bibinfo{author}{W.~J. {Wiscombe}},
  \bibinfo{author}{J.~F. {Firestone}},
\newblock \bibinfo{title}{{Spectral albedo and emissivity of CO$_{2}$ in
  Martian polar caps: model results.}},
\newblock \bibinfo{journal}{Journal of Geophysical Research}
  \bibinfo{volume}{95} (\bibinfo{year}{1990}) \bibinfo{pages}{14717--14741}.
\bibitem[{Poch et~al.(2016)Poch, Pommerol, Jost, Carrasco, Szopa, and
  Thomas}]{Poch:2016}
\bibinfo{author}{O.~Poch}, \bibinfo{author}{A.~Pommerol},
  \bibinfo{author}{B.~Jost}, \bibinfo{author}{N.~Carrasco},
  \bibinfo{author}{C.~Szopa}, \bibinfo{author}{N.~Thomas},
\newblock \bibinfo{title}{Sublimation of water ice mixed with silicates and
  tholins: Evolution of surface texture and reflectance spectra, with
  implications for comets},
\newblock \bibinfo{journal}{Icarus} \bibinfo{volume}{267}
  (\bibinfo{year}{2016}) \bibinfo{pages}{154 -- 173}.
\bibitem[{Eluszkiewicz et~al.(2005)Eluszkiewicz, Moncet, Titus, and
  Hansen}]{Eluszkiewicz:2005}
\bibinfo{author}{J.~Eluszkiewicz}, \bibinfo{author}{J.-L. Moncet},
  \bibinfo{author}{T.~N. Titus}, \bibinfo{author}{G.~B. Hansen},
\newblock \bibinfo{title}{A microphysically-based approach to modeling
  emissivity and albedo of the martian seasonal caps},
\newblock \bibinfo{journal}{Icarus} \bibinfo{volume}{174}
  (\bibinfo{year}{2005}) \bibinfo{pages}{524 -- 534}. \bibinfo{note}{Mars Polar
  Science III}.

\end{thebibliography}







\end{document}